\documentclass[pdflatex,sn-mathphys-num]{sn-jnl}% Math and Physical Sciences Numbered Reference Style 
\usepackage{silence}
\ActivateWarningFilters[hyperreflevel]

\usepackage{natbib}
\usepackage[resetlabels]{multibib}
\newcites{methods}{References}
\newcites{app}{References}
\setcitestyle{authoryear}

\usepackage{etoolbox}
\apptocmd{\sloppy}{\hbadness 10000\relax}{}{}

\let\cite\citep

\usepackage{lmodern}

\usepackage[group-separator={,},group-minimum-digits={3}]{siunitx}
\usepackage{afterpage}
\usepackage{dsfont}
\usepackage[utf8]{inputenc} % allow utf-8 input
\usepackage[T1]{fontenc}    % use 8-bit T1 fonts
\usepackage{booktabs}       % professional-quality tables
\usepackage{url}  % Allow URLS t break.

\usepackage{breakurl}
\usepackage[breaklinks]{hyperref}
\usepackage{amsfonts}       % blackboard math symbols
\usepackage{nicefrac}       % compact symbols for 1/2, etc.
\usepackage{microtype}      % microtypography
\usepackage{lipsum}
\usepackage{fancyhdr}       % header
\usepackage{graphicx}       % graphics
\graphicspath{{media/}}     % organize your images and other figures under media/ folder
\usepackage{amsmath}
\usepackage{mathtools}
\usepackage[toc]{appendix}
\usepackage{minitoc}
\usepackage{caption}
\usepackage{subcaption}
\usepackage{textcomp}  % Load this before `gensymb`.
\usepackage{gensymb}
\usepackage{booktabs}
\usepackage{float}
\usepackage{placeins}
\usepackage{tabularx}
\usepackage{makecell}
\usepackage{rotating}
\usepackage{setspace}
\usepackage{parskip}

\usepackage{xcolor} 
\usepackage{xr}
\usepackage[capitalise,noabbrev]{cleveref}

\crefname{appendix}{Supplementary}{Supplementaries}
\Crefname{appendix}{Supplementary}{Supplementaries}

\AtBeginEnvironment{appendices}{\crefalias{section}{appendix}}
\AtBeginEnvironment{appendices}{\crefalias{subsection}{appendix}}
\AtBeginEnvironment{appendices}{\crefalias{subsubsection}{appendix}}

\newcommand\ddfrac[2]{\frac{\displaystyle #1}{\displaystyle #2}}

\newcommand{\CO}{CO}
\newcommand{\NO}{NO}
\newcommand{\NOb}{NO${}_2$}
\newcommand{\SOb}{SO${}_2$}
\newcommand{\Oc}{O${}_3$}
\newcommand{\PMa}{PM${}_1$}
\newcommand{\PMb}{PM${}_{2.5}$}
\newcommand{\PMc}{PM${}_{10}$}

\usepackage{enumitem}
\setlist[itemize]{leftmargin=1cm}

\renewcommand{\paragraph}[1]{\textbf{#1}}

\makeatletter
\let\@LN\relax
\makeatother

\usepackage{geometry}  % For consistent margins
\begin{document}
\doparttoc
\faketableofcontents

\title[\textbf{A Foundation Model for the Earth System}]{\textbf{A Foundation Model for the Earth System}}


\author[1,2]{\fnm{Cristian} \sur{Bodnar}}\email{cris@silurian.ai}
\equalcont{These authors contributed equally to this work.}

\author[1]{\fnm{Wessel P.} \sur{Bruinsma}}\email{wbruinsma@microsoft.com}
\equalcont{These authors contributed equally to this work.}

\author[1,3]{\fnm{Ana} \sur{Lucic}}\email{a.lucic@uva.nl}
\equalcont{These authors contributed equally to this work.}

\author[1]{\fnm{Megan} \sur{Stanley}}\email{meganstanley@microsoft.com}
\equalcont{These authors contributed equally to this work.}

\author[4]{\fnm{Anna} \sur{Vaughan}}\email{av555@cam.ac.uk}
\equalcont{These authors contributed equally to this work.}

\author[1,5]{\fnm{Johannes} \sur{Brandstetter}}
\author[1]{\fnm{Patrick} \sur{Garvan}}
\author[1]{\fnm{Maik} \sur{Riechert}}
\author[6]{\fnm{Jonathan A.} \sur{Weyn}}
\author[6]{\fnm{Haiyu} \sur{Dong}}
\author[2,7]{\fnm{Jayesh K.} \sur{Gupta}}
\author[6]{\fnm{Kit} \sur{Thambiratnam}}
\author[4]{\fnm{Alexander T.} \sur{Archibald}}
\author[8]{\fnm{Chun-Chieh} \sur{Wu}}
\author[1]{\fnm{Elizabeth} \sur{Heider}}
\author[1,3]{\fnm{Max} \sur{Welling}}
\author[1,4,9]{\fnm{Richard E.} \sur{Turner}}
\author*[1,10]{\fnm{Paris} \sur{Perdikaris}}\email{pgp@seas.upenn.edu}

\affil[1]{Primary affiliation: \orgdiv{Microsoft Research}, \orgname{AI for Science}}
\affil[2]{\orgdiv{Silurian AI}}
\affil[3]{\orgdiv{University of Amsterdam}}
\affil[4]{\orgdiv{University of Cambridge}}
\affil[5]{\orgdiv{JKU Linz}}
\affil[6]{\orgdiv{Microsoft Corporation}}
\affil[7]{Primary affiliation: \orgdiv{Microsoft Research}}
\affil[8]{\orgdiv{National Taiwan University}}
\affil[9]{\orgdiv{Alan Turing Institute}}
\affil[10]{\orgdiv{University of Pennsylvania}}

\abstract{
Reliable forecasts of the Earth system are crucial for human progress and safety from natural disasters.
Artificial intelligence offers substantial potential to improve prediction accuracy and computational efficiency in this field, however this remains underexplored in many domains. 
Here we introduce Aurora, a large-scale foundation model for the Earth system trained on over a million hours of diverse data.
Aurora outperforms operational forecasts for air quality, ocean waves, tropical cyclone tracks, and high-resolution weather forecasting at orders of magnitude smaller computational expense than dedicated existing systems.
With the ability to fine-tune Aurora to diverse application domains at only modest computational cost,
Aurora represents significant progress
in making actionable Earth system predictions accessible to anyone.
}




% Reduce bottom margin slightly to get keywords on the title page.
% \newgeometry{top=0.5cm, bottom=2cm, left=2cm, right=2cm}
\begingroup
\thispagestyle{empty}
\setlength{\parskip}{0pt}  % Fixes the space for `\maketitle`.
\maketitle
\endgroup
% \restoregeometry

% Move the figure here to show it at the top of the second page.
\begin{figure}[t]
    \centering
    \begin{subfigure}[t]{\linewidth}
        \centering
        \includegraphics[width=.7\textwidth]{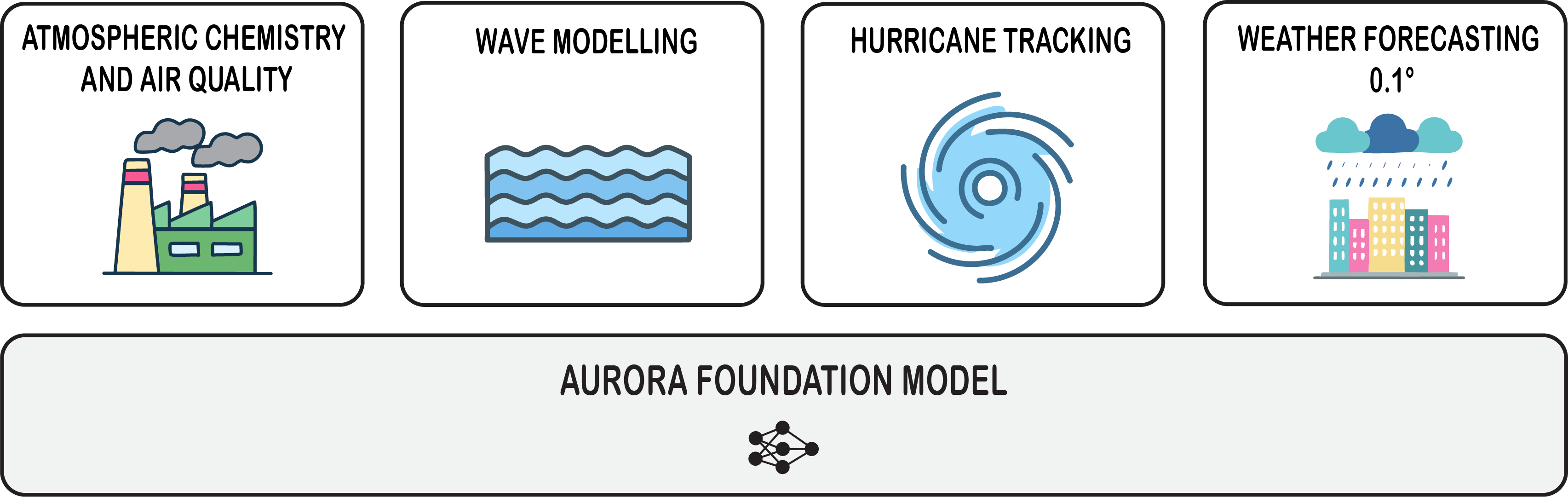}
        \vspace{-115pt}
        \caption{}
    \end{subfigure} \\
    \vspace{20pt}
    \begin{subfigure}[t]{\linewidth}
        \includegraphics[width=\textwidth]{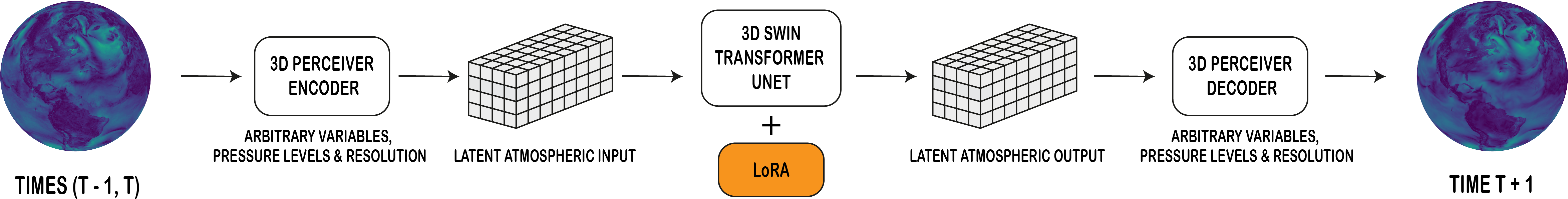}
        \vspace{-90pt}
        \caption{}
    \end{subfigure} \\
    \vspace{5pt}
    \caption{{\bf Aurora is a 1.3 billion parameter foundation model for the Earth system.}
        \textbf{a}:
            Aurora is pretrained on multiple heterogeneous datasets with different resolutions, variables, and pressure levels. The model is then fine-tuned for multiple operational forecasting scenarios at different resolutions: atmospheric chemistry and air quality at $0.4^\circ$, wave modelling at $0.25^\circ$, hurricane tracking at $0.25^\circ$, and weather forecasting at $0.1^\circ$. 
        \textbf{b}:
            Aurora is a flexible 3D Swin Transformer with 3D Perceiver-based encoders and decoders. The model is able to ingest inputs with different spatial resolutions, numbers of pressure levels, and variables. 
        }
    \label{fig:aurora_overview}
\end{figure}

\section{Introduction}

Earth system forecasts are indispensable tools for human societies, as evidenced by recent natural events such as the floods in Valencia, the air quality crisis in New Delhi, and hurricanes Helene and Milton in the eastern United States. Such systems not only provide crucial early warnings for extreme events, but are also invaluable for diverse fields ranging from agriculture to healthcare to global commerce. Modern Earth system predictions rely on complex models developed over centuries of accumulated physical knowledge, providing global forecasts of diverse variables for weather, air quality, ocean currents, sea ice, and hurricanes.

Despite their vital role, Earth system forecasting models face several limitations. They are computationally demanding, often requiring purpose-built supercomputers and dedicated engineering teams for maintenance. Their complexity, built up over years of development by large teams, complicates rapid improvements and necessitates substantial time and expertise for effective management. Finally, forecasting models incorporate numerous approximations, such as those for sub-grid scale processes, limiting accuracy. These challenges open the door for alternative approaches that may offer enhanced performance.

Machine learning provides an attractive toolbox for addressing these issues. Breakthroughs in multiple fields have shown that complex prediction systems can be streamlined with machine learning models that deliver superior outcomes \citep{abramson2024accurate,openai2024gpt4}. This concept was introduced to the Earth sciences as early as the 1990s, with pioneering work on neural networks \citep{rumelhart1986learning} applied to various Earth forecasting problems \citep{marzban1996neural,mccann1992neural,kuligowski1998experiments,kuligowski1998localized,spellman1999application,deo1998real,tangang1997forecasting,hsieh1998applying,kolehmainen2000forecasting}. However, these early models could not scale to replace full dynamical systems. In 2023, a breakthrough came with Pangu-Weather \citep{bi2023accurate}, where a neural network replaced a numerical solver, outperforming state-of-the-art forecasting systems and sparking a wave of AI-based weather prediction models \citep{chen2023fuxi,lam2023graphcast,han2024fengwu, chen2023fengwu}. These advancements have mostly centred on global medium-range weather forecasting at \SI{0.25}{\degree} resolution, leaving substantial gaps in other essential areas, including ocean dynamics, wave modelling, and atmospheric chemistry. Furthermore, the potential for machine learning to outperform complex extreme weather prediction systems, which currently rely on human analysis of a wide range of models, remains underexplored.

In this paper we introduce Aurora, a foundation model for the Earth system, capable of tackling a variety of forecasting tasks. Taking inspiration from recent successes of foundation models in other fields \citep{abramson2024accurate,openai2024gpt4}, we pretrain Aurora on over one million hours of Earth system data. We then fine-tune the model on a range of downstream tasks, demonstrating for the first time that an AI model can outperform multiple existing operational systems while also being orders of magnitude faster. Specifically, Aurora achieves state-of-the-art performance in the following critical forecasting domains:
\begingroup
\begin{itemize}
    % Make the list a little more compact. This setting only affects the list here.
    \setlength\itemsep{-.5em}
	\item 5-day global air pollution forecasts at $0.4$\degree{} resolution, outperforming resource-intensive numerical atmospheric chemistry simulations on 74\% of targets,
	\item 10-day global ocean wave forecasts at $0.25$\degree{} resolution, exceeding costly numerical models on 86\% of targets,
	\item 5-day tropical cyclone track forecasts, outperforming seven operational forecasting centres on 100\% of targets, and
	\item 10-day global weather forecasts at $0.1$\degree{} resolution, surpassing state-of-the-art NWP models on 92\% of targets while improving performance on extreme events.
\end{itemize}
\endgroup

\section{Aurora: a flexible 3D foundation model for the Earth system}
\label{section:intro}

Aurora is a machine learning model that can ingest and make forecasts for any collection of Earth system variables at any desired resolution. The model consists of three parts: (1) an encoder which converts heterogeneous inputs into a universal latent three-dimensional (3D) representation; (2) a processor which evolves the representation in time; and (3) a decoder which translates the standard 3D representation back into the desired  predictions. The processor is implemented as a 3D Swin Transformer~\cite{liu2021swin, dosovitskiy2020image} and the encoder and decoder as Perceiver-based modules~\cite{jaegle2021perceiver, jaegle2021perceiver_io} (\cref{fig:aurora_overview}). Forecasts for different lead times are generated by recursively feeding predictions back into the model as inputs. For a detailed discussion of the model, see \cref{sec:methods}.

We train Aurora on a vast body of Earth system data to learn a general-purpose representation of the dynamics that govern atmospheric and oceanic flow, and associated second-order processes. This first training phase is called \emph{pretraining} and includes a mixture of forecasts, analysis data, reanalysis data, and climate simulations (see \cref{app:dataset-inventory} for details).
After the model has been pretrained, a second training phase can leverage the learned general-purpose representations to efficiently adapt to new tasks, new datasets, and new variables. This second training phase is called \emph{fine-tuning}. Whereas pretraining is expensive and requires vast amounts of data, fine-tuning is much cheaper and  can typically be performed with little data.
We primarily pretrain on atmospheric data, because this is one of the largest sources of information about the dynamical processes underlying the Earth system.
Concretely, the pretraining objetive is to minimise the next time-step (6-hour lead time) mean absolute error (MAE) for \SI{150}{k} steps on 32 A100 GPUs, corresponding to approximately two and a half weeks of training.

Aurora is able to achieve unprecedented performance in fine-tuning tasks by simultaneously scaling the volume of data used during pretraining along with its model size. 
To evidence this scaling, we demonstrate that pretraining on more diverse data systematically improves validation performance as more datasets are added, especially for extreme values (see \cref{sec:scaling} for details).
Moreover,
we demonstrate that validation performance improves by approximately 6\% for every $10\times$ increase in model size (also \cref{sec:scaling}).
Finally, to measure the benefits of data and model scaling against existing numerical and AI models, we fine-tune Aurora for medium-range weather forecasting at \SI{0.25}{\degree} resolution, a task for which multiple state-of-the-art AI models are available for comparison.
Aurora outperforms both IFS \citep{ifs}, the state-of-the-art numerical weather prediction (NWP) system, and GraphCast \citep{lam2023graphcast} on over 91\% of all targets (see \cref{sm:0.25_extra_results} for details).

% \afterpage{
\begin{figure}[t]
    \begin{subfigure}[t]{.495\textwidth}
        \includegraphics[width=\textwidth]{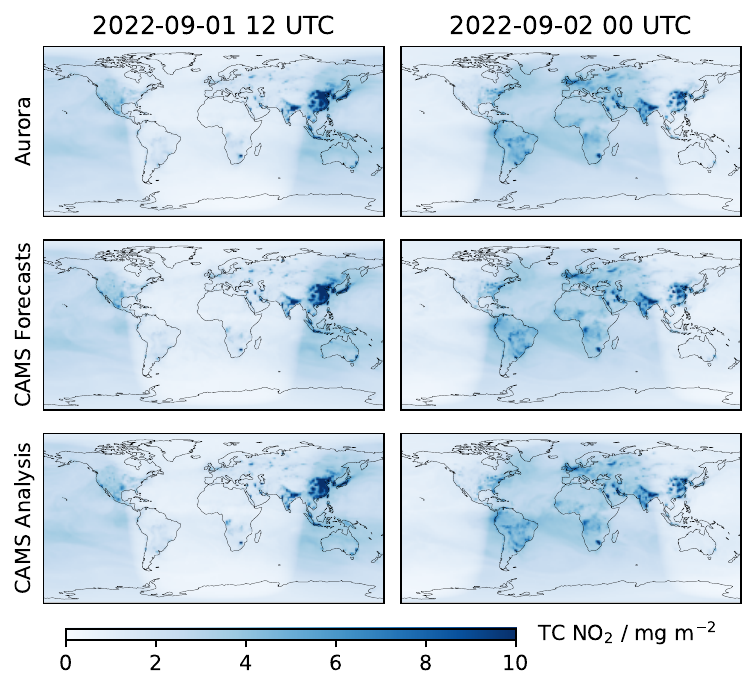}
        \vspace{-225pt}
        \caption{}
        \label{fig:cams-sample-predictions}
    \end{subfigure}\hfill%
    \begin{subfigure}[t]{.49\textwidth}
        \includegraphics[width=\textwidth]{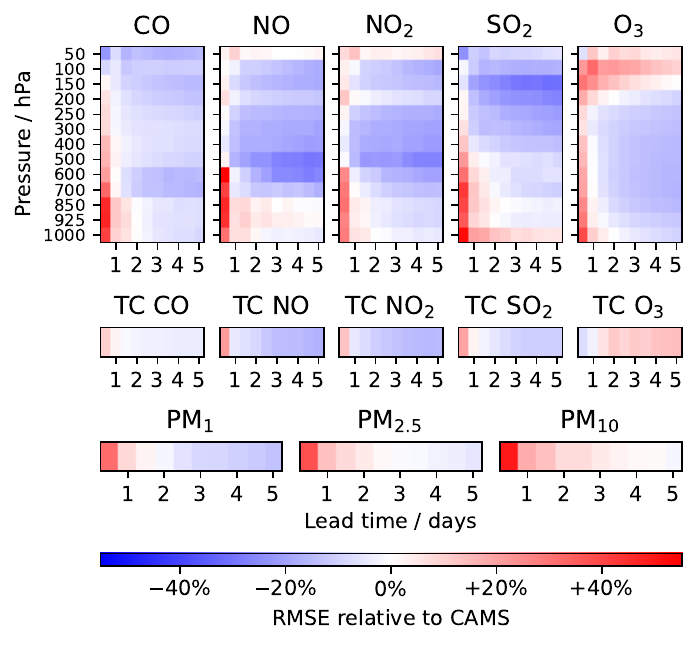}
        \vspace{-227pt}
        \caption{}
        \label{fig:cams-scorecard}
    \end{subfigure} \\
    \begin{subfigure}[t]{\textwidth}
        \centering%
        % \hspace{-25pt}
        \includegraphics[width=\columnwidth]{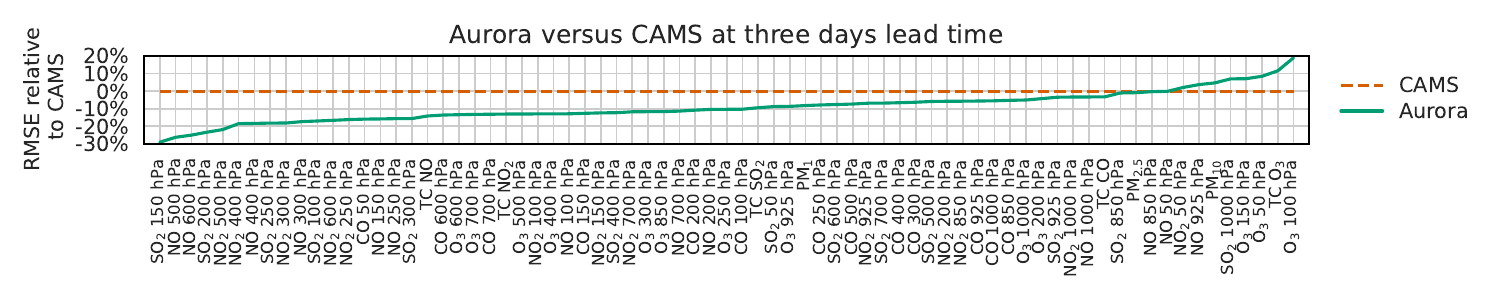}
        % \hspace{5pt}
        \vspace{-115pt}
        \caption{}
        \label{fig:cams-3-days}
    \end{subfigure}
    \caption{
        \textbf{In an operational setting, Aurora matches or outperforms CAMS in most comparisons, at orders of magnitude smaller computational expense.}
        \textbf{a}:
            Predictions for tropical cyclones \NOb~by Aurora accurately predict CAMS analysis.
            Predicting atmospheric gases correctly is extremely challenging due to their spatially heterogeneous nature. In particular, \NOb, like most air pollution variables, is skewed towards high values in areas with large anthropogenic emissions such as densely populated regions of East Asia.
            In addition, \NOb~exhibits a strong diurnal cycle; e.g., sunlight reduces background levels of \NOb~through a process called photolysis. Aurora accurately captures both the extremes and background levels. 
            Aurora and CAMS forecasts are initialised with CAMS analysis on 1 Sep 2022 at 00 UTC.
        \textbf{b}:
            Across all lead times, Aurora matches or outperforms CAMS on 74\% of all targets.
        \textbf{c}:
            At a lead time of three days, Aurora matches or outperforms CAMS on 89\% of all variables.
            See \cref{app:cams:full-results} for the full results.
    }
\end{figure}
% \clearpage
% }

\section{Modelling atmospheric chemistry and air quality}
\label{section:cams}

Air quality, a crucial factor in human health, is determined by atmospheric concentrations of specific gases and aerosols \cite{world2021global}. Accurately predicting global atmospheric composition can help mitigate the impact of air pollution events. However, forecasting atmospheric composition is significantly more complex and computationally costly than weather forecasting. It involves modelling complex chemical reactions through hundreds of stiff equations and accounting for anthropogenic emissions that drive heterogeneous pollution levels globally \citep{brasseur2017modeling}.
The Copernicus Atmosphere Monitoring Service (CAMS) takes this approach and produces global atmospheric composition forecasts and analysis products at \SI{0.4}{\degree} resolution and reanalysis products at \SI{0.75}{\degree} resolution \citep{cams}. To do this, CAMS extends the ECMWF's Integrated Forecasting System \citep[IFS;][]{ifs}, the top operational medium-range weather forecasting system in the world, with additional modules for aerosols, reactive gases, and greenhouse gases, which increases computational costs by approximately a factor of ten. To date, no AI method has attempted to produce operational predictions for global atmospheric composition at this scale.

Fine-tuning AI models on CAMS analysis data is extremely challenging for multiple reasons. First, the CAMS system is relatively new and frequently updated, so training data are limited and change in distribution. Second,  air pollution concentrations are highly heterogeneous, sparse, and have large dynamic ranges (\cref{fig:cams-sample-predictions}). Finally, pollution is driven by complex anthropogenic factors. These sources underwent complex changes during the global response to the COVID-19 pandemic, further complicating the available training data.

Six air pollutants are the main drivers of poor air quality \citep{world2021global}: carbon monoxide (\CO), nitrogen oxide (\NO), nitrogen dioxide (\NOb), sulphur dioxide (\SOb), ozone (\Oc), and particulate matter at \SI{1}{\micro m} (\PMa), \SI{2.5}{\micro m} (\PMb), and \SI{10}{\micro m} (\PMc). Air quality warnings are usually based on threshold values for \PMb~and \PMc.
Aurora models the five chemical species (\CO,~\NO,~\NOb,~\SOb, and \Oc) across atmospheric levels and as total column (TC) values and models the particulate matter variables, with CAMS analysis taken to be ground truth. We fine-tune Aurora on CAMS analysis data from Oct 2017 to May 2022 and test on CAMS analysis data from May 2022 to Nov 2022 (\cref{app:sec:splits}).
As the CAMS analysis dataset is very limited in temporal extent, we also incorporate lower resolution and CAMS reanalysis data \citep[EAC4;][]{cams-reanalysis} from Jan 2003 to Dec 2021 in the fine-tuning process. We note that CAMS reanalysis data is considered to be lower quality because it uses a lower resolution and a significantly older version of the underlying model (\cref{section:datasets}).

Aurora is competitive with CAMS (within 20\% RMSE) on 95\% of all targets, and matches or outperforms CAMS on 74\% of all targets (\cref{fig:cams-scorecard}).
At the three-day mark, Aurora is competitive with CAMS (within 20\% RMSE) on all variables, and matches or outperforms CAMS on 89\% of all variables (\cref{fig:cams-3-days}). 
CAMS outperforms Aurora on ozone in the very upper atmosphere and the twelve-hour prediction of all species in the lower part of the atmosphere. Aurora generates these prediction in approximately \SI{1.1}{s} per hour lead time on a single A100 GPU.
This yields roughly a $\times$\SI{50000}{} speed-up over CAMS \citep[see Section 2.1.5 in][for the cost of IFS]{buizza1028ifsupgrade},
representing an important  advancement in the field of atmospheric composition forecasting. Fine-tuning the pretrained model produces large gains over training a model from scratch, giving improvements for all targets with an average magnitude of 54\% (see \cref{fig:cams-scratch-scorecard}).

We conduct a case study evaluating Aurora's predictions for \PMc~on 13 June 2022, when Iraq was hit by a particularly severe sandstorm (\cref{fig:cams-iraq} in \cref{app:cams:full-results}), one of a series that led to  over \SI{5000}{} hospitalizations in the Middle East  \cite{francis2023middle}. Sandstorms involve complex interactions between particulate matter variables and atmospheric dynamics. Nevertheless,
 Aurora accurately predicts the sandstorm one day in advance with similar accuracy to CAMS, at a fraction of the computational cost. This case study shows that a foundation model approach for predicting air pollution can generalise to extreme events involving complex interactions between atmospheric dynamics and pollutants.

\begin{figure}[t]
    \begin{subfigure}[t]{\textwidth}
        \hspace{-8pt}
        \includegraphics[width=1.02\textwidth]{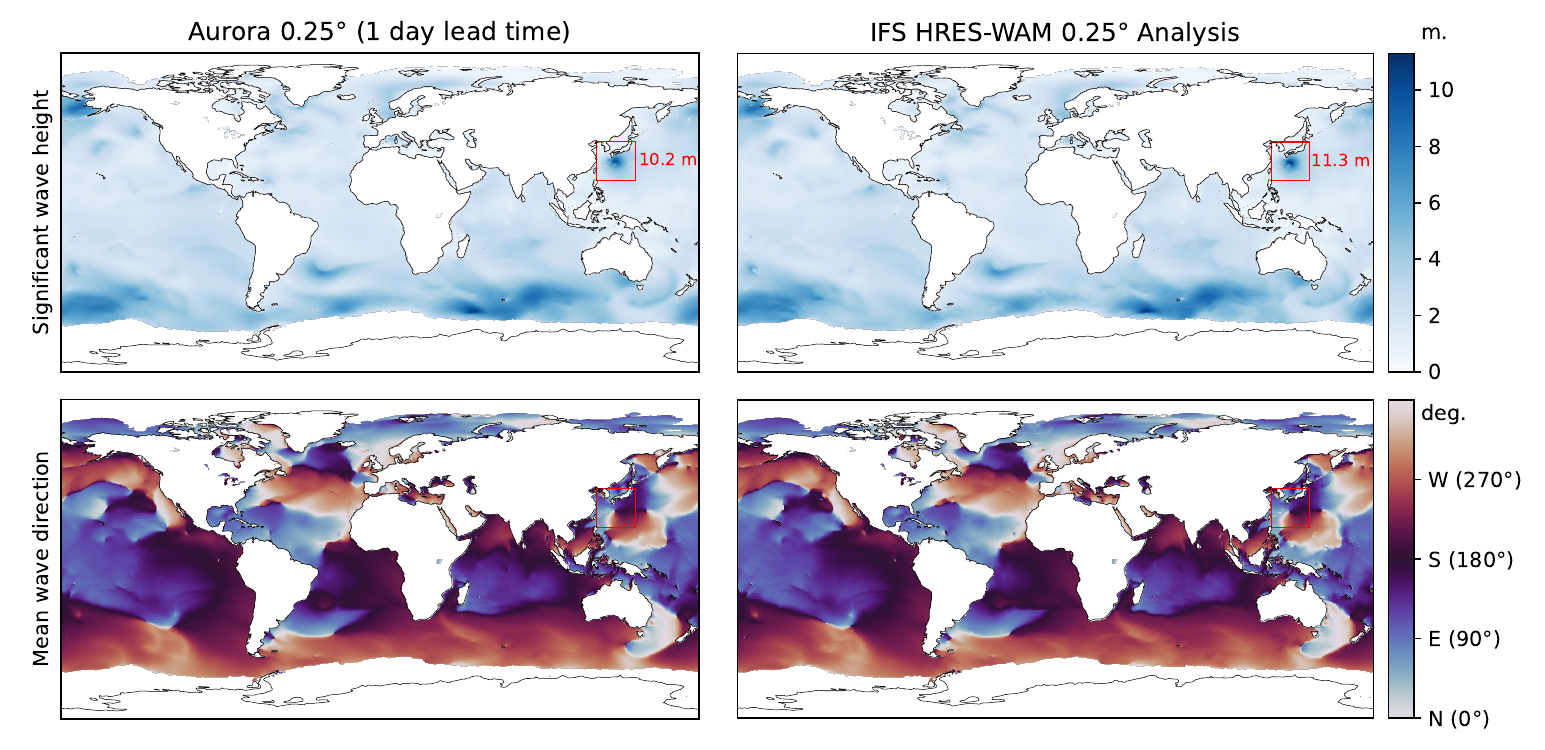}
        \vspace{-245pt}
        \caption{}
        \label{fig:wave-sample-predictions}
    \end{subfigure} \\
    \begin{subfigure}[t]{.63\textwidth}
        \hspace{6pt}\includegraphics[width=.95\textwidth]{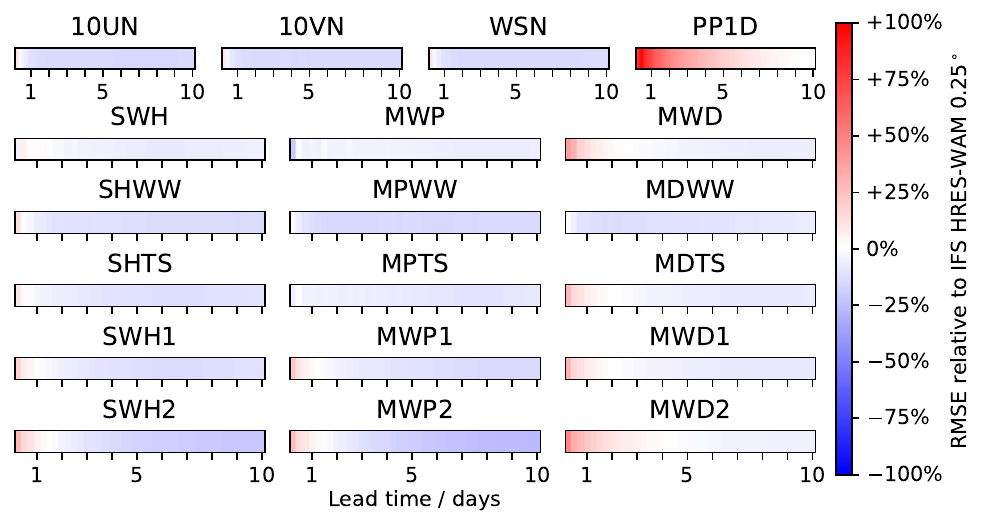}
        \vspace{-160pt}
        \caption{}
        \label{fig:wave-scorecard}
    \end{subfigure}
    \begin{subfigure}[t]{.36\textwidth}
        \hspace{-10pt}\includegraphics[width=\textwidth]{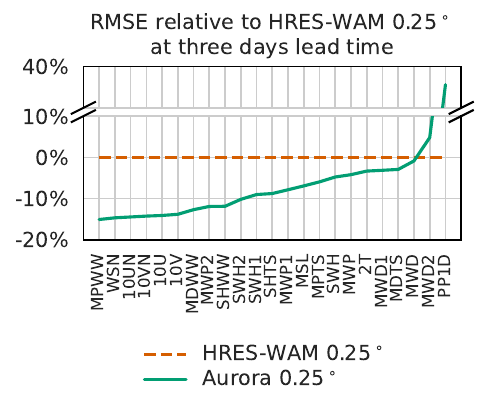}
        \vspace{-155pt}
        \caption{}
        \label{fig:wave-3-days}
    \end{subfigure}~\\
    \caption{
        \textbf{In an operational setting, Aurora matches or outperforms HRES-WAM in the majority of comparisons.}
        \textbf{a}:
            Aurora accurately predicts significant wave height and mean wave direction for Typhoon Nanmadol, the most intense tropical cyclone in 2022.
            The red box shows the location of the typhoon and the number is the peak significant wave height.
            Aurora's prediction and HRES-WAM Analysis are for 17 Sep 2022 at UTC 12, when Typhoon Nanmadol reached peak intensity.
            Aurora was initialised at 16 Sep 2022 at UTC 12.
        \textbf{b}:
            Across all lead times, Aurora matches or outperforms HRES-WAM on 86\% of all wave variables.
        \textbf{c}:
            At a lead time of three days, Aurora matches or outperforms HRES-WAM on 91\% of all surface-level variables.
            See \cref{app:wave:full-results} for the full results.
    }
\end{figure}

\section{Modeling ocean wave dynamics}
\label{section:waves}
Accurate ocean wave forecasts are critical for shipping, coastal defenses, aquaculture, off-shore energy generation and disaster preparedness. The IFS High RESolution WAve Model (HRES-WAM) system \citep{hres-wam-dataset}  produces state-of-the-art wave forecasts up to ten days lead time. IFS HRES-WAM extends the IFS by adding a coupled ocean wave module. No AI model has yet attempted to produce operational predictions for global wave forecasts at this scale.

Fine-tuning Aurora on ECMWF's HRES-WAM analysis dataset is challenging. Ocean wave variables include information about the direction, time periods, and spectral properties of waves, all of which is complex to model. 
Wave components can also be absent, meaning that the new variables can be undefined at arbitrary and variable spatial locations. Moreover, data for the variables that we consider in this experiment are only available back to 2016, a short record for such a complex task. 

The key variables in ocean wave modeling include significant wave height (SWH), mean wave period (MWP), and mean wave direction (MWD). 
 Each of these is predicted for wind waves ($\ast\ast$WW), total swell ($\ast\ast$TS), primary swell ($\ast\ast$$\ast$1) and secondary swell ($\ast\ast$$\ast$2). We also include peak wave period (PP1D) and the components of neutral wind \citep{ecmwf-neutral-wind:2017} at \SI{10}{m}, 10UN and 10VN (\cref{app:sec:processing}). For the full set of variables see \cref{tab:variables}. We simultaneously fine-tune Aurora on both wave and meteorological variables by lining up HRES-WAM analysis and HRES T0 in time. HRES T0 refers to the zero-hour forecasts of high-resolution configuration of the IFS \citep{malardel2016new}, which provides an accurate ground truth for a wide range of meteorological variables.
Both HRES-WAM analysis and HRES T0 are regridded to \SI{0.25}{\degree} spatial resolution. Since the HRES-WAM variables are undefined over land and over oceans whenever sea ice is present, we extend Aurora to support missing data \citepapp{Gordon_2020_Convolutional_Conditional_Neural_Processes} (see \cref{sec:methods:tasks} for details). We use the years 2016--2021 inclusive for fine-tuning, and evaluate on 2022 (\cref{app:sec:splits}).

Aurora is competitive with HRES-WAM (within 20\% RMSE) on 96\% of all targets, and matches or outperforms HRES-WAM on 86\% of all wave variables (\cref{fig:wave-scorecard}).
At the three day mark, Aurora is competitive with HRES-WAM (within 20\% RMSE) on all but one variable, PP1D, and matches or outperforms IFS HRES-WAM on 91\% of all variables (\cref{fig:wave-3-days}). In particular, fine-tuned Aurora shows good performance on prediction of neutral wind speeds, a critical variable for the coupling of atmospheric and wave models \citep{ecmwf-neutral-wind:2017}.

We condut a case study of Aurora's prediction of the significant wave height and mean wave direction during Typhoon Nanmadol, which struck the southern coast of Japan on September 19, 2022 (\cref{fig:wave-sample-predictions}).
Aurora generally produces strong global predictions for significant wave height and mean wave direction that follow the prevailing global wind patterns, with large waves in the typhoon accurately captured.

\section{Predicting tropical cyclone tracks}
\label{sec:tcs}

Tropical cyclones have caused more than 1.4 trillion USD in damage since 1950, and pose significant threats to lives and property \citep{wmo_tropical_cyclone}. Official forecasts of tropical cyclone tracks are vital for emergency services and the general public. These forecasts are produced by running multiple different dynamical and statistical models ranging from global ensembles such as the IFS to purpose-built tropical cyclone forecasting systems with vortex-following high-resolution grids such as the Hurricane Weather Research and Forecasting model \citep{NHC_TrackModels}. The output from these systems together with multiple consensus products is analysed by a team of human forecasters who create the final operational product. Here we demonstrate that a single, deterministic run of Aurora fine-tuned to HRES T0 at 0.25$^\circ$ (see \cref{sm:0.25_extra_results} for details) outperforms the track forecasts from these complex systems for multiple agencies on a dataset of all tropical cyclones globally in 2022--2023.

Previous comparisons of AI-based tropical cyclone forecasts to official operational forecasts have focused on forecasting track and intensity at short lead times of up to \SI{24}{h} \citep{boussioux2022hurricane,huang2022mmstn} and showed only marginal improvements at best. The analysis of other large-scale global machine learning models~\citep{lam2023graphcast,pathak2022fourcastnet,bi2022pangu} has been limited to comparisons of tracks produced by global models, with recent comparisons indicating that performance lags behind that of the official operational forecasts \citep{demaria2024}. 

\begin{figure}[t]
    \centering
    \begin{subfigure}[t]{\linewidth}
        \centering
        \includegraphics[width=\columnwidth]{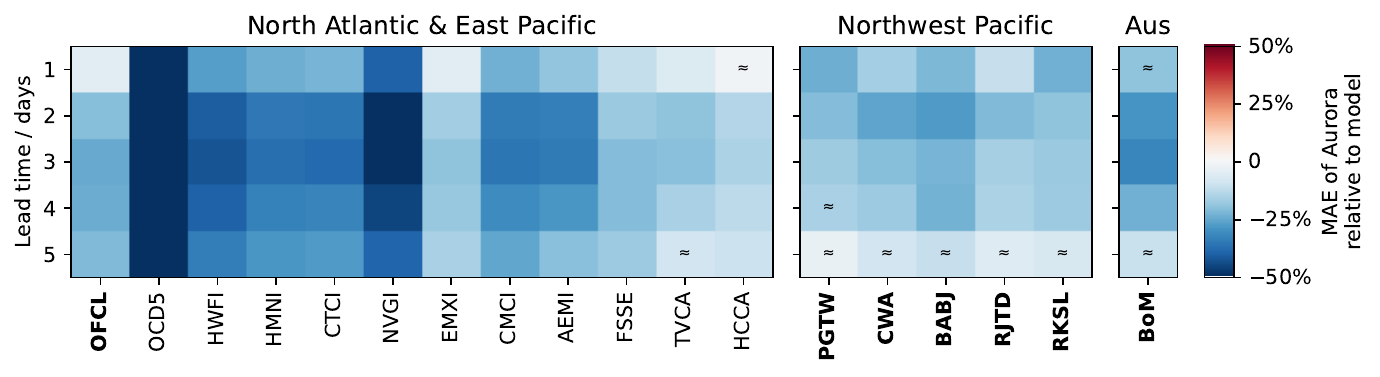}
        \vspace{-137pt}
        \caption{}
        \label{fig:tcs-scorecard}
    \end{subfigure}
    \begin{subfigure}[t]{\linewidth}
        \centering
        \includegraphics[width=\columnwidth]{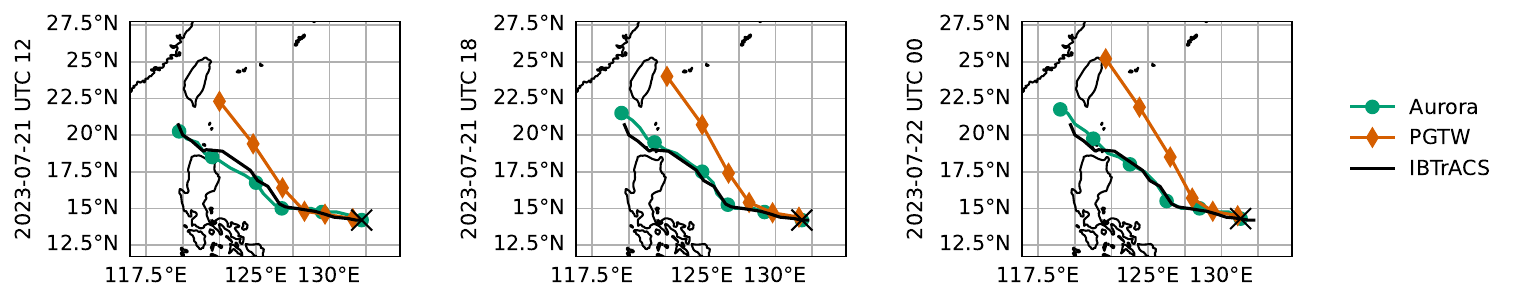}
        \vspace{-120pt}
        \caption{}
        \label{fig:tcs-prediction}
    \end{subfigure}
    \caption{
        \textbf{In an operational setting, Aurora outperforms state-of-the-art tropical cyclone prediction systems for multiple agencies and regions worldwide.}
        \textbf{a}:
            Aurora attains better track prediction MAE than multiple agencies in various regions. Official forecasts are given by OFCL, PGTW, CWA, BABJ, RJTD, RKSL, and BoM (bolded). For the North Atlantic and Eastern Pacific we additionally compare to various models utilised in creating OFCL (not bolded).
            A model does not always make forecasts, which means that different columns are computed over different data.
            Columns are therefore not indicative of model performance and only indicate the performance w.r.t.\ Aurora.
            Here ``$\approx$'' indicates that the 95\% confidence interval for the cell contains zero (see \cref{app:tcs:details-scorecard} for details).
            On average, Aurora is 20\% better than other agencies in the North Atlantic and East Pacific, 18\% in the West Pacific, and 24\% in the Australian region.
        \textbf{b}:
            On 21 July,
            a tropical depression intensified into a tropical storm and was named Typhoon Doksuri.
            Doksuri would become the costliest Pacific typhoon to date, inflicting more than 28 billion USD in damage. 
            Aurora correctly predicts that Doksuri will make landfall in the Northern Philippines, whereas PGTW predicts that it will pass over Taiwan.
    }
\label{fig:tcs}
\end{figure}

To generate the track forecasts with Aurora, we run a simple heuristic tracker labelling the centre fix of the vortex as the minimum mean sea level pressure in consecutive predictions (see \cref{app:tcs:tracker}).
We compare the Aurora track predictions to official forecasts for four basins worldwide, issued by the National Hurricane Center (North Atlantic and Eastern Pacific), China Meteorological Administration, Central Weather Administration Taiwan, Joint Typhoon Warning Centre and Japan Meteorological Agency (North West Pacific) and Australian Bureau of Meteorology (Australian region). For all agencies and lead-times Aurora outperforms the official track forecast (\Cref{fig:tcs-scorecard}). For example, in the North Atlantic and Eastern Pacific, we observe improvements of 6\% at lead time one day and 20-25\% at lead times two to five days. This is the first time that a machine learning model has surpassed full operational tropical cyclone forecasts up to five days. 

Inspection of forecasts reveals that Aurora is able to produce accurate forecasts for multiple high-impact events. For example, in the case of Typhoon Doksuri in 2023, Aurora accurately predicts landfall in the Philippines at four days lead time, in contrast to the official predictions centring the vortex off the coast of Northern Taiwan (\cref{fig:tcs-prediction}). It is additionally important to consider the performance of Aurora relative to the wider set of models available to the human forecasters to create the official forecast, as certain models outperform the official prediction at various lead times \citep{NHC_Verification_2022,NHC_Verification_2023}.
We therefore compare Aurora to the headline models in the NHC track verification report \citep{NHC_Verification_2022} for the North Atlantic and Eastern Pacific. Aurora outperforms all headline models (\Cref{fig:tcs-scorecard}), giving confidence that this is indeed a major step forward in tropical cyclone track forecast skill. 

\section{High-resolution operational weather forecasting}
\label{section:0.1}

To accurately resolve high-impact weather events such as severe storms, it is essential that weather prediction systems operate at a high spatial resolution to resolve processes occurring at smaller scales, such as convective and boundary layer effects. 
HRES \citep{malardel2016new}, 
the high-resolution configuration of the IFS,
operates on a Gaussian grid (TCo1279), which is approximately \SI{0.1}{\degree} in mid-latitudes.
In contrast, current state-of-the-art AI weather prediction models \cite{lam2023graphcast,bi2022pangu,pathak2022fourcastnet,chen2023fuxi,bonev2023spherical} can operate only at 0.25\degree~resolution.
The reason why state-of-the-art AI approaches are focussed on \SI{0.25}{\degree} is the wealth of high-quality data available at this resolution, whereas \SI{0.1}{\degree} data only goes back to 2016.
Here we demonstrate that a pretraining--fine-tuning protocol can be used to efficiently adapt Aurora to 
\SI{0.1}{\degree} and surpass the forecasting skill of IFS HRES under operational evaluation protocols.

We fine-tune Aurora to \SI{0.1}{\degree} IFS HRES analysis data, which span 2016--2022 (see \cref{sec:methods} and \cref{sec:training_methods} for details). For evaluation, we follow the operational protocol from \citet{benbouallegue2023it}, initialising Aurora with IFS HRES analysis and evaluating forecasts against IFS HRES analysis. To ensure that we do not disadvantage IFS HRES, we follow \cite{lam2023graphcast} and evaluate IFS HRES against its own so-called zero-hour forecast, referred to as HRES T0, instead of IFS HRES analysis.

\begin{figure}[t]
\centering
\begin{subfigure}[t]{.71\linewidth}
    \centering
    \includegraphics[width=\columnwidth]{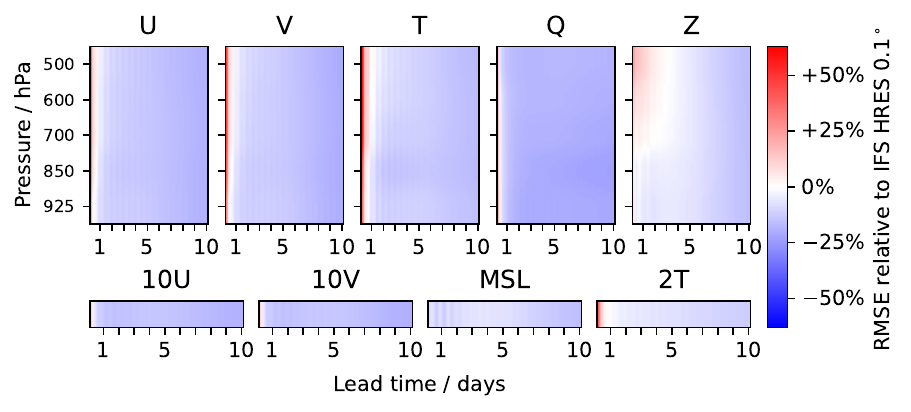}
    \vspace{-162pt}
    \caption{}
    \label{fig:0.1deg_scorecard}
\end{subfigure}
\hfill
\begin{subfigure}[t]{.28\linewidth}
    % \vspace{-10pt}
    \includegraphics[width=\linewidth]{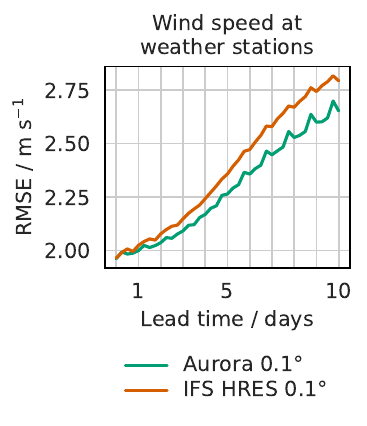}
    \vspace{-165pt}
    \caption{}
    \label{fig:weather_obs_0.1}
\end{subfigure}
\begin{subfigure}[t]{.71\linewidth}
    \centering
    \includegraphics[width=\columnwidth]{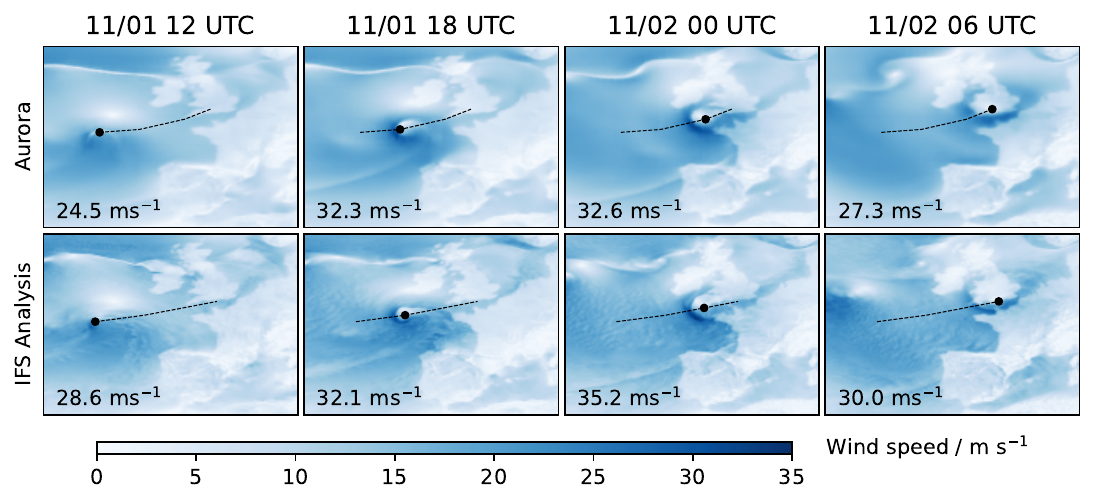}
    \vspace{-168pt}
    \caption{}
    \label{fig:ciaran-accuracy}
\end{subfigure}
\hfill
\begin{subfigure}[t]{.27\linewidth}
    \centering
    \includegraphics[width=\linewidth]{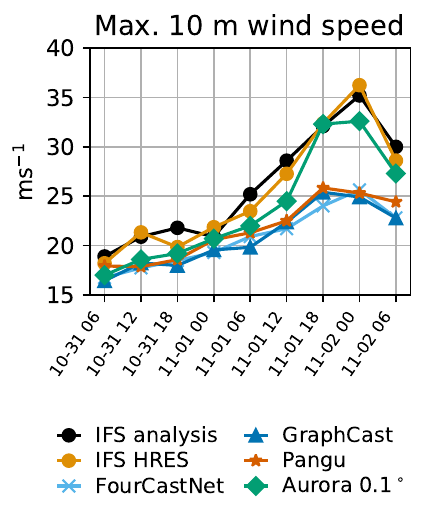}
    \vspace{-170pt}
    \caption{}
    \label{fig:ciaran-maxwindspeed-main}
\end{subfigure}
    \caption{
        \textbf{In an operational setting, Aurora outperforms IFS HRES at 0.1$\degree$ in the vast majority of comparisons. Aurora is the only AI model to accurately estimate maximum \SI{10}{m} wind speed in storm Ciar\'an.}
        \textbf{a}:
            Aurora outperforms IFS HRES on over 92\% of targets. The scorecard is limited to pressure levels lower in the atmosphere due to restricted  availability of test year data.
        \textbf{b}:
            Wind speed RMSE computed against measurements at weather stations.
            Aurora significantly outperforms IFS HRES. 
        \textbf{c}:
            Operational predictions for Storm Ciar\'an compared to IFS HRES analysis at $0.1^\circ$.
            Black dots show the location of minimum MSL and therefore trace the path of the storm. 
            The maximum \SI{10}{m} wind speed of the storm is shown in the bottom left corner of each prediction.
            To better facilitate the prediction of extreme events, Aurora was run without LoRA.
            See \cref{section:sm-ciaran}. 
        \textbf{d}:
            Operational predictions for maximum \SI{10}{m} wind speed during Storm Ciar\'an by Aurora, FourCastNet,  GraphCast, and Pangu-Weather.
            Aurora is able to predict the sudden increase in \SI{10}{m} wind speed, unlike the other AI models.
            The numbers for all AI models except Aurora have been extracted from Figure 3 by \citet{charltonperez2024ciaran}. 
    }
\label{fig:0.1-main}
\end{figure}

% \paragraph{Results.}
Aurora achieves lower RMSE than IFS HRES on  92\% of target variables, pressure levels and lead times (\cref{fig:0.1deg_scorecard}). The performance gains are most pronounced at lead times of more than 12 hours into the future, where we observe a reduction in RMSE up to 24\%. At the shortest lead times IFS HRES outperforms Aurora for many targets, as is the case for other AI models \citep{lam2023graphcast}. We additionally evaluate Aurora's forecasts on in-situ measurements of \SI{10}{m} wind speed and \SI{2}{m} temperature from the WeatherReal-ISD dataset \citeapp{weatherreal}, which includes over 13,000 weather observing stations globally. We find that Aurora outperforms IFS HRES for all lead times up to 10~days (see \cref{fig:weather_obs_0.1} and \cref{app:weather_station_obs}). Due to the limited availability of \SI{0.1}{\degree} data, we find that pretraining Aurora is essential in this application. On average, the pretrained model is better than training from scratch by 25\% (see \cref{sec:supp_0.1_results}).

We conduct a case study of Storm Ciar\'an, a high-impact mid-latitude storm which took place across North-West Europe in late 2023, resulting in the lowest recorded pressure in November in England \citepapp{ukmo2023ciaran}. Following \citet{charltonperez2024ciaran}, we initalise a selection of AIWP models at 31 Oct 00 UTC and compare them to Aurora (see \cref{fig:ciaran-maxwindspeed-main}). We observe that among the AIWP models tested \citep{pathak2022fourcastnet, lam2023graphcast,bi2022pangu} Aurora is the only one capable of accurately predicting the abrupt rise in maximum \SI{10}{m} wind speed, closely matching IFS analysis, which is taken to be the ground truth.

\section{Discussion}
\label{sec:discussion}

We have introduced Aurora, a large-scale foundation model for the Earth system that outperforms multiple specialised operational prediction systems at a fraction of the computational cost. We demonstrate state-of-the-art results for air quality, ocean waves, tropical cyclone tracks, and high-resolution weather forecasting.
From start to finish, every fine-tuning experiment took a small team of engineers and 4--8 weeks each to conceptualise, prepare the data, train the model, and process the results, compared to a development period of typically multiple years for dynamical baseline models.
It should be noted, however, that such an accelerated timeline is only possible because of the wealth of data that is available as a result of decades of research into traditional numerical approaches.

Improvements are possible along multiple axes. To begin with, Aurora can easily be extended to generate an ensemble of forecasts, which are crucial in situations where predictions are uncertain, such as for forecasts at longer lead times or for localised phenomena like precipitation \cite{price2023gencast,lessig2023atmorep,li2024generative,chen2023fengwu,zhong2024fuxi}.
Moreover, our scaling results suggest that we have not yet hit a performance ceiling, and that better fine-tuning results can be obtained by scaling pretraining to more diverse data and scaling Aurora to even larger sizes.
Finally, whereas Aurora is fully operational in all experiments,
the model does still rely on initial conditions from traditional data assimilation systems.
Following recent advances in end-to-end weather forecasting \citep{vaughan2024aardvarkweatherendtoenddatadriven}, Aurora could be extended to directly operate on observational data.

The potential implications of Aurora for the field of Earth system prediction are profound. 
Whilst in this paper we showcase the application of Aurora to four domains, it could be fine-tuned for any desired Earth system prediction task, potentially producing forecasts that outperform the current operational systems at a fraction of the cost. For example prediction of
ocean circulation,
local and regional weather,
seasonal weather,
vegetation growth and phenology,
extreme weather modalities such as floods and wildfires,
pollination patterns, agricultural productivity, renewable energy production, and sea ice extent. 
With the ability to fine-tune Aurora to diverse application domains at only modest computational cost,
Aurora represents significant progress
in making actionable predictions accessible to anyone.

\bibliography{bibliography}
\clearpage
\backmatter

% From here on, disallow the cute of \cite, \citet, and \citep. Instead,
% require \citemethods, \citetmethods, and \citepmethods.

\let\citeold\cite
\let\citemethodsold\citemethods
\renewcommand{\cite}[1]{%
    \ifcsname incitemethods\endcsname\else%
    \PackageError{main}{Use citemethods instead of cite}{}%
    \fi%
    \citeold{#1}%
}
\renewcommand{\citemethods}[1]{%
    \begingroup%
    \newcommand{\incitemethods}{}%
    \citemethodsold{#1}%
    \endgroup%
}

\let\citepold\citep
\let\citepmethodsold\citepmethods
\renewcommand{\citep}[1]{%
    \ifcsname incitepmethods\endcsname\else%
    \PackageError{main}{Use citepmethods instead of citep}{}%
    \fi%
    \citepold{#1}%
}
\renewcommand{\citepmethods}[1]{%
    \begingroup%
    \newcommand{\incitepmethods}{}%
    \citepmethodsold{#1}%
    \endgroup%
}

\let\citetold\citet
\let\citetmethodsold\citetmethods
\renewcommand{\citet}[1]{%
    \ifcsname incitetmethods\endcsname\else%
    \PackageError{main}{Use citetmethods instead of citet}{}%
    \fi%
    \citetold{#1}%
}
\renewcommand{\citetmethods}[1]{%
    \begingroup%
    \newcommand{\incitetmethods}{}%
    \citetmethodsold{#1}%
    \endgroup%
}

\section{Methods}\label{sec:methods_section}

\paragraph{Problem statement.}
We represent the observed state of the atmosphere and surface at a discrete time $t$ as a multi-dimensional array $X^t \in \mathbb{R}^{V \times H \times W}$, where $V$ is the total number of variables and $H$ and $W$ are the number latitude and longitude coordinates, respectively. The state can be split into surface ($S^t$) and atmospheric ($A^t$) components:
$X^t = (S^t, A^t)$ where $S^t \in \mathbb{R}^{V_S \times H \times W}$ and $A^t \in \mathbb{R}^{V_A C \times H \times W}$ with $V_S$ the number of surface-level variables, $V_A$ the number of atmospheric variables, and $C$ the number of pressure levels.
The goal is to predict a future state at time $t' > t$.
We learn a simulator $\Phi\colon(\mathbb{R}^{V \times H \times W})^2 \to \mathbb{R}^{V \times H \times W}$, $\Phi(X^{t-1}, X^t) = \hat{X}^{t + 1}$, which maps the observed states at the previous time $X^{t-1}$ and current time $X^{t}$ to a predicted state $\hat{X}^{t + 1}$ at the next time step.
For predictions at later time steps, we repeatedly apply the simulator, producing an autoregressive roll-out:
\begin{align*}
    \Phi(X^{t}, \hat X^{t+1}) &= \hat{X}^{t + 2}, \\
    \Phi(\hat X^{t+ 1}, \hat X^{t+2}) &= \hat{X}^{t + 3},\\
    &\;\vdots\\
   \Phi(\hat{X}^{t + k - 2}, \hat{X}^{t + k - 1}) &= \hat{X}^{t + k}.
\end{align*}
For a detailed description of the notation and problem statement, including the specific multi-dimensional array dimensions and variable definitions, please refer to \cref{sec:notation}.

\subsection{The Aurora model}

\begin{figure}[t]
    \centering
    \begin{subfigure}[t]{\linewidth}
        \includegraphics[width=\textwidth]{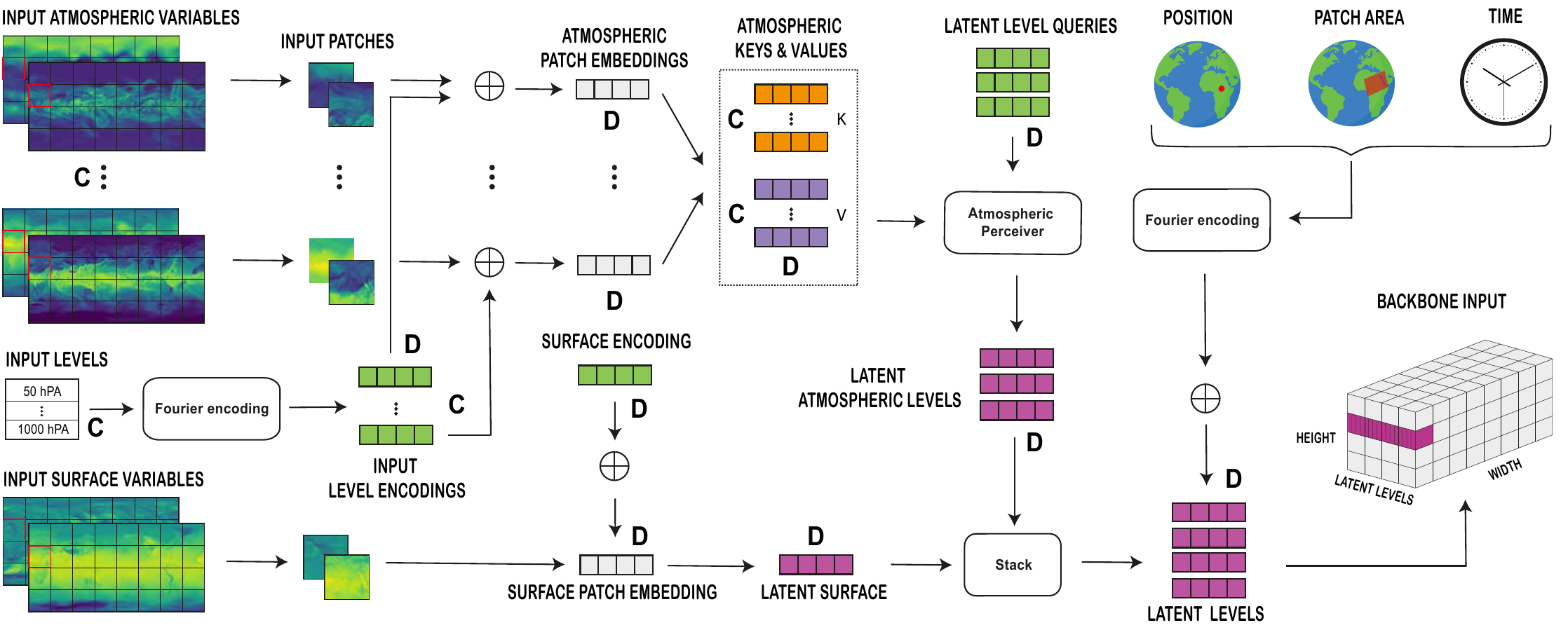}
        \vspace{-180pt}
        \caption{}
        \label{fig:aurora_encoder}
    \end{subfigure}
    \begin{subfigure}[t]{\linewidth}
        \includegraphics[width=\textwidth]{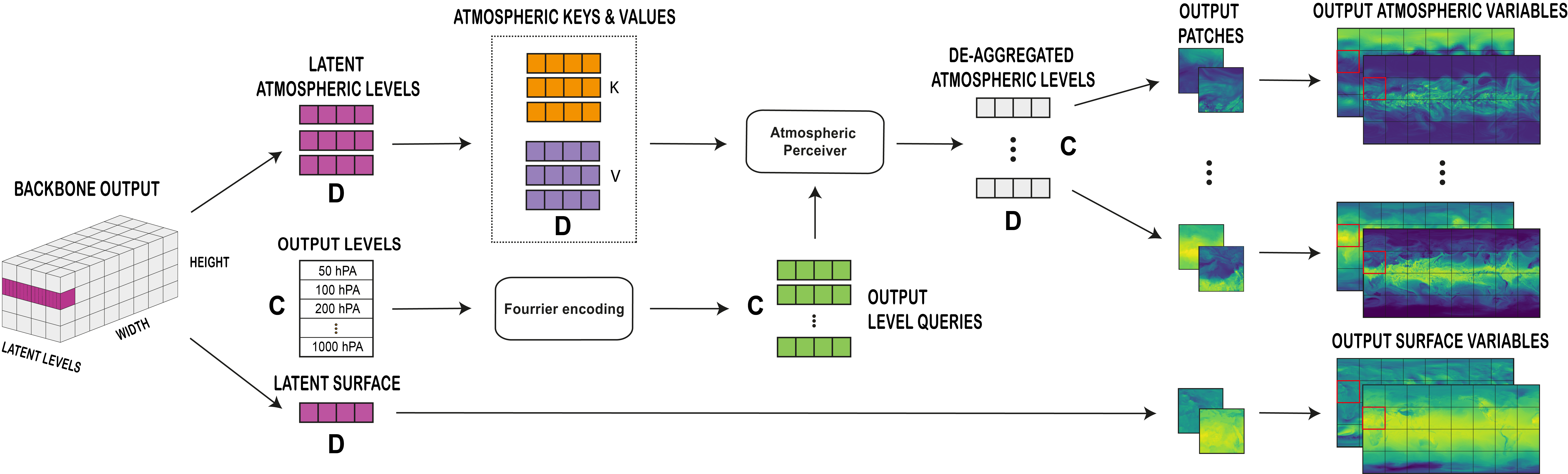}
        \vspace{-135pt}
        \caption{}
        \label{fig:aurora_decoder}
    \end{subfigure}
    \caption{
        \textbf{Aurora is an encoder--decoder model with a 3D latent representation.}
        \textbf{a}: Aurora's encoder module. Input weather states are tokenized and compressed into a 3D latent representation using Perceiver-style cross-attention blocks. The resulting latent tokens are augmented with appropriate encodings that provide spatial, temporal, and scale information.
        \textbf{b}: Aurora's decoder module. The target output variables are reconstructed in spatial patches by decoding Aurora's 3D latent state using Perceiver-style cross-attention blocks.
    }
\end{figure}

\paragraph{3D Perceiver encoder.}
To accommodate heterogeneous weather datasets with varying variables, pressure levels, and resolutions, we design a flexible encoder that maps different datasets into a standardized 3D representation for input into the model backbone; see \cref{fig:aurora_encoder}.

The encoder treats all variables as $H \times W$ images.
We incorporate static variables (orography, land-sea mask, and soil-type mask) by treating them as extra surface-level variables.
The images are split into $P \times P$ patches and the patches are mapped to embedding vectors of dimension $D$ using variable-specific linear transformations.
For the surface and every pressure level, the embeddings of different variables are summed and tagged with an additive encoding of the pressure level or a learned vector for the surface. 
A Perceiver module \citepmethods{jaegle2021perceiver} then reduces variable numbers of physical pressure levels $C$ to a fixed number $L=3$ of latent pressure levels. 
The result is a $\tfrac HP \times \tfrac WP \times L$ collection of embeddings.
This 3D representation is tagged with additive encodings for the patch position, patch area, and absolute time.
These encodings use a Fourier expansion scheme with carefully chosen minimum and maximum wavelengths to capture relevant information at appropriate scales.
The patch area encoding enables Aurora to operate at different resolutions.

For a detailed description of the encoder architecture, including specifics on input processing, pressure level aggregation, and additional encodings, please refer to  \cref{sec:encoder,sec:pos_encoding}.

\paragraph{Multi-scale 3D Swin Transformer U-Net backbone.}
The backbone of Aurora is a 3D Swin Transformer U-Net \citemethods{liu2021swin, liu2022swin}, which serves as a neural simulator (\cref{fig:backbone}). This architecture allows for efficient simulation of underlying physics at multiple scales. This architecture falls under the general family of Vision Transformers. However, unlike classical ViTs, here we employ local self-attention operations within windows and a symmetric upsampling-downsampling structure.

The backbone is characterized by the following key features: a symmetric upsampling--downsampling structure with three stages each, enabling multi-scale processing; 3D Swin Transformer layers performing local self-attention operations within windows, emulating local computations in numerical integration methods;
window shifting every other layer to propagate information between neighboring regions while accounting for Earth's spherical topology;
res-post-norm layer normalization \citemethods{liu2022swin} for increased training stability;
and a flexible design allowing operation at multiple resolutions without fixed positional biases.

Our backbone contains 48 layers across three stages, compared to the 16 layers and two stages employed in \citetmethods{bi2022pangu}. This increased depth is made possible by our efficient encoding procedure, which uses a small number of latent levels. For detailed information on the backbone architecture, including window sizes, attention mechanisms, and comparisons with previous work, please refer to \cref{sec:backbone}.

\paragraph{3D Perceiver decoder.}
The decoder reverses the operations of encoder, converting the output of the backbone, again a 3D representation, back to the regular latitude--longitude grid (see \cref{fig:aurora_decoder}). This involves deaggregating the latent atmospheric pressure levels using a Perceiver layer \citemethods{jaegle2021perceiver} to any desired collection of pressure levels and dynamically decoding into patches via variable-specific linear layers. 
For a detailed description of the decoder architecture, please refer to \cref{sec:decoder}.

\subsection{Training methods}
\label{sec:training_methods}

The overall training procedure is composed of three stages: (1) pretraining, (2) short--lead-time fine-tuning, and (3) roll-out (long--lead-time) fine-tuning. We provide an overview for each of these stages in the following paragraphs. 

\paragraph{Training objective.}
Throughout pretraining and fine-tuning, we use the mean absolute error (MAE) as our training objective $\mathcal{L}(\hat{X}^t, X^t)$. Decomposing the predicted state $\hat X^t$ and ground-truth state $X^t$ into surface-level variables and atmospheric variables, $\hat{X}^t = (\hat{S}^t, \hat{A}^t)$ and $X^t = (S^t, A^t)$ (see \cref{sec:notation}), the loss can be written as
\begin{align}
    \mathcal{L}(\hat X^t, X^t)
    &= \frac{\gamma}{V_S + V_A} \Bigg[\alpha \Bigg( \sum_{k=1}^{V_S} \frac{w^S_k}{H \times W}\sum_{i=1}^{H} \sum_{j=1}^{W} | \hat{S}^t_{k, i, j} - S^t_{k, i, j} | \Bigg) \nonumber \\
    &\hspace{-40pt}\hphantom{= \frac{1}{(V_S + V_A)} \Bigg[\alpha \Bigg( }
    + \beta \Bigg( \sum_{k=1}^{V_A} \frac{1}{C \times H \times W}  \sum_{c=1}^{C} w^A_{k,c} \sum_{i=1}^{H} \sum_{j=1}^{W} | \hat{A}^t_{k, c, i, j} - A^t_{k, c, i, j} |\Bigg)\Bigg],
\end{align}
where $w^S_k$ is the weight associated with surface-level variable $k$, $w^A_{k,c}$ is the weight associated with atmospheric variable $k$ at pressure level $c$, $\alpha$ is a weight for the surface-level component of the loss, $\beta$ is a weight for the atmospheric component of the loss, and $\gamma$ is a dataset-specific weight.
See \cref{sec:training_objective} for additional details.

\paragraph{Pretraining methods.}
All models are pretrained for \SI{150}{k} steps on 32 A100 GPUs, with a batch size of one per GPU. We use a (half) cosine decay with a linear warm-up from zero for \SI{1}{k} steps. 
The base learning rate is $5\mathrm{e}{-4}$, which the schedule reduces by a factor $10$ at the end of training.
The optimizer we use is AdamW \citepmethods{loshchilov2018decoupled}. We set the weight decay of AdamW to $5\mathrm{e}{-6}$. The only other form of regularisation we use is drop path (i.e., stochastic depth) ~\citepmethods{larsson2017fractalnet}, with the drop probability set to $0.2$. To make the model fit in memory, we use activation checkpointing for the backbone layers and we shard all the model gradients across the GPUs. The model is trained using \texttt{bf16} mixed precision.
See \cref{sec:pretraining_methods} for additional details. 

\paragraph{Short--lead-time fine-tuning.}
After pretraining Aurora, for each task that we wish to adapt Aurora to, we start by fine-tuning the entire architecture through one or two roll-out steps (depending on the task and its memory constraints), see  \cref{sec:short_leadtine_finetuning} for additional details.

\paragraph{Roll-out fine-tuning.}
To train very large Aurora models on long-term dynamics efficiently, even at high resolutions, we develop a novel roll-out fine-tuning approach.
Our approach uses Low Rank Adaptation (LoRA) \citepmethods{hu2021lora} to fine-tune all linear layers in the backbone's self-attention operations, allowing adaptation of very large models in a data and parameter efficient manner.
To save memory, we employ the ``pushforward trick'' \citepmethods{brandstetter2022message}, which propagates gradients only through the last roll-out step.
Finally, to enable training at very large numbers of roll-out steps without compromising memory or training speed, we use an in-memory replay buffer, inspired by deep reinforcement learning \citepmethods{lin_experience_replay, Mnih2015HumanlevelCT} (see \cref{fig:rollout_finetuning}).
The replay buffer samples initial conditions, computes predictions for the next time step, adds predictions back to the replay buffer, and periodically refreshes the buffer with new initial conditions from the dataset.
For detailed roll-out protocols for each fine-tuning task, please refer to \cref{sec:rollout_finetuning}.

\subsection{Datasets}
Aurora was trained and evaluated using a diverse set of weather and climate datasets, encompassing five main categories: analysis, reanalysis, forecast, reforecast, and climate simulation datasets. This variety of data sources exposes Aurora to different aspects of atmospheric dynamics, reflecting variability in initial conditions, model parametrizations, and chaotic dynamics. Key datasets used in our experiments include ERA5 reanalysis, 
HRES operational forecasts, IFS ensemble forecasts, 
GFS operational forecasts, 
GEFS ensemble reforecasts, CMIP6 climate simulations, MERRA-2 atmospheric reanalysis, as well as CAMS forecasts, analysis, and reanalysis data. For a detailed inventory of all datasets used, including specific pressure levels, resolutions, and additional context for each dataset, please refer to \cref{section:datasets}. These datasets vary in resolution, variables included, and temporal coverage, providing a comprehensive basis for training, fine-tuning and evaluating Aurora's performance across different scenarios.

\subsection{Task-specific adaptations}
\label{sec:methods:tasks}
\paragraph{Ocean wave forecasting.}
In the IFS HRES-WAM analysis data, there is a spatially-varying absence of data reflecting the distribution of sea-ice among other effects. To account for this dynamic nature of the spatial distribution of defined variables, we give each variable an additional channel to represent the presence of a measurement, so we add an additional set of density variables \citepmethods{Gordon_2020_Convolutional_Conditional_Neural_Processes} (\cref{sm:wave-extension}).

\subsection{Data infrastructure}

Training Aurora presented significant technical challenges due to the large size of individual datapoints (nearly \SI{2}{GB} for \SI{0.1}{\degree} data) and the need to handle heterogeneous datasets with varying resolutions, variables, and pressure levels.
Due to the size of datapoints, training is typically bottlenecked by data loading and not by the model.
This means that training smaller models is not always cheaper, because training costs will be dominated by data loading.
We developed a sophisticated data storage and loading infrastructure to address these technical challenges.

\paragraph{Data storage and preprocessing.}
We use Azure blob storage with several optimisations to ensure efficient data access. These optimisation include colocating data and compute to minimize latency and costs, storing datasets in appropriate chunks to avoid unnecessary data download and to minimise the number of concurrent connections, and compressing these chunks to reduce network bandwidth.

\paragraph{Data loading.}
We have developed an advanced multi-source data loading pipeline to efficiently handle heterogeneous data.
We now outline the main design principles of our pipeline.
Datasets are instantiated using YAML configuration files specifying loading parameters.
Each dataset generates a stream of lightweight \texttt{BatchGenerator} objects.
The scope of the \texttt{BatchGenerator} class is to abstract away the details and particularities of datasets by offering a common interface for generating data batches.
The streams are combined, shuffled, and sharded across GPUs.
After sharding, finally the common interface of \texttt{BatchGenerator} is used to do the work needed to download and construct batches for training and inference.

This pipeline enables efficient training on multiple heterogeneous datasets by batching only samples from the same dataset together and automatically balances workloads across GPUs by using different batch sizes for different datasets. This design offers flexibility needed to experiment with the Aurora model architecture while efficiently handling the challenges of large-scale, heterogeneous weather data processing. For a detailed description of the data loading pipeline, including the \texttt{BatchGenerator} object structure and the unpacking process, please refer to  \cref{sec:data_infrastructure}.

\subsection{Verification metrics}

We evaluate Aurora's performance using two main metrics: the root mean square error (RMSE) and the anomaly correlation coefficient (ACC). Both metrics incorporate latitude weighting to account for the non-uniform grid of the Earth. The RMSE measures the magnitude of errors between predictions and ground truth, while the ACC measures the correlation between the deviation of the prediction and ground truth from the daily climatology.

To assess performance on extreme weather events, we use a thresholded RMSE.
The thresholded RMSE uses a threshold to determine which latitude-longitude gridpoints should be included in the calculation, allowing for evaluation of model performance across different intensity levels of weather phenomena.
The thresholds are defined using the mean and standard deviation of the ERA5 reanalysis data over all training years computed separately for each latitude-longitude point. We vary these thresholds linearly for both positive and negative values to obtain RMSE curves for different intensity levels.

For a comprehensive explanation of the verification methods employed in this work, including their mathematical formulation and interpretation, please refer to  \cref{sec:verification_methods}. Taken together, these metrics employed here provide a robust framework for evaluating Aurora's performance across various weather conditions, from typical to extreme events.

\section*{Data availability}

All data used to train and evaluate Aurora can be obtained from publicly available sources. A detailed description of these sources, including download links, is provided in \cref{section:datasets}.

% \bigskip

% Springer journals and proceedings: \url{https://www.springer.com/gp/editorial-policies}

% Nature Portfolio journals: \url{https://www.nature.com/nature-research/editorial-policies}

% \textit{Scientific Reports}: \url{https://www.nature.com/srep/journal-policies/editorial-policies}

% BMC journals: \url{https://www.biomedcentral.com/getpublished/editorial-policies}
% \end{flushleft}

\bibliographystylemethods{sn-mathphys-num}
\bibliographymethods{bibliography}

\vspace{1cm}
\ActivateWarningFilters[hyperreflevel]
\bmhead{Supplementary information.}
We provide a Supplementary Materials document that contains all the methodological details required to reproduce our work, as well as supplementary text and results that supports the main claims of our paper. The Supplementary Material provides details on the following:
\begin{itemize}
\item The Aurora model.
\item Datasets.
\item Training methods.
\item Data infrastructure.
\item Verification methods.
\item Supplementary text and results including extensive comparisons against operational IFS HRES and GraphCast across various weather conditions, from typical to extreme events.
\end{itemize}

\bmhead{Acknowledgements.}
We thank ECMWF and NOAA for their commitment to open science and their major efforts to generate, curate and openly disseminate all the datasets that enabled our work,
and we thank Matthew Chantry for helpful advice on ECMWF's data sources. We thank the CAMS team at ECMWF for insightful discussions.
We thank Wenlei Shi, Yue Wang, Pipi Hu, Qi Meng from  MSR AI for Science, and Remi Tachet des Combes and Shuhang Chen from MSR for helpful inputs in the early stages of this work. We thank Divya Kumar, Weixin Jin, Sylwester Klocek, Siqi Xiang and Hongyu Sun from MSN Weather for their technical feedback throughout this project. We also thank Dieter Schwarenthorer for his help with Azure computing and licensing.
Finally, we thank Andrew Foong and Frank No\'e for constructive feedback during the writing of this manuscript. We are also grateful to Nikhil Shankar for his assistance with the HRES T0 dataset. Richard E.~Turner was funded by EPSRC Prosperity Partnership EP/T005386/1 between Microsoft Research and the University of Cambridge during the final stages of the project. 

\bmhead{Declarations.}
\begin{itemize}
\item Funding: This project was supported by Microsoft Research AI for Science.
\item Conflict of interest/Competing interests: None to report.
\item Ethics approval and consent to participate: Not applicable.
\item Consent for publication: All authors have reviewed the manuscript and consent to its publication.
\item Data availability: See section on data availability.
\item Materials availability: Not applicable.
\item Code availability: See section on code availability.
\item Author contributions: C.B., W.P.B., A.L., and M.S. were four core contributors of this project. They formulated, implemented and evaluated all aspects of Aurora, including the model architecture, the training and fine-tuning pipelines, as well as all experiments and evaluations except for the tropical cyclone results. A.V. was also a core contributor, provided critical feedback, and helped conceptualise, formulate and design the experiments. A.V. originated and carried out the tropical cyclone experiments with the assistance of W.P.B. P.G. and M.R. supported the engineering infrastructure of this project. J.B., J.K.G., and M.W. contributed to the initial development and conceptualization of this research. J.A.W., H.D., and K.T. provided regular feedback and carried out all comparisons against weather station data.  A.A. provided guidance on the CAMS experiments and model evaluation. E.H. provided assistance with program management and research timelines. R.E.T. and P.P. supervised all aspects of this project. All authors contributed to the writing and editing of this manuscript.
\end{itemize}

\DeactivateWarningFilters[hyperreflevel]

\clearpage
\begin{appendices}
\setcounter{page}{1}
% Reset the line count by resetting the internal counter of `vruler`.
%\rulercount 1\relax  

% From here on, disallow the use of \cite, \citet, and \citep. Instead,
% require \citeapp, \citetapp, and \citepapp.

\renewcommand{\citemethods}[1]{\PackageError{main}{Use citeapp instead of citemethods}{}}
\renewcommand{\citetmethods}[1]{\PackageError{main}{Use citetapp instead of citetmethods}{}}
\renewcommand{\citepmethods}[1]{\PackageError{main}{Use citepapp instead of citepmethods}{}}

\let\citeappold\citeapp%
\renewcommand{\cite}[1]{%
    \ifcsname inciteapp\endcsname\else%
    \PackageError{main}{Use citeapp instead of cite}{}%
    \fi%
    \citeold{#1}%
}
\renewcommand{\citeapp}[1]{%
    \begingroup%
    \newcommand{\inciteapp}{}%
    \citeappold{#1}%
    \endgroup%
}

\let\citepappold\citepapp
\renewcommand{\citep}[1]{%
    \ifcsname incitepapp\endcsname\else%
    \PackageError{main}{Use citepapp instead of citep}{}%
    \fi%
    \citepold{#1}%
}
\renewcommand{\citepapp}[1]{%
    \begingroup%
    \newcommand{\incitepapp}{}%
    \citepappold{#1}%
    \endgroup%
}

\let\citetappold\citetapp
\renewcommand{\citet}[1]{%
    \ifcsname incitetapp\endcsname\else%
    \PackageError{main}{Use citetapp instead of citet}{}%
    \fi%
    \citetold{#1}%
}
\renewcommand{\citetapp}[1]{%
    \begingroup%
    \newcommand{\incitetapp}{}%
    \citetappold{#1}%
    \endgroup%
}

\pagestyle{empty}
\begin{center}
{\Large
Supplementary Materials for \\ \vspace{3mm}
A Foundation Model for the Earth System
}
\end{center}

\addcontentsline{toc}{section}{Appendix}
\part{}

\noptcrule
\mtcsetoffset{parttoc}{0em}
\setlength{\ptcindent}{0em}
\section*{Table of Contents}
\vspace{-1em}
\mtcsettitle{parttoc}{}
\parttoc
\clearpage
\pagestyle{fancy}

\newcommand{\megan}[1]{\textcolor{red}{Megan: #1}}
\newcommand{\wessel}[1]{\textcolor{red}{Wessel: #1}}
\newcommand{\cristian}[1]{\textcolor{red}{Cristian: #1}}
\newcommand{\ana}[1]{\textcolor{red}{Ana: #1}}

\section{Notation and problem statement}
\label{sec:notation}

We denote the observed state of the atmosphere (including the surface) at a certain time $t$ via a tensor $X^t$ with dimensions $V \times H \times W$, where $V$ denotes the total number of variables, $H$ specifies the number of latitude coordinates (i.e., the height) and $W$ denotes the number of longitude coordinates (i.e., the width). Thus, we use the $X^t_{v, i, j}$ indexing scheme to refer to the state of variable $v$ at time $t$ and latitude-longitude coordinates given by $(i, j)$. It is sometimes convenient to split this observed state $X^t$ into its surface ($S^{t})$ and atmospheric components ($A^t)$. The surface state $S^t$ is a $V_S \times H \times W$ tensor, where $V_S$ denotes the number of surface variables. The atmospheric state is a $V_A \times C \times H \times W$ tensor where $V_A$ denotes the number of atmospheric variables and $C$ represents the number of pressure levels throughout the atmosphere. Thus, the total number of variables is given by $V = V_S + V_A \times C$. 

Given a system state at time $t$, our goal is to learn to predict the next state at a time $t' > t$. For simplicity, we operate with discrete time units of $\Delta t$ (which is sufficient for any practical purposes), and consider $t' = t + k \Delta t$ for some positive integer $k$. To predict the state at the next time step, we learn a simulator $\Phi\colon (X^{t-\Delta t}, X^t) \mapsto \hat{X}^{t + \Delta t}$, which maps the observed state of the world at the previous time and the current time to a predicted state $\hat{X}^{t + \Delta t}$ at future time ${t + \Delta t}$. Since $\Delta t$ might vary from one task to another, we will often simplify this notation to $\Phi\colon (X^{t-1}, X^t) \mapsto \hat{X}^{t + 1}$.

To generate predictions for later time increments, the predictions of the learned simulator can be stacked together in an autoregressive manner. Assuming $\hat{X}^{t} = X^{t}, \hat{X}^{t-1} = X^{t-1}$, this can be described recursively as
\begin{align}
    \hat{X}^{t + k} = \Phi(\hat{X}^{t + k - 2}, \hat{X}^{t + k - 1}).
\end{align}
We call this an autoregressive roll-out. It is obtained by applying the function $\Phi$ iteratively, a total of $k$ times. Finally, for a certain prediction $\hat{X}^{t'}$, we denote the surface and atmospheric components of that prediction by $\hat{S}^{t'}$ and $\hat{A}^{t'}$, respectively.  

\section{The Aurora model}
\label{sec:methods}

In this section, we provide a more detailed description of each component of the Aurora model architecture.

\subsection{3D Perceiver encoder}
\label{sec:encoder}

Compared to textual data, training large models on a diverse collection of weather datasets is a substantially more challenging task due to the heterogeneous nature of the data. While language is generally homogeneous, weather datasets include different variables, pressure levels and resolutions. 
Moreover, due to the costs of storing the complete outputs of a simulation, many sources only make subsets of the total data available by reducing the dataset size along one or more of the axes mentioned above. 
To accommodate such heterogeneous datasets, we design a flexible encoder that maps datasets with varying structure into a standardised 3D tensor that enters the  model backbone.

\paragraph{Inputs.} The encoder treats all the variables as $H \times W$ images on a regular latitude-longitude grid. For each variable, we include the state at the current time $t$ and the state at time $t - 1$. This results in a $T \times H \times W$ tensor, where $T=2$ is the time dimension. For a dataset with $C$ pressure levels and $V_A$ atmospheric variables, the state of the atmosphere is represented as a $V_A \times C \times T \times H \times W$. Similarly, for a dataset with $V_S$ surface variables, the state of the surface level is represented as a $V_S \times T \times H \times W$ tensor. In practice, all computations are batched, which adds an additional batch dimension in front of these tensors, which we omit for simplicity. 

\paragraph{Static variables.}
Aurora includes three \text{static variables}, which stay constant over time:
(1) geopotential at the surface (Z), which provides information about local orography, 
(2) a land--sea mask (LSM), and
(3) soil-type (SLT).
Internally in the model, the static variables are incorporated by appending them to the collection of surface-level variables.
The static variables are part of ERA5's invariant surface-level variables.

\paragraph{Level embeddings.}  As in standard ViTs, we split each of the $H \times W$ images in $P \times P$ patches. 
The patches at each level are then mapped into a vector in $\mathbb{R}^D$ by a linear layer, i.e. $C \times V_A \times T \times P \times P \mapsto C \times D$ and $V_S \times T \times P \times P \mapsto 1 \times D$. 
In order to accommodate datasets with different variables, this linear transformation is constructed dynamically for each variable $v$ using a set of weights $W_v$ specific to that variable. 
Each of the level embeddings is then tagged with an additive level encoding. For the atmospheric levels, the level encoding is computed via a sine/cosine encoding of the atmospheric pressure associated with each level (e.g., 150 hPa). For the surface level, we use a fully-learned vector of dimension $D$. 

\paragraph{Level aggregation.} The next step is to reduce the number of physical pressure levels in the atmosphere, which can vary across datasets, to a fixed set of latent pressure levels. \citeapp{nguyen2023climax} introduced a simple aggregation mechanism using attention pooling. 
Here, we adopt a more expressive scheme that leverages Perceiver modules \citeapp{jaegle2021perceiver} consisting of one cross-attention layer followed by a residual MLP. 
The inputs to the Perceiver are $C_L = 3$ latent query vectors, along with $C$ keys and values vectors computed from the level embeddings via a linear transform. 
The output of the Perceiver is a $C_L \times D$ tensor, encoding the latent state of the atmosphere. 
Concurrently, the surface level embedding is simply passed through a residual MLP. This latent state of the surface is then concatenated with the latent state of the atmosphere across the vertical dimension, which yields a $(C_L + 1) \times D$ latent representation of the weather state at the location of each patch.

\paragraph{Scale, position and absolute time embeddings.}
By gathering the latent representations at each patch, one obtains a 3D tensor $C_L \times H / P \times W / P$, which can be seen as a latent grid on which the backbone will perform the simulation. Therefore, we equip these tokens with information about their latitude and longitude coordinates as well as their physical size (i.e., km$^2$ per patch -- which differs depending on the latitude). 
Finally, we include absolute time information for each of the patches. For more information on how these encodings are computed, see \cref{sec:pos_encoding}. 

\subsection{Multi-scale 3D Swin Transformer U-Net backbone}
\label{sec:backbone}

\begin{figure}[t]
    \centering
    \includegraphics[width=1.0\textwidth]{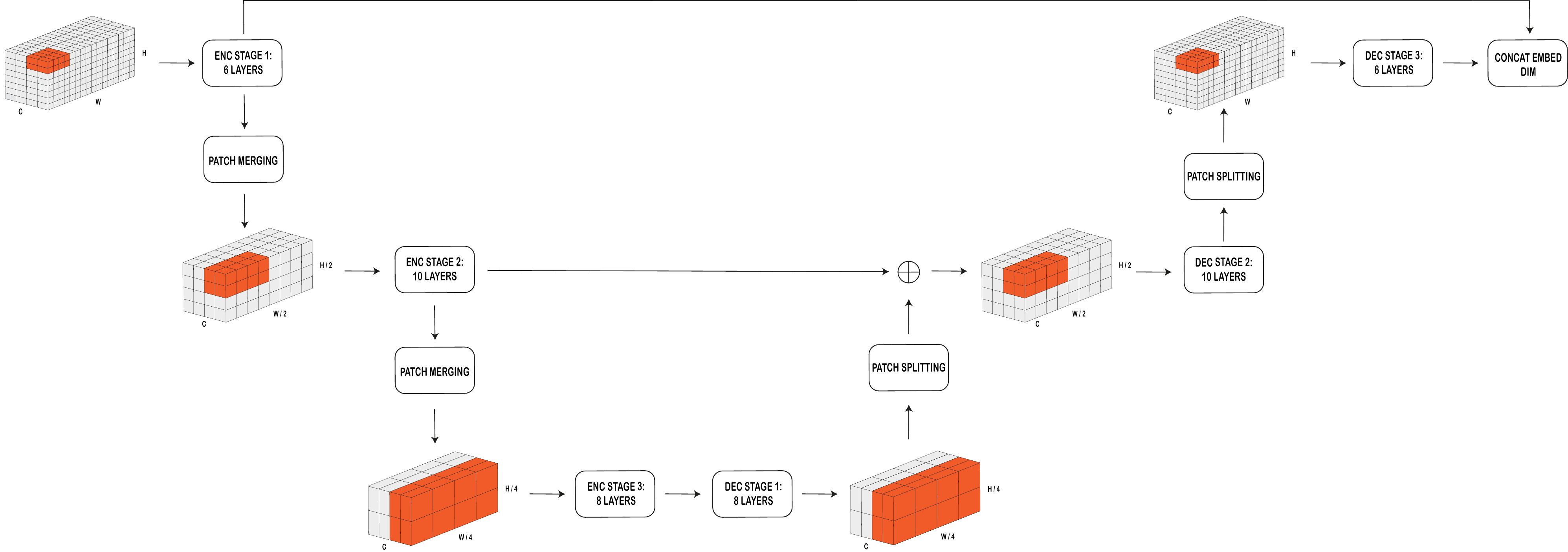}
    \caption{Overview of the 3D Swin Transformer U-Net which is the backbone of Aurora. The U-Net is formed by an encoder (\textit{left}) and a decoder (\textit{right}). Each of these is composed of three stages with (6, 10, 8) and (8, 10, 6) Swin 3D Transformer layers, respectively. In total, the backbone has 48 layers. The highlighted tokens show a single 3D self-attention window of size (2, 4, 2) along the depth, width and height dimensions, respectively. The patch  merging layers halve the spatial resolution of the 3D tensor, while the patch splitting layers double it. This allows the backbone to simulate the underlying physics at multiple scales. Although not shown in the figure, the embedding dimension of each token is doubled after each patch merging layer and halved after each patch splitting layer.}
    \label{fig:backbone}
\end{figure}

If the 3D tensor that the backbone receives as input can be viewed as a latent 3D mesh on which the simulation is performed, then the backbone can be viewed as a (neural) simulator. Transformers, due to their proven scaling properties~\citeapp{kaplan2020scaling} and connections with numerical integration~\citeapp{brandstetter2022message}, are a natural architectural choice for this simulation engine. 
To this end, we opt for a 3D Swin Transformer U-Net~\citeapp{liu2021swin, liu2022swin} architecture (\cref{fig:backbone}), which has also been successfully employed by~\citeapp{bi2022pangu}.

The backbone itself is also composed of an encoder and a decoder, each made out of three stages. After each stage of the encoder, the spatial resolution of the 3D tensor is halved. After each decoder stage, the resolution is doubled and the output is combined with the corresponding outputs of the encoder. This structure enables the backbone to simulate the underlying physics at multiple scales.

Each layer of the backbone is a 3D Swin Transformer layer performing local self-attention operations between tokens in the same regions (called windows), which can be seen as a form of message passing~\citepapp{gilmer2017neural}. One such window (of size (2, 4, 2)) is highlighted in \cref{fig:backbone}. 
In every other layer, all the windows shift across all dimensions by half of the window size along that dimension. In this case, the window would shift by (1, 2, 1) along the three axes, which allows the model to propagate information between neighbouring regions. Due to the spherical topology of the Earth, when the window is shifted, we allow the left and right side of the images to communicate directly as in \citetapp{bi2022pangu}. Overall, this procedure emulates the local computations performed by numerical integration methods, while avoiding the quadratic complexity of vanilla Transformers, which is impractical for the resolutions Aurora operates at. In practice, we use a window size of (2, 12, 6) for all of our experiments, similarly to \citetapp{bi2022pangu}. 

To increase stability throughout training, we use the res-post-norm layer normalisation introduced in \citetapp{liu2022swin}, but opt for a standard dot-product attention operation instead of the cosine attention procedure from Swin v2 \citeapp{liu2022swin}. 
To obtain a simple and flexible backbone that can operate at multiple resolutions, we do not use any form of positional bias in the attention operations as previously done by \citetapp{bi2022pangu, liu2021swin, liu2022swin}, which requires inputs with a fixed spatial resolution. Instead, we opt for positional and scale encodings in the encoder, as described in the previous section, which does not impose any constraints on the inputs. 

Finally, we note that the encoding procedure we employ allows us to flexibly use a small number of latent levels, which in turn frees up memory for massively scaling the number of parameters in the backbone. Compared to the backbone in~\citetapp{bi2022pangu}, which has 16 layers and two stages, our backbone has an additional stage and contains 48 layers.

\subsection{3D Perceiver decoder}
\label{sec:decoder}

The decoder maps the standardised outputs of the latent simulation back into images on the regular lat-lon grid. Its structure mirrors that of the encoder. The three latent atmospheric pressure levels are de-aggregated into $C$, $D$-dimensional embeddings (i.e., one per output atmospheric level). 
As in the encoder, this is done via a Perceiver layer \citeapp{jaegle2021perceiver}, which uses the sine/cosine embeddings of the output levels' pressure as queries. These vectors are then decoded into $P \times P$ patches via a linear layer. The latent surface level is decoded directly. Analogously to the patch embedding layer in the encoder, the linear layer creating the output patches is constructed dynamically by selecting the weights associated with each variable. This overall architecture allows the decoder to output predictions at arbitrary pressure levels, for an arbitrary set of variables.  

\subsection{Position, scale, level, and time encodings}
\label{sec:pos_encoding}

Given the minimum and maximum wavelengths $\lambda_\text{min}$ and $\lambda_\text{max}$ we want to capture, we use a Fourier encoding of the following form to encode a value $x$ into a $D$-dimensional vector, 
\begin{align}
    \text{Emb}(x) = \left[\cos{\frac{2 \pi x}{\lambda_i}},  \sin{\frac{2 \pi x}{\lambda_i}}\right] \ \text{for } 0 \leq i < D / 2 ,
\end{align}
where the $f_i$ are log-spaced values between the minimum and maximum wavelength, i.e.,  
\begin{align}
    \lambda_i = \exp\left(\log \lambda_\text{min} + i \cdot \frac{\log \lambda_\text{max} - \log \lambda_\text{min}}{D/2 - 1}\right).
\end{align}
In the following, we describe how the value $x$ is computed for all the encodings and how $\lambda_\text{min}$ and $\lambda_\text{max}$ are chosen. 

\paragraph{Positional encoding.} A two-dimensional positional encoding is used to account for the spatial position of each token that enters the Transformer backbone. For an input token of dimension $D$, the first $D/2$ dimensions are used for latitude information while the next $D/2$ dimensions for longitude information. In contrast to standard positional encodings used in ViT \citeapp{dosovitskiy2020image, nguyen2023climax}, where the coordinate IDs of the patch are used, we use a geometry-aware positional encoding given by the average latitude and longitude of each patch. We set $\lambda_\text{min} = 0.01$ and $\lambda_\text{max} = 720$. 

\paragraph{Scale encoding.} A separate, non-trainable, two-dimensional positional encoding is added to account for the physical scale of each patch token. Specifically, give a patch with minimum and maximum latitudes $\phi_1, \phi_2$, and minimum and maximum longitudes $\theta_1, \theta_2$ (measured in radians), we compute the area of the patch using the formula
\begin{align}
    A = R^2 (\sin \phi_2 - \sin \phi_1) (\theta_2 - \theta_1),
\end{align}
where $R = 6371$ km is the radius of the Earth. Therefore, patches that originate from a high-resolution input dataset (e.g., IFS HRES data) will yield small areas, while patches that originate from coarser datasets (e.g., CAMS) will be assigned encodings with larger magnitude. We set $\lambda_\text{min}$ to a sufficiently small value and $\lambda_\text{max}$ to the surface area of the entire Earth. 

\paragraph{Pressure level encoding.} Here, we simply encode the value of the atmospheric pressure associated with each level (e.g., 500 hPa). We set $\lambda_\text{min} = 0.01$ and $\lambda_\text{max} = 10,000$. 

\paragraph{Absolute time encoding.} We encode the absolute time associated with a certain input as the number of hours since January 1, 1970. We set $\lambda_\text{min}$ to one hour and $\lambda_\text{max}$ to the number of hours in a year. Therefore, this encoding is able to capture important information such as time of day, time of the week, month of the year, etc.

\subsection{Data normalisation}

Aurora normalises all variables before processing them in the encoder and unnormalises the output of the decoder to produce the final predictions.
Every surface-level variable and every pressure level of every atmospheric variable is normalised separately by a spatially constant scale and centre:
\begin{equation}
    X_{v,i,j,\text{normalised}}^t = \frac{X_{v,i,j}^t - \text{centre}_v}{\text{scale}_v}.
\end{equation}
We collectively call the scales and centres the normalisation statistics.
The centres are estimated by the empirical means computed over the whole ERA5 training data, and the scales are estimated by the empirical standard deviations computed over the whole ERA5 training data.
These normalisation statistics are then used for all datasets.
Final predictions are produced by unnormalising the output of the decoder:
\begin{equation}
    \hat X_{v,i,j}^t = \text{scale}_v \cdot \hat X_{v,i,j,\text{normalised}}^t + \text{centre}_v,
\end{equation}
where $\hat X_{v,i,j,\text{normalised}}^t = \Phi( X_{v,i,j,\text{normalised}}^{t-1},  X_{v,i,j,\text{normalised}}^{t-2})$ represents the raw, unnormalised output of the decoder.

\subsection{Extensions for 0.1\degree{} weather forecasting}
\label{app:sec:model:extensions-0.1}

Due to the increase in resolution, it becomes extremely challenging to make a large vision model like Aurora fit in the available GPU memory, which is why we perform a few minor modifications to the architecture at fine-tuning time. 

\paragraph{Increase patch size.} We increase patch size by a factor of $2.5\times$, from $4$ (used in pretraining) to $10$. Since this corresponds to the increase in resolution from $0.25\degree{}$ to $0.1\degree$, the spatial resolution of the backbone input remains the same as in pretraining. Therefore, there is no memory increase in the backbone computations, where most of the parameters reside. To ensure maximum transfer of skills from pretraining to fine-tuning, we use a bilinear interpolation to interpolate from the smaller patch size to the larger patch size, then we scale the magnitude of the interpolated weights by the ratio of the two patch sizes in order to preserve the magnitude of the inputs~\citepapp{beyer2023flexivit}. This ensures that the pretrained model can make reasonable predictions at 0.1\degree{} even before it is fine-tuned. We find that this also makes fine-tuning significantly faster.

\paragraph{Remove backbone layers.} We remove two layers from the two middle stages of the backbone encoder and decoder, respectively. Thus, the number of layers for all stages becomes (6, 8, 8) for the backbone encoder and (8, 8, 6) for the backbone deocoder. Since the model is pretrained using stochastic depth, the model is robust to this change and we do not notice any significant drop in the initial loss before fine-tuning.

\subsection{Extensions for air pollution forecasting}
\label{app:sec:cams-extension}

In \cref{section:cams}, we fine-tune Aurora to predict concentrations of air pollutants in addition to a standard collection of meteorological variables.
CAMS uses spatially and temporally disaggregated emissions data as inputs.
These inputs allow natural factors (i.e., wildfires, vegetation, etc.) and anthropogenic factors (i.e., vehicle combustion, energy production, etc.) to influence the levels of the air pollutants simulated by CAMS.
Aurora does not use any emissions data as input, except for some additional static variables which are fixed across all times and experiments. 
Instead, Aurora must implicitly learn to account for these effects, but likely would improve if direct access to CAMS emissions factors, or other similar emissions data, were available.
We therefore slightly adapt Aurora to better predict these variables.
All adaptations are carefully described in this section.

\paragraph{12-hour model.}
Some pollutants, such as \NO, show clear diurnal cycles (\cref{fig:cams-sample-predictions}).
We therefore fine-tune a version of Aurora that was pretrained with time step equal to 12 hours instead of 6 hours.
A 12-hour model which takes in two previous time steps can use the state at time $t-{}$\SI{24}{h} to make predictions for time $t$.
The 12-hour model was pretrained exactly like the 6-hour model, but only for \SI{80.5}{k} steps, corresponding to roughly a week and a half of training.

\paragraph{Differencing.}
Since some pollutants show clear diurnal cycles,
we build the ability to predict the difference with respect to a previous state explicitly into the model.
Only for the new pollution variables,
Aurora explicitly predicts the difference with respect to either time $t - {}$\SI{12}{h} or time $t - {}$\SI{24}{h}.
Variables \NO, TC \NO, \NOb, TC \NOb, \SOb, TC \SOb, \PMa, \PMb, and \PMc~show clear diurnal cycles, so for these variables we predict the difference with respect to time $t - {}$\SI{24}{h}.
For variables \CO, TC \CO, \Oc, and TC \Oc~the behaviour is less clear, so for these variables we predict the difference with respect to time $t - {}$\SI{12}{h}. 

We do not restrict Aurora to necessarily predict the difference.
In the decoder, we introduce a modulation factor, initialised to one, produced by an additional head:
\begin{equation}
    \hat{X}^{t} = \underbracket[0.5pt]{\big[\Phi_{\text{mod}}(X^{t - 1}, X^{t- 2}) + 1\big]}_{\text{modulation factor}} X^{a} + \underbracket[0.5pt]{\Phi_{\text{pred}}(X^{t  - 1}, X^{t - 2})}_{\text{delta or direct prediction}},
\end{equation}
where $a$ is either $t  -1$ or $t - 2$, depending on the variable.
$\Phi_{\text{mod}}$ and $\Phi_{\text{pred}}$ are two different heads of the decoder, both initialised to zero.
Therefore, at initialisation, Aurora predicts the appropriate persistence prediction only for the new pollution variables. 

\paragraph{Normalisation for concentrations of air pollutants.}
Fine-tuning on CAMS analysis happens in two stages: 
an initial fine-tuning stage on CAMS reanalysis data,
and the main fine-tuning on CAMS analysis data (see \cref{table:datasets-eval,sec:short_leadtine_finetuning}).
For the initial fine-tuning stage on CAMS reanalysis,
normalisation statistics for the new variables are computed over the whole CAMS reanalysis fine-tuning data;
and for the main fine-tuning on CAMS analysis,
normalisation statistics for the new variables are computed over the whole CAMS analyis training data.
Due to the heterogeneous and skewed nature of air pollution variables, the empirical standard deviation is not a good estimate of the scale.
Instead, for all air pollution variables $v$, we set $\text{centre}_v = 0$, and estimate $\text{scale}_v$  by half the spatial maximum averaged over time,
\begin{equation}
    \text{scale}_v = \frac12 \cdot \frac{1}{T}\sum_{t=1}^T \max\,\{X^t_{v,i,j} : i=1,\ldots,H\,j=1,\ldots,W\}.
\end{equation}
By construction of this normalisation, the normalised air pollution variables will typically be in the range $[0, 2]$.

\paragraph{Transformation of concentration variables.}
Variables which are concentration values have a large dynamic range by exhibiting important structure on varying orders of magnitude.
Including these variables without any modification makes the network sensitive to high-magnitude values (i.e., spikes) and insensitive to low-magnitude values.
One solution is to use the logarithm of these variables.
However, including the logarithm of positive variables makes the network overly sensitive to extremely-low-magnitude values (i.e., zero becomes negative infinity).
We therefore include a linear combination of the original variable $x$ and $\log(x)$ which clips away the weaknesses of $x$ and $\log(x)$:
\begin{equation} \label{eq:log-transform}
    x_{\text{transformed}} = c_1 \min(x, 2.5) + c_2 \frac{\log(\max(x, 10^{-4})) - \log(10^{-4})}{\log(10^{-4})},
\end{equation}
where $c_1$ and $c_2$ are initialised to $\tfrac12$.
The particular constants in the above transformation assume that $V$ is normalised to usually be in the range $[0, 2]$.
For $10^{-4} \le x \le 2.5$, which is the most interesting range, $x_{\text{transformed}}$ depends both on $x$, which is sensitive to normal-magnitude values, and on $\log(x)$, which is sensitive to low-magnitude values.
For $x \ge 2.5$, $\min(x, 2.5)$ clips $x$ to $2.5$, so $x_{\text{transformed}}$ then only depends on $x$ through $\log(x)$.
For $x \le 10^{-4}$, the $c_2$-term is zero exactly, so $x_{\text{transformed}}$ then only depends on $x$ through $x$.

\paragraph{Additional static variables.}
To help Aurora better understand diurnal cycles and human behaviour, we extend the static variables to include the time of day (normalised to $[0, 1)$), the day of week (normalised to $[0, 1)$), and the day of the year (divided by $365.25$).
These normalised values $x$ are included as two constant masks $\cos(2 \pi x)$ and $\sin(2 \pi x)$.
We also include more pollution-related static variables:
the monthly average anthropogenic emissions in the year 2019 summed across all industries for (1) ammonia, (2) \CO, (3) the sum of \NO~and \NOb, and (4) \SOb.
These monthly averages $x$ are included in two forms:
as
\begin{equation}
    \frac{x}{\max(x)},
    \qquad
    \text{and}
    \qquad
    \Big[\log\Big(\max\Big(\frac{x}{\max(x)}, 10^{-10}\Big)\Big) - \log(10^{-10})\Big]/10.
\end{equation}
The monthly averages can be downloaded from CAMS global emission inventories at \url{https://ads.atmosphere.copernicus.eu/cdsapp#!/dataset/cams-global-emission-inventories} and are regridded to the resolution of CAMS analysis data.
In total, there are $3 \cdot 2 + 4 \cdot 2 = 14$ additional static variables.

Finally, whereas Aurora originally only includes the static variables as additional surface-level variables, we now also include them as additional atmospheric variables by replicating them for every pressure level.

\paragraph{Variable-specific clipping.}
\SOb~at 850 hPa 
and nearer to the surface tends to be particularly spikey.
To help stabilise roll-outs, during and after LoRA roll-out fine-tuning, we clip the predictions only for \SOb~at \SI{850}{hPa} and above at 1.
This clipping happens before unnormalisation.

\paragraph{Patch size decrease.} 
The lower resolution of CAMS analysis data ($0.4\degree$ instead of $0.25\degree$) allows us to use a smaller patch size of 3 instead of patch size 4.
To ensure maximum transfer, we initialise the size 3 patches following the procedure described in \cref{app:sec:model:extensions-0.1}.

\paragraph{Pressure-level-specific patch embeddings.}
Near the surface, air pollution variables are strongly affected by human behaviour. 
However, near the top of the atmosphere, air pollution variables evolve according to complicated atmospheric chemistry dynamics.
To deal with the increased variety of behaviours across pressure levels for the atmospheric variables, we make the patch embeddings in the 3D Perceiver encoder and the patch reconstructions in the 3D Perceiver decoder dependent on the pressure level.
The pressure-level-dependent patches are initialised with existing pressure-level-independent patches wherever possible and otherwise with uniform random values on the encoder side and with zeros on the decoder side.

\paragraph{Additional 3D Perceiver decoder.}
To help Aurora with the difference in behaviour between meteorological and air pollution variables, we use a separate 3D Perceiver decoder for the air pollution variables.
This separate 3D Perceiver decoder is initialised to the one learned during pretraining.

\paragraph{32-bit floating point computation.}
To better handle low-magnitude values, all computations before and after the backbone are performed in \texttt{float32}.
The backbone operates in \texttt{bf16} mixed precision mode.

\subsection{Extensions for wave forecasting}
\label{sm:wave-extension}

In \cref{section:cams}, we fine-tune Aurora to forecast ocean surface waves.
All adaptations of Aurora for wave forecasting are described in this section.

\paragraph{Initialisation for new wave variables.}
All new wave variables are modelled as surface-level variables.
The new patch embeddings are all initialised to zero, except for the patch embeddings of 10UN and 10VN, which are initialised to the patch embeddings of 10U and 10V learned during pretraining.

\paragraph{Angle-valued variables.}
All angle-valued variables, which are MWD, MDWW, MDTS, MWD1, and MWD2, are transformed with $x \mapsto (\sin(x), \cos(x))$ prior to running the encoder.
The model has separate patch embeddings for the sine and cosine components of angle-valued variables.
After running the decoder, sine and cosine components are transformed back to angles using $(\sin, \cos) \mapsto \operatorname{atan2}(\sin, \cos)$.
The angle-valued variables have normalisation mean zero and normalisation scale one.

\paragraph{Density channels for missing data.}
Wave variables are undefined above land.
In addition, wave variables can be missing above water, for example in the case that there is sea ice or a swell component is just absent.
The model therefore must be able to handle data missing in the inputs and predict missing values in the outputs.
We accomplish this by, for every wave variable, incorporating a so-called \emph{density channel} \citepapp{Gordon_2020_Convolutional_Conditional_Neural_Processes}.

For a variable, the density channel is one if data are present and zero if data are absent. 
In the variable, missing values are then simply set to zero after normalisation.
We need to include such a density channel, as otherwise the model would not be able to distinguish between a zero in the variable and a missing value.
The model treats the variable and the associated density channel as separate surface-level variables with separate patch embeddings.
The model also predicts the density channels, which is how the model can predict that a variable is absent at a specific location.
To bound the density channels to the range $[0, 1]$, we use the sigmoid function. 

When running the model autoregressively, it will take its own predictions for density channels as inputs.
In this case, to avoid any distribution  mismatch, we set all density channels to one wherever it is greater than $\tfrac12$, and we set all density channels \emph{and} associated data channels to zero wherever the density channels are less than $\tfrac12$.

\paragraph{Additional static variables.}
We include two additional static variables: the bathemetry, and a mask which is one wherever HRES-WAM models wave variables (between latitudes -78\degree~and 90\degree) and zero elsewhere.
The additional static variables are normalised to the range $[0, 1]$.

\paragraph{Additional layer normalisation.}
When fine-tuning Aurora to the wave data, to stabilise training, we apply a layer normalisation to the keys of the first level aggregation attention block and another layer normalisation to the queries of the first level aggregation attention block.
These layer normalisations are applied before the keys and queries are split across multiple heads.

\subsection{Model hyperparameters and configurations}
\label{sm:hyperparameters}

The 1.3B parameter Aurora model which we use in the main experiments instantiates the architecture as follows. The embedding dimension in the encoder and first stage of the backbone is 512. This dimension doubles at every subsequent stage of the backbone. The number of attention heads in the backbone is selected such that the embedding dimension per head is 64 throughout the backbone. Due to the concatenation at the end of the backbone, the embedding dimension in the decoder is $1024$. In the Perceiver layers of the encoder and decoder, we use an increased number of cross-attention heads (16), in order to give the model fine-grained control over how the latent state of the atmosphere is constructed.

\begin{table}[t]
    \centering
    \caption{The four Aurora model configurations used for the model scaling experiments in \cref{sec:scaling}. For the \SI{290}{M} model, the attention head dimensionality of the first layer of the encoder and last layer of the decoder was 80 instead of 64.}
    \begin{tabular}{llllll}
        \toprule
         Model size & \makecell[l]{Backbone \\encoder layers} & \makecell[l]{Backbone \\ decoder layers}  & Embed dim. & \makecell[l]{Attention \\ head dim.} & Perceiver heads \\ \midrule
         \SI{117}{M} & (2, 6, 2) & (2, 6, 2) & 256 & 64 & 8 \\
         \SI{290}{M} & (4, 8, 4) & (4, 8, 4) & 320 & 64 & 8 \\
         \SI{660}{M} & (6, 8, 8) & (8, 8, 6) & 384 & 64 & 12 \\
         \SI{1.3}{B} & (6, 10, 8) & (8, 10, 6) & 512 & 64 & 16 \\ \bottomrule
    \end{tabular}
    \label{tab:model_configurations}
\end{table}

For the model scaling experiment in section \cref{sec:scaling}, we instantiate smaller versions of this model by reducing the number of backbone layers, the embedding dimension and the number of (cross) attention heads, while always preserving the attention head dimension of 64 (\cref{tab:model_configurations}).

\section{Datasets}
\label{section:datasets}

We use several different datasets throughout our experiments with Aurora. This section provides a detailed inventory of all these datasets and the various details behind them.  

\subsection{Types of datasets}

The datasets used in the paper can be classified into five categories: analysis, reanalysis, forecasts, reforecasts, and climate simulation datasets.
Observational data such weather station observations and satellite data only measure the world in very select ways in particular regions, namely where the weather stations and satellites are located. 
To produce a gridded estimate of the state of the world throughout the entire atmosphere, numerical models are used to fill in the missing gaps.
The different kind of datasets differ in precisely which observational data they incorporate and how they use a numerical model.

\emph{Reanalysis data} is regarded as the best estimate of the state of the weather at a particular point in time. 
It uses a fixed numerical model to assimilate observations over a longer time window, including information from both the past as well as the future, to make an estimate about the present. 
As a result, reanalysis is not available in real time since it incorporates information the future, and it therefore cannot be used to initialise operational NWP forecasts.

To initialise operational NWP forecasts, one instead maintains \emph{analysis data}.
Analysis data is produced in a recursive manner:
just before NWP forecasts are scheduled to be run, the analysis assimilation system combines the most recent observational data with previous NWP forecasts to produce an estimate of the NWP initial condition.
Compared to reanalysis data, analysis data is available in real time because it only incorporates observational data from the past.

In addition, one can make after-the-fact {\it reforecasts} by using reanalysis data as the initial condition.
These initial conditions are reanalysis data, so they are not available when making forecasts operationally. 
Nevertheless, the purpose of reforecasts is to indicate how NWP models would have performed if they had access to the best possible global state at initialisation time.

Finally, {\it climate simulations} model the physics, chemistry, and biology of the atmosphere to generate potential climate scenarios under different forcing factors (e.g. various emission levels).   

\subsection{Dataset inventory}
\label{app:dataset-inventory}

We enumerate all the datasets used in the paper below. 
Unless otherwise specified, we use the 13 pressure levels from WeatherBench2 \citepapp{rasp2023weatherbench}: $50$, $100$, $150$, $200$, $250$, $300$, $400$, $500$, $600$, $700$, $850$, $925$, and $1000$ hPa. 

\begin{itemize}
    \item \textbf{ERA5} \citepapp{era5-single-dataset,era5-atmos-dataset} is a global reanalysis dataset from ECMWF for weather. It is widely regarded as the best estimate we have of the total state of the weather at 0.25\degree.
    
    \item \textbf{HRES forecasts} \citepapp{hres-dataset} refers to the high-resolution version of the operational NWP forecasting model run by ECMWF. It is considered to be the most accurate NWP forecasting model and it runs at 0.1\degree~resolution. We use it both at its native resolution and regridded to 0.25\degree.
    
    \item \textbf{HRES T0} provides the initial conditions used to initialise the HRES forecasts and it is often considered as the ground truth against which to evaluate the quality of the forecasts~\citepapp{rasp2023weatherbench}. 
    
    \item \textbf{HRES analysis} represents the official analysis product of ECWMF. It contains an additional assimilation step on top of HRES T0. In this sense, it provides a slightly more accurate source of truth. 
    
    \item \textbf{IFS ENS} \citepapp{ifs-ens-dataset} is the ensemble version of IFS, with 50 members run at a slightly coarser resolution of 18~km (prior to June 27 2023, after which the resolution increased to 9~km with model cycle 48R1). The 50 ensemble members are generated with perturbed initial conditions and stochastic model physics within the IFS model. We use the dataset from the WeatherBench2 repository \citepapp{rasp2023weatherbench}, where there are only 3 pressure levels available: \SI{500}{}, \SI{700}{}, and \SI{850}{hPa}.
    
    \item \textbf{IFS ENS mean} contains the mean predictions for each variable based on IFS ENS. It is provided by WeatherBench2 \citepapp{rasp2023weatherbench} and contains the same 3 pressure levels as IFS ENS. 
    
    \item \textbf{GFS forecasts} \citepapp{gfs-dataset} provides operational forecasts with a base resolution of 18 km. Here we use the data re-gridded to 0.25\degree. The zero time (initialisation) of these forecasts is the real-time operational analysis, derived analogously to HRES-T0.
    
    \item \textbf{GFS T0} \citepapp{gfs-dataset} refers to the real-time operational analysis, obtained from the zero time (initialisation) of the GFS forecasts. 
    
    \item \textbf{GEFS reforecasts} \citepapp{gefs-dataset} is based on 21 ensemble members to address underlying uncertainties in the input data such limited coverage, instruments or observing systems biases, as well as the limitations of the model itself. In practice, such large quantities of data are only archived for a few preceding months. Therefore, we use the {\it reforecast} data, spanning 2000-2019, initialised with reanalysis initial conditions at 00 UTC each day. In this setting, there are only five ensemble members, all of which are included. GEFS has 6 pressure levels, we use the 3 that line up with those from WeatherBench2: $850$, $925$, $1000$ hPa. 
    
    \item \textbf{CMIP6} is a climate model inter-comparison project, combining various climate modelling experiments, which include land, sea, atmosphere, and aerosol variables. We have two datasets from CMIP6: CMCC-CM2-VHR4 \citepapp{cmip6-cmcc-dataset} and ECMWF-IFS-HR \citepapp{cmip6-ifs-hr-dataset}, which each have 7 pressure levels: $50$, $250$, $500$, $600$, $700$, $850$, $925$ hP. 
    
    \item \textbf{MERRA-2} \citepapp{merra2-dataset} is an atmospheric reanalysis dataset from NASA'S Global modelling and Assimilation Office, incorporating space-based observations of aerosols. We use pressure levels corresponding to the 13 levels from WeatherBench2. 

    \item \textbf{HRES-WAM} \citepapp{hres-wam-dataset} refers to analysis and forecast data from ECMWF's ocean wave model called HRES-WAM.
    This model is coupled to HRES, operates at \SI{0.1}{\degree} resolution, and models surface-level variables describing ocean waves.
    We regrid this data to \SI{0.25}{\degree} resolution.
    
    \item \textbf{CAMS} \citepapp{cams-analysis-dataset} refers to analysis and forecast data from the Copernicus Atmospheric Monitoring Service. The data has \SI{0.4}{\degree} resolution and includes meteorological variables as well as variables describing the composition of the atmosphere, such as concentrations of air pollutants. CAMS undergoes frequent model updates. We use the 13 pressure levels from WeatherBench2. 
    
    \item \textbf{CAMS reanalysis} \citepapp{cams-reanalysis} refers to the fourth generation ECMWF global reanalysis of atmospheric composition (EAC4) from the Copernicus Atmospheric Monitoring Service \citepapp{camsra-dataset-ads}. The data has \SI{0.75}{\degree} resolution and, like CAMS, includes meteorological and variables as well as variables describing the composition of the atmosphere. However, unlike CAMS, CAMS reanalysis is produced with a single, albeit considerably older, IFS model cycle: CY42R1. 
    CAMS reanalysis is considered to be less accurate than recent CAMS data because the model cycle is so much older and the resolution is coarser; a quantitive comparison of CAMS and CAMS reanalysis against station observations can be found at \url{https://aeroval.met.no/}. We use the 13 pressure levels from WeatherBench2.
\end{itemize}

Collectively, these data sources open different windows onto the underlying atmospheric dynamics, and expose Aurora to different  factors that reflect variability with respect to initial conditions, model parameterizations, and chaotic dynamics. The range of variables used, along with their definitions, are detailed in \cref{tab:variables}.

In \cref{table:datasets-pretraining}, \cref{table:datasets-fine-tuning}, and \cref{table:datasets-eval}, we detail the dataset details used for pretraining, fine-tuning, and evaluating Aurora, respectively.

\afterpage{
\begin{table}[ht]   
    \centering
    \captionsetup{width=\textwidth}
    \caption{
        Naming of the variables used from the datasets, including units.
        Surface-level variables are available as a single value per latitude--longitude point.
        Atmospheric variables are instead available at a range of levels of constant air pressure throughout the atmosphere, therefore there are multiple values per latitude--longitude point.
        Atmospheric pollutants are measured as concentrations throughout the layers of the atmosphere with units of \SI{}{\kilogram\per\cubic\metre}.
        Atmospheric pollutants are also included as their total column (TC) value, which become 2D variables with units \SI{}{\kilogram\per\square\metre} and are here categorised as surface-level variables.
    }
    \begin{tabular}{lll}
        \toprule
        Variable & Units & Description \\ \midrule
        \multicolumn{3}{l}{\sc Surface-level meteorological variables} \\
        2T & K & Temperature at \SI{2}{m} above surface of land or sea \\
        U10 & \SI{}{\metre\per\second} &  Eastward component of wind at \SI{10}{m}\\
        V10 & \SI{}{\metre\per\second} &  Southward component of wind at \SI{10}{m}\\
        WS & \SI{}{\metre\per\second} & Wind speed at \SI{10}{m}; equal to $(\text{U10}^2 + \text{V10}^2)^{1/2}$\\
        MSL & Pa & Air pressure at mean sea level \\[.25em]
        \multicolumn{3}{l}{\sc Atmospheric meteorological variables} \\
        U& \SI{}{\metre\per\second} & Eastward component of wind\\
        V& \SI{}{\metre\per\second} & Southward component of wind\\
        T& K&  Temperature\\
        Q& \SI{}{\kilogram\per\kilogram}& Specific humidity  \\
        Z& \SI{}{\square\metre\per\square\second} & Geopotential \\[.25em]
        \multicolumn{3}{l}{\sc Surface-level pollution variables} \\
        PM$_{1}$ & \SI{}{\kilogram\per\cubic\metre} & Concentration of particulate matter with at most diameter \SI{1}{\micro m} \\
        PM$_{2.5}$ & \SI{}{\kilogram\per\cubic\metre} & Concentration of particulate matter with at most diameter \SI{2.5}{\micro m} \\
        PM$_{10}$ & \SI{}{\kilogram\per\cubic\metre} & Concentration of particulate matter with at most diameter \SI{10}{\micro m} \\
        TC $x$ & \SI{}{\kilogram\per\square\metre} & Total column value of $x$ \\[.25em]
        \multicolumn{3}{l}{\sc Atmospheric pollution variables} \\
        CO & \SI{}{\kilogram\per\cubic\metre}& Concentration of carbon monoxide\\
        NO & \SI{}{\kilogram\per\cubic\metre}& Concentration of nitrogen monoxide\\
        NO$_{2}$ & \SI{}{\kilogram\per\cubic\metre}& Concentration of nitrogen dioxide\\
        SO$_{2}$ & \SI{}{\kilogram\per\cubic\metre}& Concentration of sulphur dioxide\\
        O$_{3}$ & \SI{}{\kilogram\per\cubic\metre}& Concentration of ozone\\[.25em]
        \multicolumn{3}{l}{\sc Surface-level wave variables} \\
        10UN & \SI{}{\metre\per\second}& Eastward component of natural wind at \SI{10}{m}\\
        10VN & \SI{}{\metre\per\second}& Southward component of natural wind at \SI{10}{m}\\
        WSN & \SI{}{\metre\per\second}& Natural wind speed at \SI{10}{m}; equal to $(\text{10UN}^2 + \text{10VN}^2)^{1/2}$\\
        PP1D & \SI{}{\second}& Peak wave period \\
        SWH & \SI{}{\metre}& Significant wave height of total wave (wind waves plus swell)\\
        MWP & \SI{}{\second}& Mean wave period  of total wave \\
        MWD & degrees & Mean wave direction of total wave (w.r.t.\ north)\\
        SHWW & \SI{}{\metre}& Significant height of wind waves\\
        MPWW & \SI{}{\second}& Mean period of wind waves\\
        MDWW & degrees & Mean direction of wind waves (w.r.t.\ north)\\
        SHTS & \SI{}{\metre}& Significant height of total swell\\
        MPTS & \SI{}{\second}& Mean period of total swell\\
        MDTS & degrees & Mean direction of total swell (w.r.t.\ north)\\
        SWH1 & \SI{}{\metre}& Significant height of first swell partition\\
        MWP1 & \SI{}{\second}& Mean wave period of first swell partition\\
        MWD1 & degrees & Mean wave direction of first swell partition (w.r.t.\ north)\\
        SWH2 & \SI{}{\metre}& Significant height of second swell partition\\
        MWP2 & \SI{}{\second}& Mean wave period of second swell partition\\
        MWD2 & degrees & Mean wave direction of second swell partition (w.r.t.\ north)\\
        \bottomrule
    \end{tabular}
    \label{tab:variables}
\end{table}
\FloatBarrier
}

\afterpage{
\begin{sidewaystable}[ph!]
\caption{
    Summary of the ten datasets used to pretrain Aurora in various configurations presented in this work.
    For every dataset, lists the resolution, timeframe that the dataset spans, the surface-level variables, the atmospheric variables, the number of pressure levels for the atmospheric variables, the number of time steps, and the total size.
}
\begin{tabular}{@{}llllllrr@{}}
\toprule
Name
    & Resolution
    & Timeframe
    & Surf.\ variables
    & Atmos.\ variables
    & Levels
    & Steps
    & Size
    \\ \midrule
ERA5
    & \SI{0.25}{\degree} $\times$ \SI{0.25}{\degree}
    & 1979--2020
    & 2T, 10U, 10V, MSL
    & U, V, T, Q, Z
    & 13
    & \SI{368.18}{k}
    & \SI{105.50}{TB}
    \\
HRES-0.25 forecasts
    & \SI{0.25}{\degree} $\times$ \SI{0.25}{\degree}
    & 2016--2020
    & 2T, 10U, 10V, MSL
    & U, V, T, Q, Z
    & 13
    & \SI{149.81}{k}
    & \SI{42.93}{TB}
    \\
IFS-ENS-0.25
    & \SI{0.25}{\degree} $\times$ \SI{0.25}{\degree}
    & 2018--2020
    & 2T, 10U, 10V, MSL
    & U, V, T, Q, Z
    & 3
    & \SI{6.69}{M}
    & \SI{527.54}{TB}
    \\
IFS-ENS-0.25 mean
    & \SI{0.25}{\degree} $\times$ \SI{0.25}{\degree}
    & 2018--2020
    & 2T, 10U, 10V, MSL
    & U, V, T, Q, Z
    & 3
    & \SI{133.71}{k}
    & \SI{10.55}{TB}
    \\
GFS forecasts
    & \SI{0.25}{\degree} $\times$ \SI{0.25}{\degree}
    & Feb 2015--2020
    & 2T, 10U, 10V, MSL
    & U, V, T, Q, Z
    & 13
    & \SI{354.40}{k}
    & \SI{101.56}{TB}
    \\
GFS T0
    & \SI{0.25}{\degree} $\times$ \SI{0.25}{\degree}
    & Feb 2015--2020
    & 2T, 10U, 10V, MSL
    & U, V, T, Q, Z
    & 13
    & \SI{8.64}{k}
    & \SI{2.48}{TB}
    \\
GEFS reforecasts
    & \SI{0.25}{\degree} $\times$ \SI{0.25}{\degree}
    & 2000--2019
    & 2T, MSL
    & U, V, T, Q, Z
    & 7
    & \SI{2.96}{M}
    & \SI{454.61}{TB}
    \\
CMCC-CM2-VHR4
    & \SI{0.25}{\degree} $\times$ \SI{0.25}{\degree}
    & 1950--2014
    & 2T, 10U, 10V, MSL
    & U, V, T, Q
    & 7
    & \SI{94.96}{k}
    & \SI{12.62}{TB}
    \\
ECMWF-IFS-HR
    & \SI{0.45}{\degree} $\times$ \SI{0.45}{\degree}
    & 1950--2014
    & 2T, 10U, 10V, MSL
    & U, V, T, Q, Z
    & 7
    & \SI{94.96}{k}
    & \SI{4.75}{TB}
    \\
MERRA-2
    & \SI{0.625}{\degree} $\times$ \SI{0.50}{\degree}
    & 1980--2020
    & 2T, 10U, 10V, MSL
    & U, V, T, Q
    & 13
    & \SI{119.81}{k}
    & \SI{5.58}{TB}
    \\ \midrule
    &
    &
    &
    & 
    & Total
    & \SI{10.97}{M}
    & \SI{1268.12}{TB}
    \\ \bottomrule
\end{tabular}
\label{table:datasets-pretraining}
\end{sidewaystable}

% Finetuning table
\begin{sidewaystable}[ph!]
\caption{
    Summary of the datasets used to fine-tune Aurora in various experiments presented in this work.
    For every dataset, lists the resolution, timeframe that the dataset spans, the surface-level variables, the atmospheric variables, the number of pressure levels for the atmospheric variables, the number of time steps, and the total size.
    Note that, in the wave experiment in \cref{section:waves}, Aurora is fine-tuned on both wave and meteorological variables by lining up HRES-WAM-0.25 and HRES-0.25 T0.
}
\begin{tabular}{@{}llllllrr@{}}
\toprule
Name
    & Resolution
    & Timeframe
    & Surf.\ variables
    & Atmos.\ variables
    & Levels
    & Steps
    & Size
    \\ \midrule
CAMS reanalysis
    & \SI{0.75}{\degree} $\times$ \SI{0.85}{\degree}
    & 2003--2021
    & \begin{tabular}[t]{@{}l@{}}
        2T, 10U, 10V, MSL, \\
        TC CO, TC NO, TC NO${}_{2}$, \\
        TC SO${}_{2}$, TC O${}_{3}$, \\
        PM${}_{1}$, PM${}_{2.5}$, PM${}_{10}$
    \end{tabular}
    & \begin{tabular}[t]{@{}l@{}}
        U, V, T, Q, Z, \\
        CO, NO, NO${}_{2}$, \\
        SO${}_{2}$, O${}_{3}$
    \end{tabular}
    & 13
    & \SI{55.52}{k}
    & \SI{3.65}{TB}
    \\
CAMS analysis
    & \SI{0.40}{\degree} $\times$ \SI{0.40}{\degree}
    & Oct 2017--May 2022
    & \begin{tabular}[t]{@{}l@{}}
        2T, 10U, 10V, MSL, \\
        TC CO, TC NO, TC NO${}_{2}$, \\
        TC SO${}_{2}$, TC O${}_{3}$, \\
        PM${}_{1}$, PM${}_{2.5}$, PM${}_{10}$
    \end{tabular}
    & \begin{tabular}[t]{@{}l@{}}
        U, V, T, Q, Z, \\
        CO, NO, NO${}_{2}$, \\
        SO${}_{2}$, O${}_{3}$
    \end{tabular}
    & 13
    & \SI{3.41}{k}
    & \SI{0.79}{TB}
    \\
HRES-WAM-0.25 an.\
    & \SI{0.25}{\degree} $\times$ \SI{0.25}{\degree}
    & 2016--2021
    & \begin{tabular}[t]{@{}l@{}}
        % 2T, 10U, 10V, MSL, \\
        10UN, 10VN, WSN, PP1D, \\
        SWH, MWP, MWD, \\
        SHWW, MPWW, MDWW, \\
        SHTS, MPTS, MDTS, \\
        SWH1, MWP1, MWD1, \\
        SWH2, MWP2, MWD2
    \end{tabular}
    & None
    & N/A
    & \SI{8.77}{k}
    & \SI{0.73}{TB}
    \\
HRES-0.25 T0
    & \SI{0.25}{\degree} $\times$ \SI{0.25}{\degree}
    & 2016--2021
    & 2T, 10U, 10V, MSL
    & U, V, T, Q, Z
    & 13
    & \SI{8.77}{k}
    & \SI{2.51}{TB}
    \\
HRES-0.1 analysis
    & \SI{0.10}{\degree} $\times$ \SI{0.10}{\degree}
    & 2016--2022
    & 2T, 10U, 10V, MSL
    & U, V, T, Q, Z
    & 13
    & \SI{10.23}{k}
    & \SI{18.30}{TB}
    \\ \midrule
    &
    &
    &
    &
    & Total
    & \SI{86.69}{k}
    & \SI{25.98}{TB}
    \\ \bottomrule
\end{tabular}
\label{table:datasets-fine-tuning}
\end{sidewaystable}

% Eval table
\begin{sidewaystable}[ph!]
\caption{
    Summary of the datasets used to evaluate Aurora in various fine-tuning experiments presented in this work.
    For every dataset, lists the resolution, timeframe that the dataset spans, the surface-level variables, the atmospheric variables, the number of pressure levels for the atmospheric variables, the number of time steps, and the total size.
    Note that, in the wave experiment in \cref{section:waves}, Aurora is evaluated on both wave and meteorological variables by lining up HRES-WAM-0.25 and HRES-0.25 T0.
}
\begin{tabular}{@{}llllllrr@{}}
\toprule
Name
    & Resolution
    & Timeframe
    & Surf.\ variables
    & Atmos.\ variables
    & Levels
    & Steps
    & Size
    \\ \midrule
CAMS analysis
    & \SI{0.40}{\degree} $\times$ \SI{0.40}{\degree}
    & June 2022--Nov 2022
    & \begin{tabular}[t]{@{}l@{}}
        2T, 10U, 10V, MSL, \\
        TC CO, TC NO, TC NO${}_{2}$, \\
        TC SO${}_{2}$, TC O${}_{3}$, \\
        PM${}_{1}$, PM${}_{2.5}$, PM${}_{10}$
    \end{tabular}
    & \begin{tabular}[t]{@{}l@{}}
        U, V, T, Q, Z, \\
        CO, NO, NO${}_{2}$, \\
        SO${}_{2}$, O${}_{3}$
    \end{tabular}
    & 13
    & 366
    & \SI{0.08}{TB}
    \\
HRES-WAM-0.25 an.\
    & \SI{0.25}{\degree} $\times$ \SI{0.25}{\degree}
    & 2022
    & \begin{tabular}[t]{@{}l@{}}
        % 2T, 10U, 10V, MSL, \\
        10UN, 10VN, WSN, PP1D, \\
        SWH, MWP, MWD, \\
        SHWW, MPWW, MDWW, \\
        SHTS, MPTS, MDTS, \\
        SWH1, MWP1, MWD1, \\
        SWH2, MWP2, MWD2
    \end{tabular}
    & None
    & N/A
    & \SI{2.92}{k}
    & \SI{0.24}{TB}
    \\
HRES-0.25 T0
    & \SI{0.25}{\degree} $\times$ \SI{0.25}{\degree}
    & 2022
    & 2T, 10U, 10V, MSL
    & U, V, T, Q, Z
    & 13
    & \SI{2.92}{k}
    & \SI{0.84}{TB}
    \\
HRES-0.1 analysis
    & \SI{0.10}{\degree} $\times$ \SI{0.10}{\degree}
    & 2023
    & 2T, 10U, 10V, MSL
    & U, V, T, Q, Z
    & 13
    & \SI{2.92}{k}
    & \SI{5.23}{TB}
    \\ \midrule
    &
    &
    &
    &
    & Total
    & \SI{9.13}{k}
    & \SI{6.39}{TB}
    \\ \bottomrule
\end{tabular}
\label{table:datasets-eval}
\end{sidewaystable}
\FloatBarrier
}

\subsection{Motivation for CMIP6 dataset selection}
\label{app:sec:cmip-selection}

We selected the CMCC-CM2-VHR4 and ECMWF-IFS-HR from CMIP6 for several reasons:
\begin{enumerate}
    \item {\bf Resolution compatibility:} CMCC-CM2-VHR4 provides data at $0.25\degree$, matching ERA5, while ECMWF-IFS-HR at $0.45\degree$ offers a complementary intermediate scale which is well matched for the atmospheric chemistry setting. As detailed in Table C4, this resolution diversity helps Aurora learn scale-aware representations while maintaining sufficient fidelity for our target applications.
    \item {\bf Variable coverage:} Both datasets include our core set of meteorological variables (2T, U10, V10, MSL, U, V, T, Q) across 7 pressure levels (50, 250, 500, 600, 700, 850, 925 hPa), providing good vertical coverage of the atmosphere while maintaining consistency with our other data sources.
    \item {\bf Temporal extent:} These datasets span 1950-2014, offering exposure to long-term climate variability that complements our weather datasets. As shown in \cref{sec:scaling}, this temporal diversity particularly benefits the prediction of extreme events, result in up to 35\% error reduction in the tails of the distribution.
\end{enumerate}

\subsection{Training, validation, and test splits}
\label{app:sec:splits}
For validation while pretraining, we use one year of IFS HRES at 0.25\degree~resolution: 2020. 
Our test years are 2022 and 2023, depending on the dataset. 
Details of dataset splits can be found in \cref{table:datasets-pretraining}, \cref{table:datasets-fine-tuning}, and \cref{table:datasets-eval}.

\paragraph{Motivation for CAMS analysis train--test split.}
CAMS undergoes frequent updates that dramatically affect the data distribution.
Particularly notable are the updates in Sep 2017, when an issue was fixed that caused \PMb~and~\PMc~to be zero;
July 2019, when the vertical resolution was increased;
May 2021, which significantly alters the behaviour of the PMs;
Dec 2022, which alters the background error covariances and significantly alters the distribution of the data;
and June 2023, when CAMS is updated to the latest cycle C48R1 and is officially included as a chapter in the IFS documentation \citepapp{cams}.
Note that there were many more updates, but the mentioned ones were observed to be the most significant.
Between these updates, the period May 2021 to Nov 2022 inclusive appears to be the most stable period.
From this period, we use the last five months, Jun 2022 to Nov 2022 inclusive, for testing; and we include the first twelve months, Jun 2021 to May 2022 inclusive, for the fine-tuning data.
We additionally include in the fine-tuning data all data going back to Oct 2017, after the issues with \PMb~and \PMc~were fixed.

\subsection{Dataset-specific postprocessing}
\label{app:sec:processing}
For some datasets, like HRES-WAM, more careful postprocessing of the data is required.
We detail all additional postprocessing steps here.

\paragraph{Postprocessing for HRES-WAM.}
In \cref{section:waves} we predict the neutral wind velocities. The neutral wind is related in a complex manner to the true wind velocity and represents the wind speed under neutral atmospheric conditions \citepapp{ecmwf-neutral-wind:2017}. It is critical in coupling wave and atmospheric models, and assimilation of measurements of wind speed over oceans, so we consider it an important variable for Aurora to predict after fine-tuning on this data. 

HRES-WAM does not actually contain 10UN and 10VN, but instead the magnitude and the direction of the neutral wind speed, WIND and DWI respectively.
We obtain 10UN and 10VN from WIND and DWI according to the following transformation:
\begin{align}
    \text{10UN} = -\sin(\text{DWI}) \cdot \text{WIND}, \qquad
    \text{10VN} = -\cos(\text{DWI}) \cdot\text{WIND}.
\end{align}
Note the minus sign, which is necessary because the direction is where the wind \emph{comes from}, not where it \emph{goes towards}.

Absence of waves, e.g.\ above land or wherever there is sea ice, is represented with NaNs in the data.
In addition, we sometimes find that significant wave heights and/or mean wave periods can be zero.
Since zero significant wave height also means that there is no wave (or a particular wave component is absent), whenever significant wave height is zero, we set it and the corresponding mean wave period and mean wave direction to NaN.
We do this separately for the total wave and every wave component, i.e.\ wind waves, total swell, swell one, and swell two.
We verify that no spurious zeros remain after this processing step.

\section{Training methods}
% \label{sec:training_methods}

The overall training procedure is composed of three stages: (1) pretraining, (2) short-lead-time fine-tuning, (3) roll-out fine-tuning. We describe each of these stages in detail in the following subsections.

\subsection{Training objective}
\label{sec:training_objective}

Throughout pretraining and fine-tuning, we use the mean absolute error (MAE) as our training objective $\mathcal{L}(\hat{X}^t, X^t)$. Decomposing the predicted state $\hat X^t$ and ground truth state $X^t$ into surface-level variables and atmospheric variables, $\hat{X}^t = (\hat{S}^t, \hat{A}^t)$ and $X^t = (S^t, A^t)$ (\cref{sec:notation}), the loss can be written as
\begin{align}
    \mathcal{L}(\hat X^t, X^t)
    &= \frac{\gamma}{V_S + V_A} \Bigg[\alpha \Bigg( \sum_{k=1}^{V_S} \frac{w^S_k}{H \times W}\sum_{i=1}^{H} \sum_{j=1}^{W} | \hat{S}^t_{k, i, j} - S^t_{k, i, j} | \Bigg) \nonumber \\
    &\hspace{-60pt}\hphantom{= \frac{1}{(V_S + V_A)} \Bigg[\alpha \Bigg( }
    + \beta \Bigg( \sum_{k=1}^{V_A} \frac{1}{C \times H \times W}  \sum_{c=1}^{C} w^A_{k,c} \sum_{i=1}^{H} \sum_{j=1}^{W} | \hat{A}^t_{k, c, i, j} - A^t_{k, c, i, j} |\Bigg)\Bigg],
\end{align}
where $w^S_k$ denotes the weight associated with surface-level variable $k$ and $w^A_{k,c}$ denotes the weight associated with atmospheric variable $k$ at pressure level $c$.
The overall surface loss is weighted by $\alpha = \frac14$, while the overall atmospheric loss is weighted by $\beta = 1$. Finally, the entire loss for a particular example is weighted by a dataset weight $\gamma$, which allows us to upweight the datasets with higher fidelity such as ERA5 and GFS-T0. Specifically, we use $\gamma_{\text{ERA5}} = 2.0$, $\gamma_{\text{GFS-T0}} = 1.5$, and set the rest of the dataset weights to $1$. When training, we minimise the expected value of this loss computed over a mini-batch of samples. 

\paragraph{Variable weighting for pretraining.}
During pretraining, we set
$w^S_{\text{MSL}} = 1.5$,
$w^S_{\text{U10}} = 0.77$,
$w^S_{\text{V10}} = 0.66$, and
$w^S_{\text{2T}} = 3.0$.
For the atmospheric variables, for all pressure levels $c$, we set
$w^A_{\text{Z},c} = 2.8$,
$w^A_{\text{Q},c} = 0.78$,
$w^A_{\text{T},c} = 1.7$,
$w^A_{\text{U},c} = 0.87$, and
$w^A_{\text{V},c} = 0.6$.
The weights are chosen to balance the losses of the individual variables and have been inspired by the weights used by \citetapp{bi2022pangu}.

\paragraph{Variable weighting for concentrations of air pollutants.}
Air pollutants are extremely sparse.
To balance these variables with the meteorological variables, we require a radically different approach.
For the air pollution variables, we specify the weights such that the per-variable normalised MAE is roughly one.

For all air pollution variables, we compute the MAE for the persistence prediction of \SI{12}{h} or \SI{24}{h}, depending on whether the model predicts the difference with respect to the state \SI{12}{h} ago or \SI{24}{h} ago (\cref{app:sec:cams-extension}).
These persistence errors are computed on CAMS reanalysis data.
We then set
\begin{equation}
    w^S_v = \frac{\text{scale}_v}{\text{persistence MAE}_v},
    \qquad
    w^A_{v,c} = \frac{\text{scale}_{v,c}}{\text{persistence MAE}_{v,c}}.
\end{equation}
Intuitively, multiplication by the scale first undoes the data normalisation and then dividing by the persistence MAE brings the normalised MAE to have a value of approximately one. 

\paragraph{Variable weighting for wave variables.}
When fine-tuning Aurora for wave forecasting, we tune the weights to emphasise the more important and the more difficult variables.
Specifically, we set
$s^S_{\text{SWH}} = 2.0$,
$s^S_{\text{MWP}} = 2.0$,
$s^S_{\text{MWD}} = 2.0$,
$s^S_{\text{PP1D}} = 4.0$,
$s^S_{\text{MPTS}} = 2.0$, and
$s^S_{\text{MDTS}} = 2.0$.

\paragraph{Variable weighting for IFS T0 and analysis fine-tuning.}
During fine-tuning on IFS T0 \SI{0.25}{\degree} and IFS analysis \SI{0.1}{\degree}, we slightly adjust the pretraining weights.
For the surface-level variables, we set
$w^S_{\text{MSL}} = 1.6$,
$w^S_{\text{U10}} = 0.77$,
$w^S_{\text{V10}} = 0.66$, and
$w^S_{\text{2T}} = 3.5$.
For the atmospheric variables, for all pressure levels $c$, we set
$w^A_{\text{Z},c} = 3.5$,
$w^A_{\text{Q},c} = 0.8$,
$w^A_{\text{T},c} = 1.7$,
$w^A_{\text{U},c} = 0.87$,and
$w^A_{\text{V},c} = 0.6$.

\paragraph{Adjustments of the training loss for wave variables.}
For angle-valued variables, the MAE training loss is computed over the sine and cosine components of the angle (see \cref{sm:wave-extension}).
The sine and cosine components inherit the weighting of the original variables.

To deal with missing data, Aurora predicts the original variable along with a density channel (see \cref{sm:wave-extension}). 
We compute the MAE over both the original variable and the density channel by also computing the density channel for the target variable following exactly the procedure outlined \cref{sm:wave-extension}.
The density channels inherit the weighting of the original variables.
Note that the procedure sets all NaNs to zero, meaning that the MAE can be computed without risking the loss to become NaN. 

\subsection{Pretraining methods}
\label{sec:pretraining_methods}

All models are pretrained for \SI{150}{k} steps on 32 A100 GPUs, with a batch size of one per GPU. Our model sees about 4.8 million frames after 150k steps of pretraining, which takes approximately two and a half weeks. This roughly corresponds to 3 epochs over the C4 pretraining configuration that is outlined in \cref{sec:scaling}. We use a (half) cosine decay with a linear warm-up from zero for \SI{1}{k} steps. 
The base learning rate is $5\mathrm{e}{-4}$, which the schedule reduces by a factor $10\times$ at the end of training.
The optimizer we use is AdamW~\citepapp{loshchilov2018decoupled}. We set the weight decay of AdamW to $5\mathrm{e}{-6}$. The only other form of regularisation we use is drop path (i.e., stochastic depth) ~\citepapp{larsson2017fractalnet}, with the drop probability set to $0.2$. In order to make the model fit in memory, we use activation checkpointing for the backbone layers and we shard all the model gradients across GPUs. The model is trained using \texttt{bf16} mixed precision. 

\paragraph{Aurora pretraining in comparison.} The training of Aurora is comparable to GraphCast (four weeks on 32 Cloud TPU v4 devices, \citepapp{lam2023graphcast}) and faster than Pangu-Weather (about seven weeks using 64 GPUs, \citepapp{bi2022pangu}), and roughly comparable to AIFS (about one week with 64 GPUs in total \citepapp{lang+al:2024}). This demonstrates that Aurora achieves its improved performance and additional capabilities while maintaining training costs in line with other state-of-the-art models.

\paragraph{12-hour air pollution model.} The 12-hour model used in the air pollution experiments was trained in exactly the same manner as described above, but for \SI{80.5}{k} steps instead of \SI{150}{k} steps. 

\subsection{Short lead-time fine-tuning}
\label{sec:short_leadtine_finetuning}

For each task we wish to adapt the pretrained Aurora model to, we start by fine-tuning the entire architecture through one or two roll-out steps (depending on the task and its memory constraints). In all cases, we use a task-dependent hyperparameter selection, which we describe below.

\paragraph{HRES 0.25\degree{} T0.} We fine-tune the weights of the entire model for \SI{8}{k} training steps across 8 GPUs, with a batch size of 1 per GPU. At each iteration, we perform two roll-out steps and backpropagate through both of these steps. The model is optimised to minimise the MAE loss averaged across both roll-out steps. For this regime, we use a \SI{1}{k} step learning rate warm-up, followed by a constant learning rate of $5\mathrm{e}{-5}$. We use the same weight decay as in pretraining and disable drop path. To ensure the model fits in memory for two roll-out steps, we also use activation checkpointing for the encoder and the decoder, along with gradient sharding as in pretraining.

\paragraph{HRES 0.1\degree{} analysis. } We fine-tune the weights of the entire model for 12.5k steps across 8 GPUs, with a batch size of 1 per GPU. Due to the increased memory constraints at this higher resolution, we train the model only through a single step prediction. We use a 1k step learning rate warm-up, followed by a constant learning rate of $2\mathrm{e}{-4}$. We set the weight decay to zero and disable drop path. To accommodate the higher memory requirements, we use activation checkpointing for all the layers of the model and use sharding for both the weights and the gradients. 

\paragraph{CAMS 0.4\degree{} analysis.}
We train with single-step prediction, \SI{12}{h} in this case, and the batch size is fixed to 1 per GPU.
We use a linear warmup of 100 steps from zero, but we do not use a learning rate schedule after that. 
We also use no weight decay, disable drop path, use activation checkpointing for all the layers of the model, and use sharding for the weights and gradients. 

Fine-tuning on CAMS analysis data proceeds in two steps.
In the first step, we fine-tune on CAMS reanalysis data using 16 GPUs for \SI{22}{k} steps at the high learning rate and then for \SI{14.5}{k} steps at the low learning rate.
The high learning rate is $1\mathrm{e}{-3}$ for the encoder patch embeddings of \emph{only} the new pollution variables and $1\mathrm{e}{-4}$ for the rest of the the network.
The low learning rate is $1\mathrm{e}{-4}$ for the encoder patch embeddings of only the new pollution variables and $1\mathrm{e}{-4}$ for the rest of the network.
To ensure maximum transfer from the CAMS reanalysis data to the CAMS analysis data, the CAMS reanalysis data is regridded to the resolution of CAMS analysis data, \SI{0.4}{\degree}.
In the second step, we fine-tune on CAMS analysis data using 8 GPUs for \SI{7.5}{k} steps at the high learning rate and finally for \SI{5.5}{k} steps at the low learning rate.
The final model is fine-tuned for \SI{49.5}{k} steps in total.

\paragraph{HRES-WAM 0.25\degree{} analysis.}
We fine-tune the weights of the entire model for \SI{14}{k} steps across 8 GPUs, with a batch size of 1 per GPU.
For the learning rate, we use a linear warmup of 500 steps to $1\mathrm{e}{-3}$ for the encoder and decoder parameters of the new wave variables and $3\mathrm{e}{-4}$ for the rest of the network.
The learning rate is then annealed to $1\mathrm{e}{-5}$ according to a cosine schedule for \SI{30}{k} steps (the run was stopped at \SI{14}{k} steps).
Afterwards, the training data was restricted to the range Jul 2018--2021:
on 5 June 2018, IFS cycle 45r1 was implemented, which significantly changed the distribution of the data by coupling of the three-dimensional ocean and atmosphere.
For this adjusted training data range, we further fine-tune the weights of the entire model for \SI{10}{k} steps across 8 GPUs, with a batch size of 1 per GPU.
For the learning rate, we again use a linear warmup of 500 steps to $5\mathrm{e}{-5}$ for all parameters.
The learning rate is then annealed to $1\mathrm{e}{-5}$ according to a cosine schedule for \SI{10}{k} steps.
The final model is fine-tuned for \SI{24}{k} steps in total.

\subsection{Roll-out fine-tuning} 
\label{sec:rollout_finetuning}

To ensure long-term multi-step dynamics, AI models typically fine-tune the model specifically for roll-outs. Backpropagating through the autoregressive roll-outs for a large number of steps is unfeasible for a 1.3B parameter model such as Aurora. This is particularly true at 0.1\degree{} resolution, where even a single step roll-out requires close to the memory limit of an A100 GPU with 80GB of memory. 

\begin{figure}[t]
    \centering
    \includegraphics[width=0.8\textwidth]{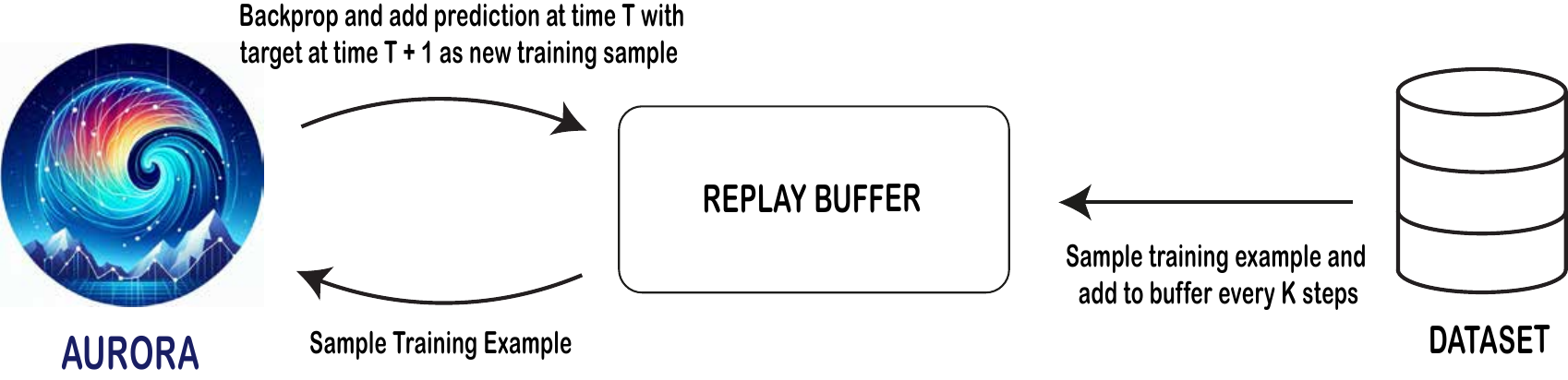}
    \caption{Diagram of the roll-out fine-tuning procedure. The replay buffer is initially populated with samples from the dataset. At each fine-tuning step, the model fetches a training sample from the replay buffer, performs a training step, and then it adds this new prediction (together with its next step target from the dataset) to the replay buffer. Every $K$ steps, the replay buffer is refreshed with a new training sample from the dataset. Since the replay buffer is generally much smaller than the dataset size, this ensures that enough samples from the dataset are seen. Overall, this procedure allows the model to train fast on an evolving distribution of autoregressive roll-out steps coming from a mixture of model versions. This avoids the need to run expensive roll-out procedures at each step.}
    \label{fig:rollout_finetuning}
\end{figure}

We use Low Rank Adaption (LoRA)~\citepapp{hu2021lora} layers for roll-out fine-tuning all the linear layers involved in the self-attention operations of the backbone. This allows us to take advantage of the large model size and the fact that it can be easily adapted once pretrained. 
That is, for each linear transformation $W$ involved in the Swin self-attention layers, we learn low-rank matrices $A, B$ to modulate the outputs of $W$ for an input $x$ via $Wx + BAx$. For more details, see \citetapp{hu2021lora}.

Furthermore, to avoid any memory increases compared to single-step fine-tuning, we use the ``pushforward trick'' introduced in \citetapp{brandstetter2022message}, where gradients are propagated only through the last roll-out step. 
We run this at scale by using an in-memory replay buffer to avoid delays with generating long roll-outs on each training step, similarly to how it is used in deep reinforcement learning~\citepapp{lin_experience_replay, Mnih2015HumanlevelCT} (\cref{fig:rollout_finetuning}). 
At each training step, the model samples an initial condition from the replay buffer, computes a prediction for the next time step,  then adds this new prediction back to the replay buffer. We periodically fetch fresh initial conditions from the dataset and add them to the replay buffer (i.e., the dataset sampling period).
This procedure allows the model to train at all roll-out steps without extra memory or speed penalties. 

\paragraph{HRES 0.25\degree{} analysis.} We use 20 GPUs to fine-tune the LoRA layers for \SI{13}{k} steps, each with a buffer size of 200. 
This results in a total replay buffer size of 4000 samples. We use a dataset sampling period of 10 steps. 
To ensure the model learns to predict the early steps well (i.e., shorter lead-times) before attempting to predict the later time steps, we use a schedule where for the first \SI{5}{k} steps, we only keep predictions up to 4 days ahead in the buffer. The 4--10 day lead times are allowed in the buffer only after \SI{5}{k} steps. We use a constant learning rate of $5\mathrm{e}{-5}$.

\paragraph{HRES 0.1\degree{} analysis.} Since the 0.1\degree{} data is $6.25\times$ larger than 0.25\degree{} data, we use 32 GPUs with a buffer size of 20 on each GPU. 
This is the maximum we can fit in the CPU memory of each node (i.e., \SI{880}{GB}). We use a dataset sampling period of 10 steps. We train the LoRA weights of the model for \SI{6.25}{k} steps using a constant learning rate of $5\mathrm{e}{-5}$.

\paragraph{CAMS 0.4\degree{} analysis.}
We use 16 GPUs with a buffer size of 200 on each GPU and a dataset sampling period of 10 steps.
We train the LoRA weights of the model for \SI{6.5}{k} steps using a constant learning rate of $5\mathrm{e}{-5}$.

\paragraph{HRES-WAM 0.25\degree{} analysis.}
We use 8 GPUs with a buffer size of 100 on each GPU and a dataset sampling period of 10 steps.
We train the LoRA weights of the model for \SI{6}{k} steps using a constant learning rate of $5\mathrm{e}{-5}$.

\section{Data infrastructure}
\label{sec:data_infrastructure}

Data loading represented one of the major technical challenges when training Aurora. First, due to the sheer size of one datapoint (close to 2GB for 0.1\degree{} data), loading the data efficiently on the GPUs becomes extremely challenging. Second, training Aurora on many heterogeneous datasets with different resolutions, variables, and pressure levels, makes the task even more difficult as batching together samples from different datasets becomes nontrivial and the workload of the GPUs must be balanced. In what follows, we describe the data storage and data loading infrastruture we built to handle these two problems.

\paragraph{Data storage and preprocessing.} Because of the large size of the datasets, the data cannot be stored locally, so must be stored with a cloud solution.
We use Azure's blob storage.
To ensure that data can be efficiently downloaded from blob storage, we perform several optimizations in how the data is stored:
\begin{enumerate}
    \item We co-locate data and compute to minimise latency and costs. 
    \item We preprocess each dataset into small chunks/files in order to make sure that a worker does not have to download substantially more data than necessary to access a given sample from the dataset. This is important because the raw form of many datasets stores a month (or more) of data into a single file. We found the Zarr format to particularly useful for this. 
    \item For large datasets (like ERA5), we compress the files in order to minimise network bandwidth when downloading samples.
    \item For large datasets, we also make sure that all the variables for a given timestep are stored in the same file. This minimises the number of concurrent file downloads that have to be performed to fetch one sample.
\end{enumerate}

\begin{figure}[t]
    \centering
    \includegraphics[width=1.0\textwidth]{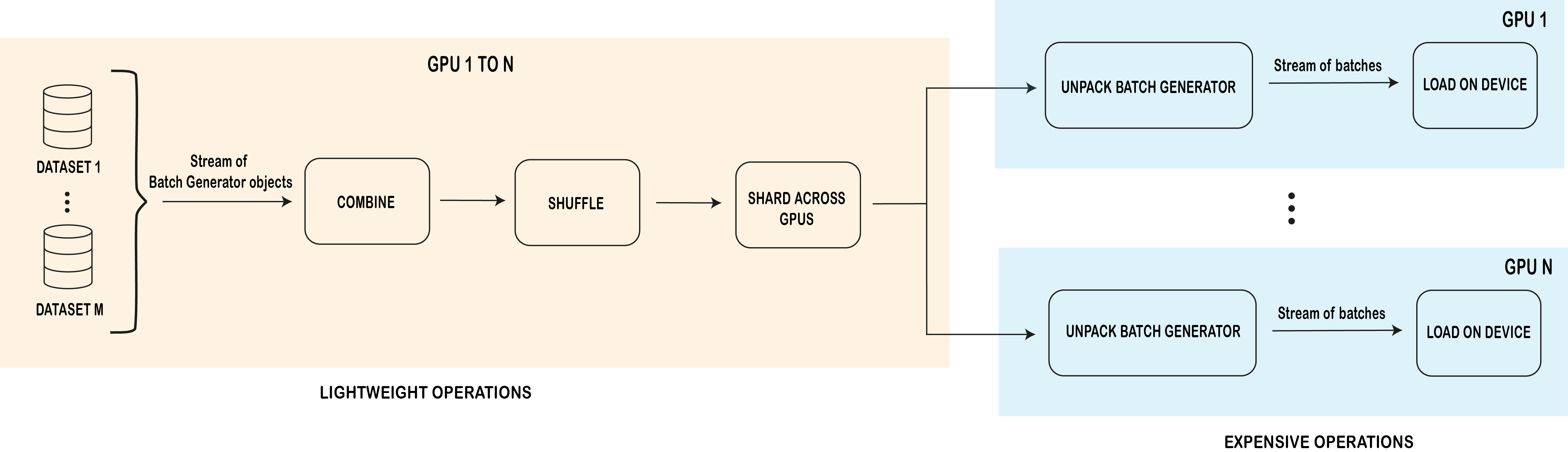}
    \caption{
        Overview of the multi-source dataloader. Each dataset produces a stream of lightweight \texttt{BatchGenerator} objects. The streams are combined, shuffled, and sharded across GPUs. After sharding, each GPU receives a fraction of the \texttt{BatchGenerator} objects, which are then unpacked. This step performs the expensive procedures like downloading the data associated with the object from Azure blob storage. This procedure ensures that only samples from the same together are batched together and provides the flexibility to perform custom per-dataset operations. 
    }
    \label{fig:dataloader}
\end{figure}

\paragraph{Data loading.} To load heterogoenous data efficiently, we developed an advanced multi-source data loading pipeline that can satisfy these requirements (\cref{fig:dataloader}). Each dataset is instantiated by a \texttt{yaml} config specifying how the data from that dataset should be loaded (e.g., dataset location, which lead time should be used, the number of past steps to be supplied as input, etc.). Then, each dataset generates a stream of \texttt{BatchGenerator} objects. These are lightweight objects containing the necessary metadata (e.g., file paths) to download and generate at least one batch from the associated dataset. The streams of \texttt{BatchGenerator} objects from all the datasets are combined into one stream via sampling, which allows us to adjust the relative proportions of the datasets within each epoch. This is followed by shuffling all the \texttt{BatchGenerator} objects and sharding them across GPUs. Up to sharding, all these lightweight operations are run identically on all the GPUs using the same random seeds. 

The expensive operations of the data loading pipeline follow the sharding. The \texttt{BatchGenerator} objects sent to each GPU are unpacked into actual batches of data that can be used for training and inference. This unpacking process, consists of a series of sequential transformations that are specific to each dataset and the particular way it instantiates its \texttt{BatchGenerator} objects. In general, the sequence of transformation includes: downloading some (compressed) dataset files, decompressing them, reading the data from the files, generating the samples, mapping the variable names to a canonical set of names, performing data checks (e.g., handling NaNs), batching, and, finally, loading batches to the device. All of these operations are performed across multiple workers.

This data loading pipeline enables efficient training on multiple heterogeneous datasets. 
Performing batched computations is straightforward since only samples from the same dataset are batched together, and therefore there are no any additional constraints imposed on the model. 
To balance workloads across GPUs and workers, one can use bigger batch sizes for the smaller datasets and smaller batch sizes for the bigger datasets, since samples from different datasets have different sizes. 

\citetapp{dehghani2024patch} approach the problem of training ViTs on multi-resolution images at the model layer level, by concatenating tokens from different images and using masking to prevent tokens from different images from communicating during self-attention operations. 
However, their approach imposes a constraint on the model architecture and it is difficult to implement in more sophisticated ViT architectures such as the Swin Transformer~\citepapp{liu2021swin,liu2022swin}. Masked self-attention, even if implemented with sparse matrix multiplications, is not able to achieve the expected theoretical performance due to the lack of semi-structured sparsity support on current hardware. Therefore, another advantage of our data loading pipeline is that it allows for the freedom to experiment on the model architecture side.

\section{Verification metrics}
\label{sec:verification_methods}

\subsection{Main evaluation metrics}
\label{sec:main_eval_metrics}

The main metrics we use to measure the performance of Aurora against other methods are the root mean square error (RMSE) and the anomaly correlation coefficient (ACC).
These metrics use a latitude weighting because the variables and targets live on a non-uniform grid, 
\begin{align}
    w(i) = \frac{\cos(\text{lat}(i))}{\frac{1}{H}\sum_{i'=1}^{H} \cos(\text{lat}(i')) },
\label{eq:lat-w}
\end{align}
which downweights smaller cells (closer to the poles) and upweights bigger cells (closer to the equator). The weights are normalised to unit mean. While various versions of this weighting are used in practice, the form used here is the one from \citetapp{rasp_wb1}.

\paragraph{Root mean square error (RMSE).} The latitude-weighted RMSE measures the magnitude of the errors between the predictions and the ground truth,
\begin{align} \label{eq:rmse}
    \text{RMSE} = \frac{1}{T}\sum_{t = 1}^T\sqrt{\frac{1}{H \times W}\sum_{i=1}^{H}\sum_{j=1}^{W} w(i) (\hat{X}^t_{i,j} - {X}^t_{i, j})^2},
\end{align}
where $t$ indexes over the sample datasets, $i, j$ index over the latitude and longitude of each image, and $w(i)$ is the latitude weighting factor.

\paragraph{Adjustments of the RMSE for wave variables.}
For angle-valued variables, when computing the difference between two angles, we need to account the fact that $0\degree$ is equal to $360\degree$, meaning that e.g.\ the absolute difference between $10\degree$ and $350\degree$ should be $20\degree$ and not $340\degree$.
We account for this by replacing $|d| = |\hat{X}^t_{i,j} - {X}^t_{i, j}|$ with $\min(|d|, 360\degree - |d|)$.

In addition, when computing the RMSE, because wave variables have missing data and Aurora can also predict the absence of data (see \cref{sm:wave-extension}), either or both $\hat{X}^t_{i,j}$ and ${X}^t_{i,j}$ may be undefined.
In \cref{eq:rmse}, we compute the average only over points where the target ${X}^t_{i,j}$ is defined.
When doing this, we correctly renormalise the latitude weights to average to one.
Wherever the target ${X}^t_{i,j}$ is defined but the prediction $\hat{X}^t_{i,j}$ is undefined, we replace $\hat{X}^t_{i, j}$ with the normalisation mean.

\paragraph{Anomaly correlation coefficient (ACC).} The ACC measures the correlation between the deviation of the prediction and the ground truth from the daily climatology (i.e., the daily mean of a variable for that day of the year). It takes the form
\begin{align}
    \text{ACC} = \frac{1}{T}\sum_{t =1}^T \ddfrac{
        \sum_{i=1}^{H}\sum_{j=1}^{W} w(i) (\hat{X}^t_{i,j} - C^t_{i, j})(X^k_{i,j} - C^t_{i, j})
    }{
        \sqrt{\left[\sum_{i=1}^{H}\sum_{j=1}^{W} w(i) (\hat{X}^t_{i,j} - C^t_{i, j})^2\right]\left[\sum_{i=1}^{H}\sum_{j=1}^{W} w(i) (X^t_{i,j} - C^t_{i, j})^2\right]}
    },
\end{align}
where $C^t_{i,j}$ is the daily climatology for location $i, j$ for the day of the year corresponding to time $t$. Our daily climatology is computed based on the climatology provided by \citetapp{rasp2023weatherbench}. 

\subsection{Extreme weather metrics}
\label{app:extreme_metrics}
\paragraph{Thresholded RMSE.} The thresholded RMSE measures the magnitude of the errors between the predictions and the ground truth. We apply an additional threshold here to indicate which latitude-longitude gridpoints should be included in the sum. This can be written as:
\begin{align}
\label{eq:threshold_rmses}
\text{RMSE}_{g}= \frac{1}{T}\sum_{t=1}^{T}\sqrt{\sum_{i=1}^{H}\sum_{j=1}^{W} \tilde{w}^{g, t}_{i, j} (\hat{X}^{t}_{i,j} - {X}^{t}_{i, j})^2}.
\end{align}
When computing RMSEs above a threshold, the weighting factor for the right-sided thresholds (plotted on the right side of the relevant plots) is then defined by:
\begin{align}
\label{eq:rhs}
\tilde{w}^{g, t}_{i, j} = \frac{\mathds{1}({X}^{t}_{i, j} > b^{g}_{i, j})w(i)}{\sum_{i=1}^{H}\sum_{j=1}^{W}\mathds{1}(X^{g}_{i, j} > b^{g}_{i, j})w(i)}.
\end{align}
For the left-sided thresholds, this is instead defined by:
\begin{align}
\label{eq:lhs}
\tilde{w}^{g, t}_{i, j} = \frac{\mathds{1}({X}^{t}_{i, j} < b^{g}_{i, j})w(i)}{\sum_{i=1}^{H}\sum_{j=1}^{W}\mathds{1}(X^{g}_{i, j} < b^{g}_{i, j})w(i)}.
\end{align}
This incorporates the threshold $b$ defined as:
\begin{align}
    b_{i, j} = \mu_{i, j} + g \cdot \sigma_{i, j},
\end{align}
where $\mu$ and $\sigma$ are the mean and standard deviation of the ERA5 reanalysis data over all training years, computed for each latitude-longitude point $i, j$. Here, $g$ is a factor that is varied linearly for both positive and negative values and is used to obtain the thresholded RMSEs at a number of points, and in the figures we plot the resulting curves. 
The latitude weighting factor $w(i)$ is as defined in~\cref{eq:lat-w}. 
The value is computed for every example frame, and this set of values, sampled with replacement, is used to bootstrap 95\% confidence intervals by recomputing~\cref{eq:threshold_rmses}. For $g > 0$, we plot $\text{RMSE}_{g}$ computed using~\cref{eq:rhs}, and for $g < 0$, we plot $\text{RMSE}_{g}$ computed using~\cref{eq:lhs}.

The plots show a discontinuity at zero; this is to improve their readability. 
We remove the lower half of the right-hand side of the curve and the upper half of the left-hand side of the curve.  
However, it is important to point out that $\text{RMSE}_{g}$ is directly comparable to $\text{RMSE}$: $\text{RMSE}_{g}$ computed using~\cref{eq:lhs} follows $\text{RMSE}_{g} \rightarrow \text{RMSE}$ as $g \rightarrow \infty$, while $\text{RMSE}_{g}$ computed using~\cref{eq:rhs} follows $\text{RMSE}_{g} \rightarrow \text{RMSE}$ as $g \rightarrow -\infty$.

\section{Data diversity and model scaling improve atmospheric forecasting}
\label{sec:scaling}

A key tenet of the foundation model paradigm is that performance improves in a significant and predictable fashion as the data and the model are scaled up \citeapp{hoffmann2022training, kaplan2020scaling}.
We provide evidence that these findings from natural language processing and computer vision also hold for atmospheric forecasting.  

\begin{figure}[t]
\begin{subfigure}[t]{\linewidth}
    \centering
    \includegraphics[width=\textwidth]{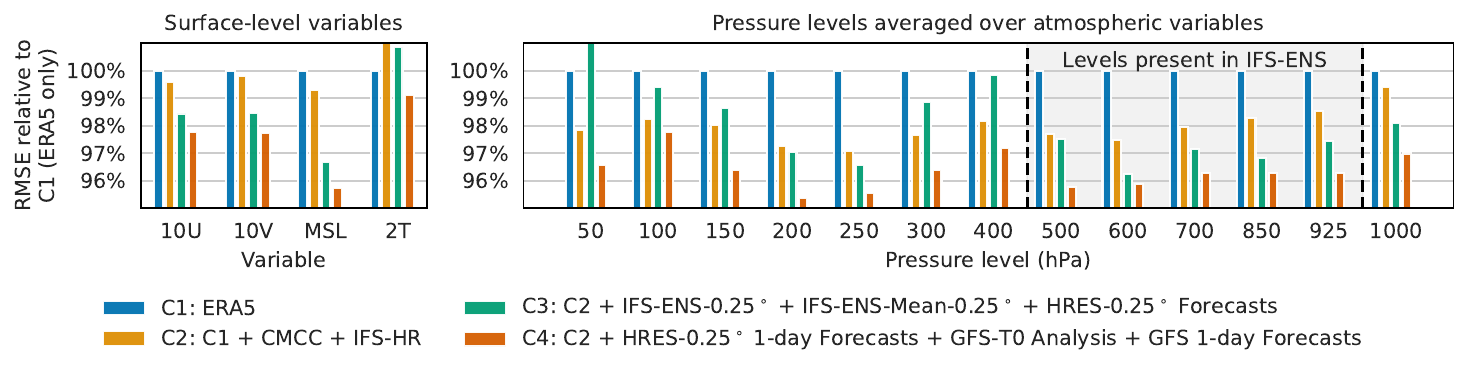}
    \vspace{-140pt}
    \caption{}
    \label{fig:data-scaling-era5-main}
\end{subfigure}
\begin{subfigure}[t]{\textwidth}
    \centering
    \includegraphics[width=\linewidth]{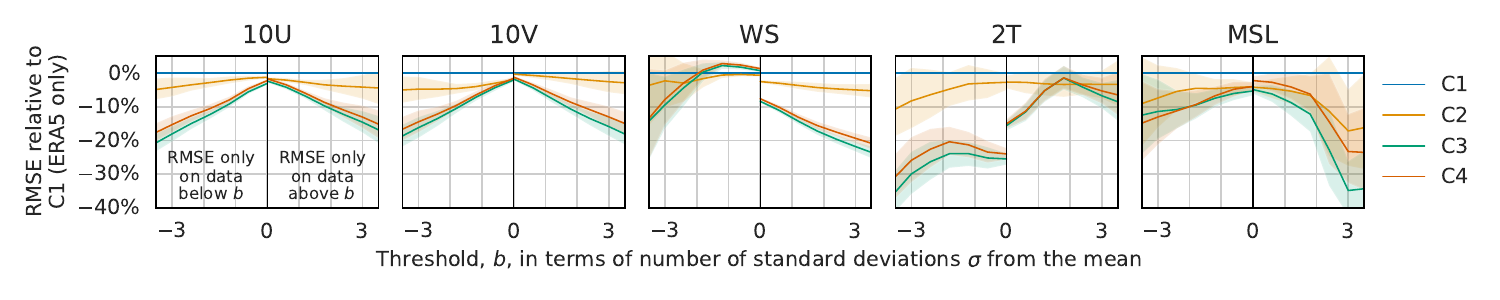}
    \vspace{-115pt}
    \caption{}
    \label{fig:data-scaling-hres-extremes}
\end{subfigure}
\begin{subfigure}[t]{\linewidth}
    \centering
    \includegraphics[width=\textwidth]{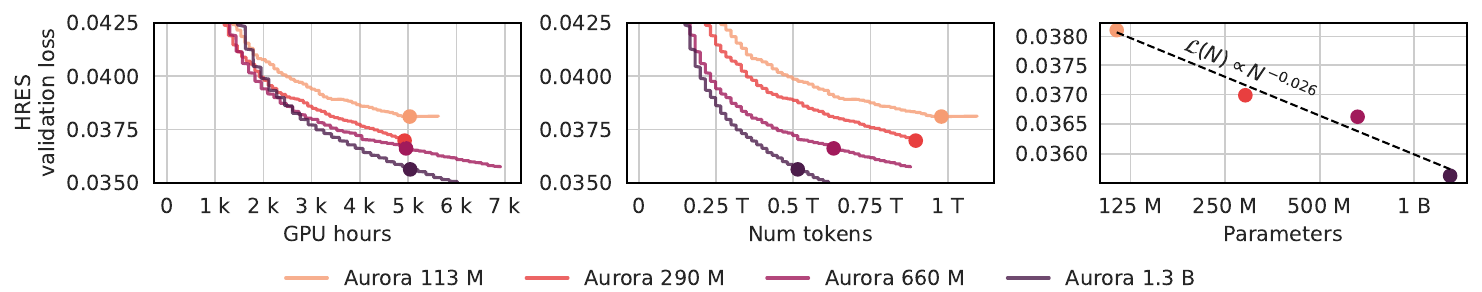}
    \vspace{-115pt}
    \caption{}
    \label{fig:model-scaling}
\end{subfigure}
    \caption{
        \textbf{Pretraining on diverse data and increasing model size improves performance.}
        \textbf{a}:
            Performance on ERA5 2021 at \SI{6}{h} lead time for models pretrained on different dataset configurations, labeled C1--C4, without fine-tuning.
            Adding low-fidelity simulation data from CMIP6 (i.e., CMCC and IFS-HR) improves performance almost uniformly (C2). Adding even more simulation data improves performance further on most surface variables and for the atmospheric levels present in this newly added data (C3). Finally, configuration C4, which includes comprehensive atmospheric coverage and analysis data from GFS, achieves the best overall performance with improvements across the board.  
        \textbf{b}:
            For the same configurations considered in \textbf{a}, performance for extreme values on IFS HRES 2022 at \SI{6}{h} lead time.
            Shows RMSEs computed only on data below (left panes) or above (right panes) a threshold $b$.
            Pretraining on many diverse data sources also improves the forecasting of extreme values.
        \textbf{c}:
            Bigger models obtain lower IFS HRES validation loss for the same number of GPU hours.
            At \SI{5}{k} GPU hours, we find that the validation loss behaves like $\mathcal{L}(N) \propto N^{-0.026}$ where $N$ is the number of parameters, which corresponds to a $6\%$ reduction in validation loss for every $10\times$ increase in model size.
    }
\end{figure}

% \paragraph{Data scaling.}
Most medium-range AIWP approaches have employed a single dataset in their training strategy, usually ERA5 at 0.25\degree~resolution. 
Training Aurora on multiple datasets allows us %, for the first time,
to examine the benefits of including diverse datasets in pretraining. To that end, we pretrain the main Aurora model on four different dataset configurations (labelled C1--C4) and compare (1) the RMSE of the models across different variables and levels on ERA5 2021 (\cref{fig:data-scaling-era5-main}) and (2) the RMSE of the models for extreme values at the surface on HRES 2022 (\cref{fig:data-scaling-hres-extremes}). Neither of these two years of data were included in pretraining and no further fine-tuning was applied to these models. 

% \afterpage{
% \clearpage
% }

Adding climate simulation data consisting of CMIP6-CMCC and CMIP6-IFS-HR alongside high-quality ERA5 reanalysis data (C2) results in lower RMSEs than the ERA5-only model (C1) across all the groups of variables (except for 2T) (\cref{fig:data-scaling-era5-main}). 
Although these data are of lower fidelity compared to reanalysis datasets like ERA5, including climate simulations increases the diversity of the data and, as our results suggest, generalization performance.
When also including IFS ensemble data, IFS HRES forecast data, and the IFS ensemble mean (C3), the performance improves even further at the surface and around the levels present in the IFS-ENS data, which are highlighted in the figure. 
Since the improvements seen with C3 diminish in other parts of the atmosphere, we also test an alternative configuration (C4), which provides better coverage of the entire atmosphere. Additionally, it limits the samples used in training for the forecasts datasets to one-day lead-times to increase the data quality. This last configuration performs best overall and, for this reason, it is also the configuration used for the main results reported in this paper. 
Crucially, notice that the downstream validation dataset we use here is a test year of ERA5 itself, so we see positive transfer when including additional datasets even in the case where the downstream tasks overlaps perfectly with ERA5.

Including additional data in pretraining not only improves the RMSEs in aggregate, but it also leads to significant improvements (up to 35\%)  in the tails of the distribution at the surface on HRES 2022. 
\cref{fig:data-scaling-hres-extremes} shows 
RMSEs computed across all cases where the underlying variable is higher than a threshold value (see \cref{app:extreme_metrics}) for five surface-level variables. 
As before, C2 improves over C1, while C3 and C4 further improve upon C2. In all instances, the improvements compared to the ERA5-pretrained model (C1) are mostly in the extremes. 
We postulate that this is because climate simulations and diverse data sources expose the model to more rare events than historical data can provide, which  ultimately improves the model robustness in these regimes. 

Overall, these results provide evidence that using a heterogeneous group of weather and climate datasets, of mixed fidelity, can improve forecasting performance. In \cref{sec:app_data_scaling} we verify that these findings also hold across the individual atmospheric variables and include additional results on HRES 2022. Finally, we discuss one other dataset configuration that does not yield performance improvements. 

Orthogonal to the data scaling dimension, we test the impact of model scaling on four pretrained Aurora configurations, each corresponding to a different model size: \SI{113}{M}, \SI{290}{M}, \SI{660}{M} and \SI{1.3}{B} parameters (see \cref{sm:hyperparameters}). All models are pretrained on the dataset mixture from C4 and continuously validated on IFS analysis data. After the initial training phase, bigger models achieve lower validation losses for the same number of GPU hours (\cref{fig:model-scaling}). The bigger models achieve this whilst being trained on fewer data samples (measured in number of tokens in the figure). When fitting a power law to the performance of the three models at the \SI{5}{k} GPU hours mark, we obtain a scaling law suggesting that validation performance improves by approximately 6\% for every $10\times$ increase in model size.
Note that the validation performance on IFS analysis data is computed without fine-tuning on IFS analysis data, to limit computational costs.

\FloatBarrier

\begin{figure}[t]
    \begin{subfigure}[t]{\linewidth}
        \centering
        \includegraphics[width=\textwidth]{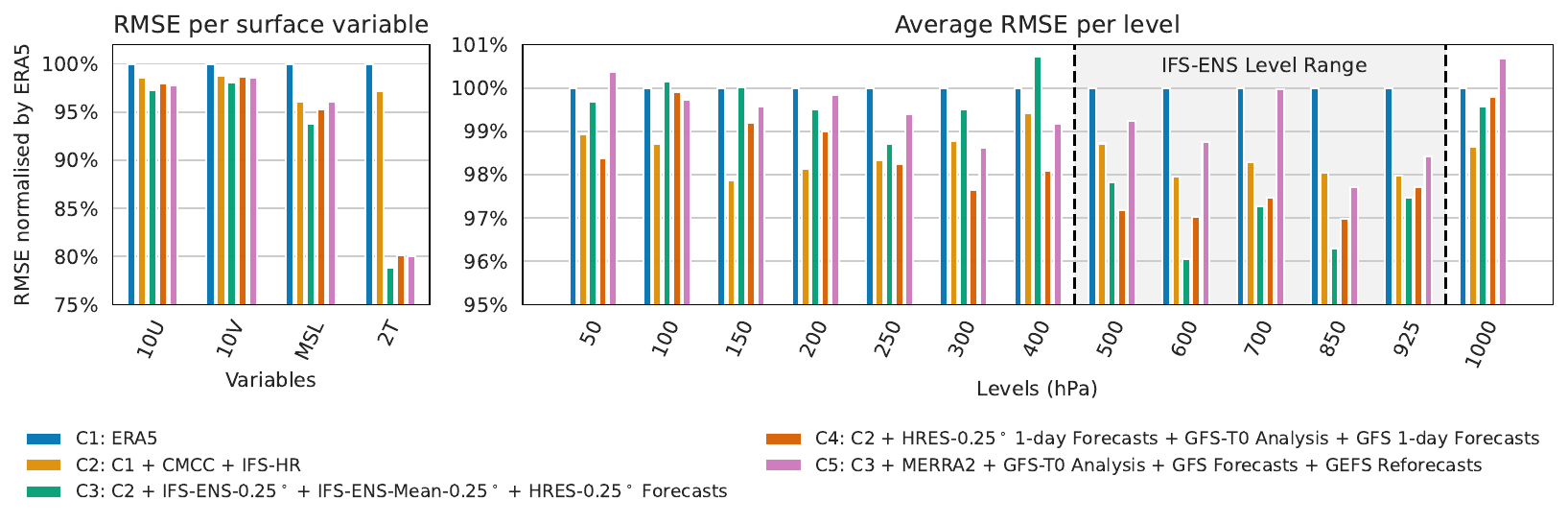}
        % \vspace{-180pt}
    \end{subfigure}
    \begin{subfigure}[t]{\linewidth}
        \centering
        \includegraphics[width=\textwidth]{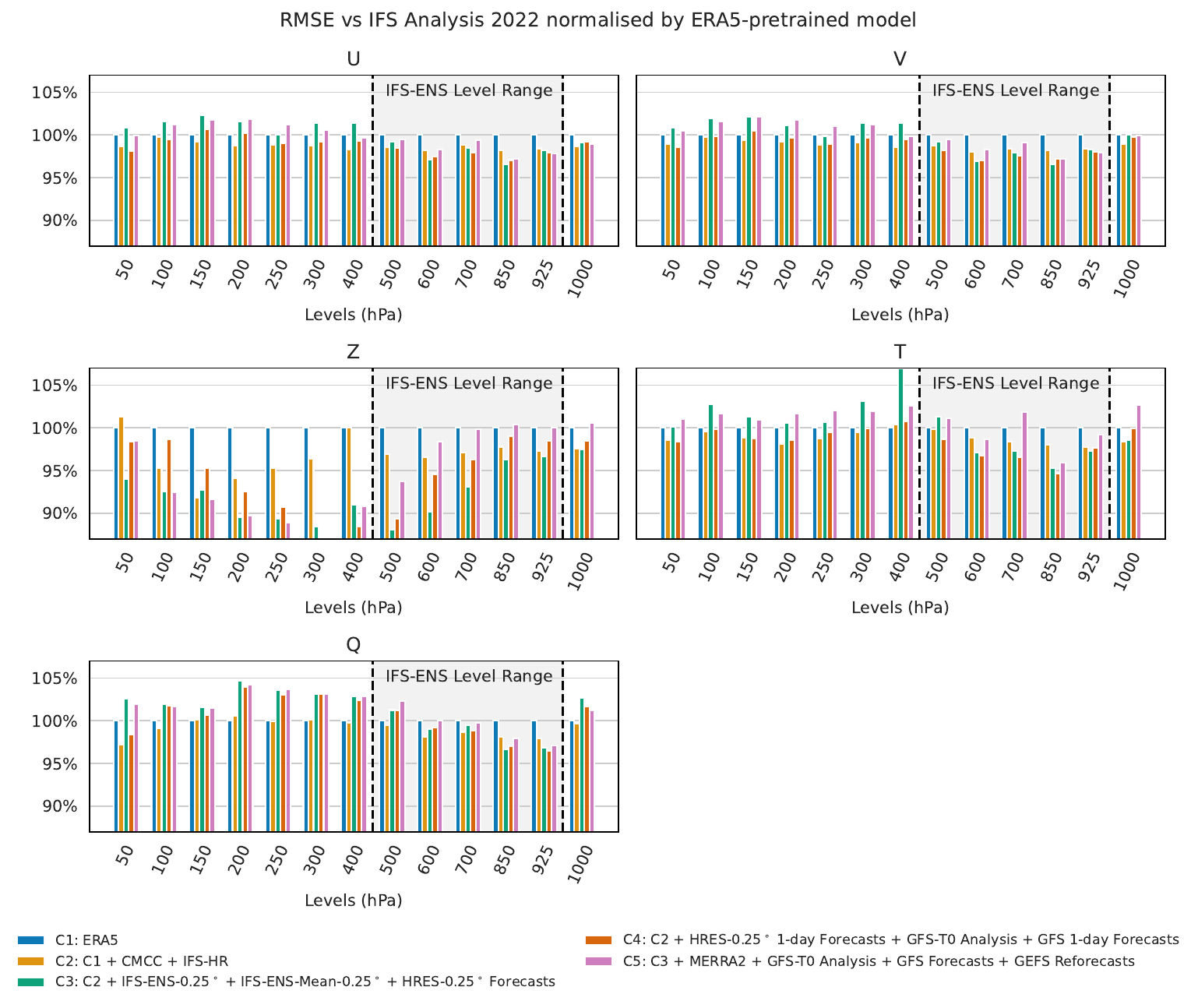}
        % \vspace{-180pt}
    \end{subfigure}
    \caption{Effects of data scaling measured vs HRES-T0 analysis 2022 across five dataset configurations. The top part of the figure shows aggregate RMSE per level normalised by the performance of the ERA5 pretrained model. The bottom part shows performance across all the individual atmospheric variables and levels.}
    \label{fig:data-scaling-hres}
\end{figure}

\subsection{Effects of data diversity}
\label{sec:app_data_scaling}

The following two subsections offer an overview of the main reasons why pretraining on multiple datasets can have beneficial effects and a more detailed discussion of our results together with additional experiments. 

\subsubsection{Why increase the data diversity?}

The use of several different dataset to train Aurora can help along several dimensions:

\begin{enumerate}
    \item Climate simulations expose Aurora to entirely new simulated weather patterns not present in observational datasets, effectively increasing the variety and size of our training data. These simulations capture potential future atmospheric states and extreme events that may be underrepresented in historical data.
    \item Forecast datasets provide additional examples of realistic weather patterns that did not necessarily occur, again increasing the diversity of training examples. These forecasts help the model learn about plausible atmospheric evolution paths beyond what is captured in analysis data alone. 
    \item While GFS differs from IFS in its implementation, it models similar underlying physics with different parameterizations. This variation helps Aurora learn more robust representations of atmospheric dynamics that can avoid inheriting the biases of an specific NWP model.
    \item Different datasets have varying spatial resolutions, numbers of pressure levels, and quality of physics models. By training on this heterogeneous mix, Aurora learns to extract the best features from each source, e.g., high-resolution details from some datasets, better vertical structure from others, and improved physical consistency from the most sophisticated models.
\end{enumerate}

Overall, pretraining on the output of numerical simulators can be seen as an analogous practice to the training of large language models on synthetically generated data~\citeapp{gunasekar2024textbooks}, a standard practice in the field. 

\subsubsection{Detailed analysis}

\begin{figure}[t]
    \begin{subfigure}[t]{\linewidth}
        \centering
        \includegraphics[width=\textwidth]{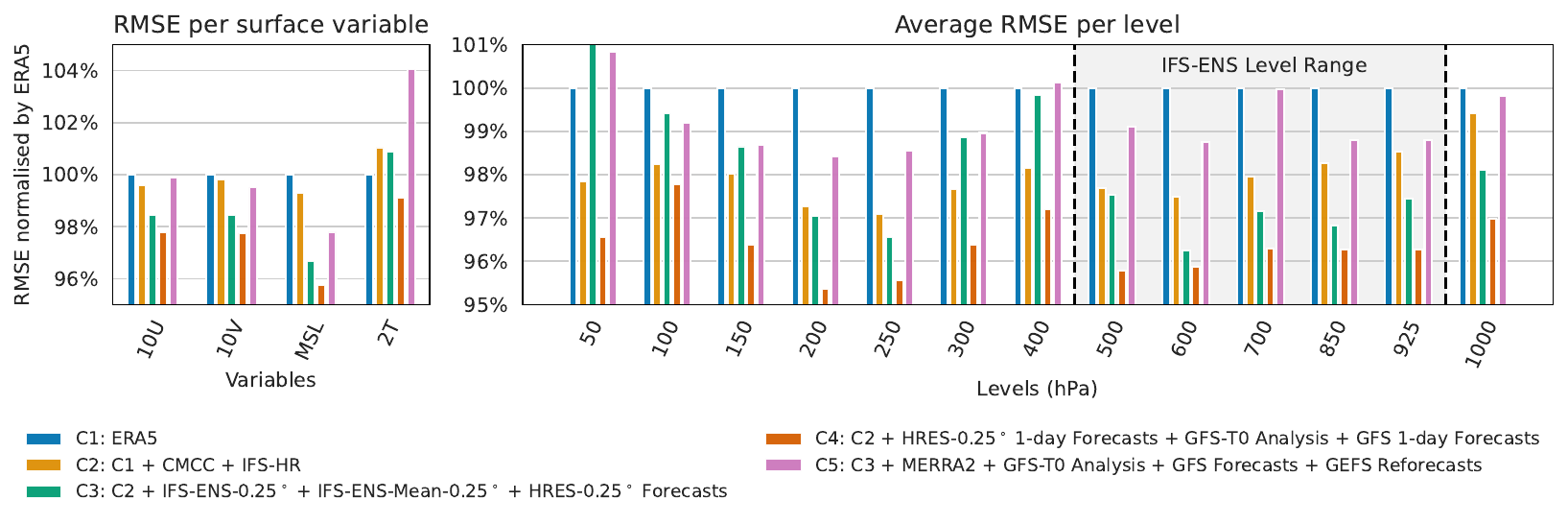}
        % \vspace{-180pt}
    \end{subfigure}
    \begin{subfigure}[t]{\linewidth}
        \centering
        \includegraphics[width=\textwidth]{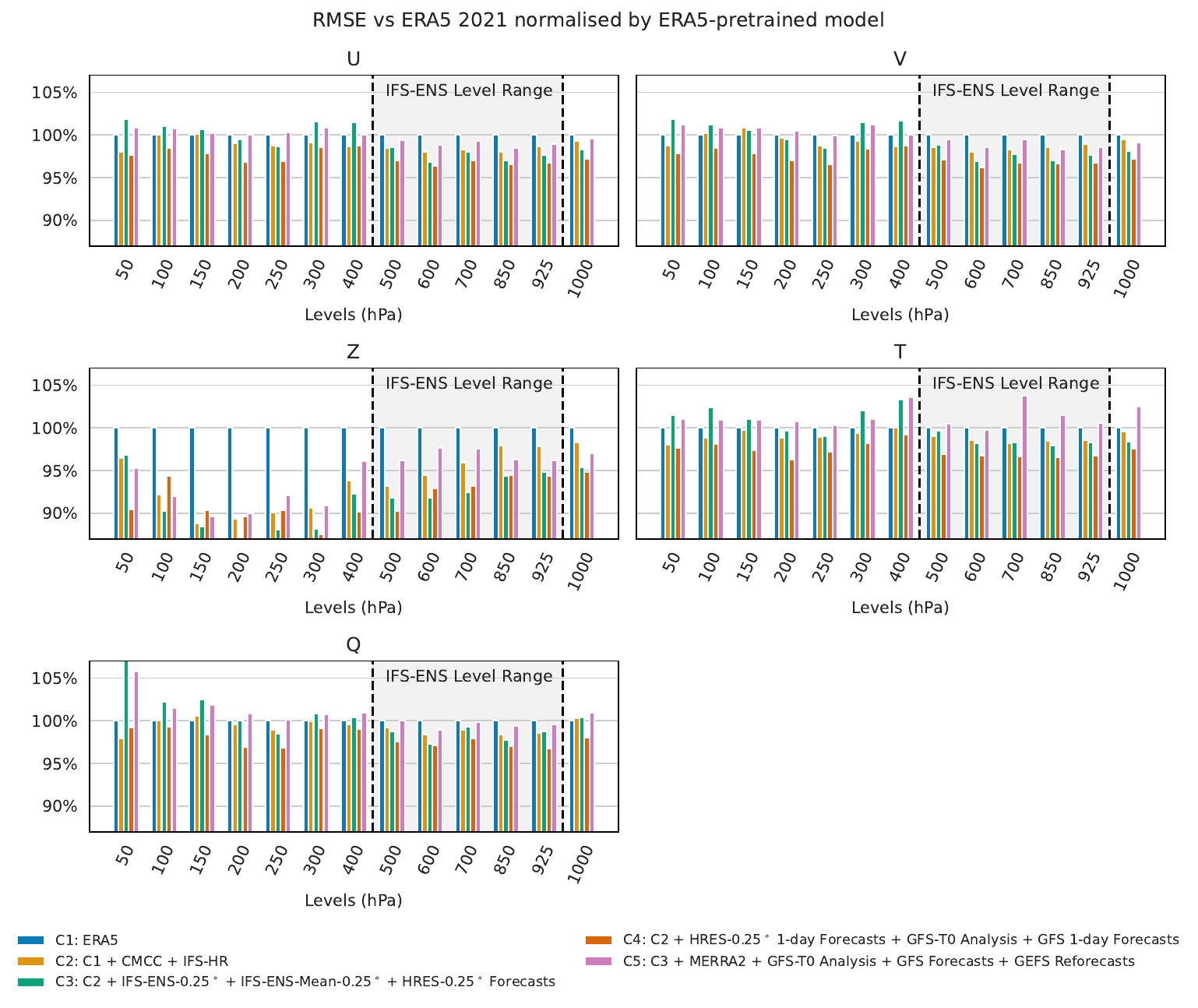}
        % \vspace{-180pt}
    \end{subfigure}
    \caption{Effects of data scaling measured vs ERA5 2021 across five dataset configurations. The top part of the figure shows aggregate RMSE per level normalised by the performance of the ERA5 pretrained model. The bottom part shows performance across all the individual atmospheric variables and levels.}
    \label{fig:data-scaling-era5}
\end{figure}

Here we examine the effects of increasing data diversity across all variables and levels in more detail. 
Besides the four dataset configurations C1--C4 discussed before, we also consider one additional dataset configuration based on all the datasets described in \cref{table:datasets-pretraining}, which we label as C5. This latter configuration contains all the datasets included in C3 and additionally Merra-2, GFS-T0 analysis, GFS forecasts and GEFS reforecasts. 
We evaluate all pretrained jobs both on ERA5 2021 (\cref{fig:data-scaling-era5}) and HRES-T0 analysis 2022 (\cref{fig:data-scaling-hres}). No further fine-tuning was performed after the pretraining.

On ERA5, we observe that the trend described before holds across all the individual atmospheric variables. C2 improves over C1 almost universally, C3 improves further over C2 on most atmospheric variables for those levels that are covered by the IFS-ENS and IFS-ENS-Mean datasets (which are the biggest datasets in C3). Similarly, C4 improves over C2, as expected due to the inclusion relationship between them. Furthermore, C4 also generally performs better than C3. 

Perhaps surprisingly, we notice that C5, which is a superset of all the other configurations, does not perform as well as other configurations and often does worse than the ERA5 job. There are couple of potential hypothesis behind this. A first possibility is that some of the newly introduced datasets in C5 do not align well with the evaluation dataset (ERA5). For instance, Merra-2 is the lowest resolution dataset in the entire pretraining mix and its addition might harm performance on a high resolution dataset like ERA5. At the same time, the GEFS reforecast dataset only includes three levels in the lower part of the atmosphere (850, 925, 1000 hPa), so it does not provide a good coverage of the atmosphere. Another possibility is that including such a large amount of data likely requires significantly more training than \SI{150}{k} steps to yield good performance. 

On HRES T0 2022, the trend for the surface variables looks looks similar to ERA5, with the exception of 2T. Since HRES T0 is missing an extra assimilation step for the surface, 2T in HRES T0 exhibits different biases compared to 2T in ERA5. We believe including any kind of IFS forecast data in pretraining helps alleviate this distribution mismatch, leading to substantial improvements. In terms of the atmosphere, performance also generally improves across most variables with the addition of more data, but the order of the configurations fluctuates more. One interesting case is specific humidity (Q), where for unknown reasons, C3 and C4 perform significantly worse than the ERA5-pretrained model on some levels. This could be a potential explanation for why Aurora performs slightly worse than GraphCast at certain lead times and levels of Q in the scorecard from \cref{fig:0.25deg_scorecard}.  

Finally, we note that on both test datasets, the biggest improvements from additional data on the atmospheric variables are obtained for the geopotential (Z). 
This is likely explained by the fact that the physical equations describing the evolution of the geopotential are well-understood and, therefore, the NWP-simulations we add in pretraining provide very high-quality data for this variable. Interestingly, the geopotential also improves universally on C2, despite being absent from the new datasets in this configuration. This demonstrates the ability of the model to learn useful patterns even from incomplete variable sets. 

\subsection{Model scaling laws for surface-level variables}
\label{app:model_scaling_surf_vars}

\begin{figure}[b]
    \includegraphics[width=\textwidth]{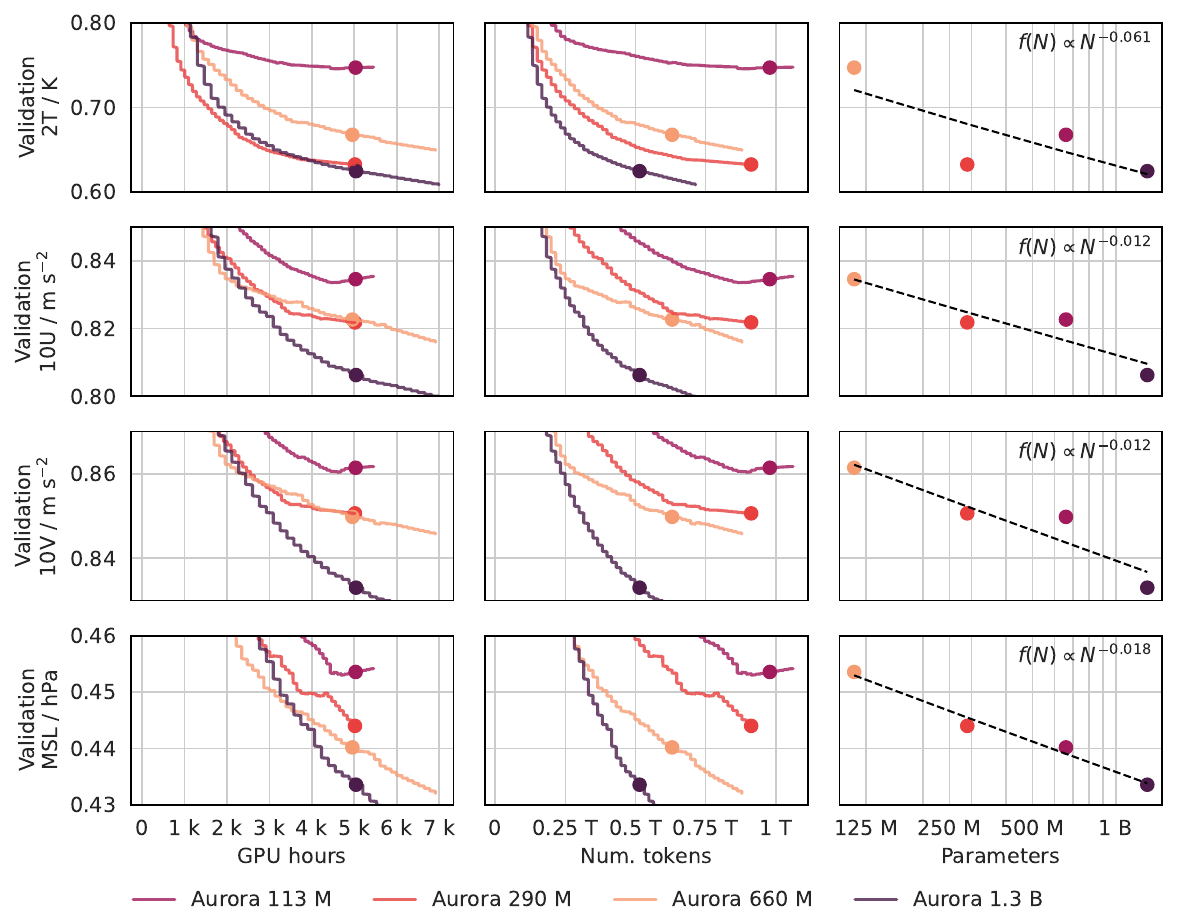}
    \hspace{5pt}
    \caption{
        Validation loss for all surface-level variables.
        For every surface-level variable, at \SI{5}{k} GPU hours, we find that the validation loss roughly behaves like $f(N) \propto N^{-\alpha}$ where $N$ is the number of parameters and $\alpha > 0$ is an estimated parameter.
    }
    \label{fig:model-scaling-surf-vars}
\end{figure}

In \cref{fig:model-scaling}, we
saw that bigger models obtain lower validation loss for the same number of GPU hours.
In \cref{fig:model-scaling-surf-vars}, we show that the same conclusion is also true for every surface-level variable individually.
Because the training loss is an aggregate of approximately 70 different variables (four surface-level variables and five atmospheric variables at thirteen different pressure levels), how individual variables scale as model size increases is more noisy than the aggregate loss.
Nevertheless, we observe clear scaling behaviour for every surface-level variable individually.

\FloatBarrier
\section{Comparison against GraphCast, Pangu, and IFS HRES at 0.25\texorpdfstring{\degree{}}{ degrees} resolution}
\label{sm:0.25_extra_results}

The vast majority of AI models for weather prediction have been (pre)trained at 0.25\degree{} resolution on a single dataset. As an additional validation of the benefits of fine-tuning a large model pretrained on many datasets, we compare Aurora against operational GraphCast~\citeapp{lam2023graphcast}, which is pretrained only on ERA5 and fine-tuned to IFS-HRES analysis, rendering it the closest setup to Aurora. Moreover, operational GraphCast is currently considered the state-of-the-art in operational AI weather forecasting at 0.25\degree~and lead times up to 5 days~\citepapp{rasp2023weatherbench}. As an additional datapoint, we include in the comparison Pangu weather \citepapp{bi2022pangu}, another AI model which is widely considered competitive. For consistency we apply the same evaluation procedure to all three AIWP models.
We also include IFS HRES forecasts in this comparison, which is the gold standard in operational numerical weather prediction.
We show that Aurora outperforms both when measured against IFS analysis, weather station observations, and extreme values. 

\subsection{Experimental setup}
This comparison uses the HRES-T0 2022 dataset as a test year and is carried out at the standard 0.25\degree{} resolution used in AIWP. To this end, Aurora is fine-tuned on HRES-T0 ``analysis'' data for years 2016-2021 (see Supplementary~\ref{app:dataset-inventory}), the same years that operational GraphCast was fine-tuned on. 
In order to ensure a consistent comparison against operational NWP systems, both GraphCast and Aurora are evaluated using daily 00 and 12 UTC initialisation times from the HRES-T0 dataset provided by WeatherBench2. As recommended, we treat HRES-T0 as the ground truth targets for this evaluation.  
To facilitate a complete comparison between models across different variables, pressure levels and lead times, we compute and visualize RMSE skill scores in the form of \emph{relative RMSE difference between the models that are compared}, normalized by the RMSE score of the model we compare against. We also include comparisons based on the ACC metric in \cref{sec:verification_studies}.

\paragraph{Comment on the evaluation dataset.}
It is important to clarify that this evaluation procedure does not compare against ERA5 (as seen in GraphCast) for two primary reasons: 
\begin{enumerate}
    \item This procedure evaluates the system under operational conditions following the protocol established by ECMWF \citepapp{benbouallegue2023it}. This represents a real-world operational evaluation. Note that GraphCast \citeapp{lam2023graphcast} had to use a rather \emph{ad hoc} procedure to attempt to compensate for the fact that ERA5 was used on the input and output side for their model (see section 5.2.1 of the GraphCast supplementary material). 
    \item The IFS HRES analysis is closer to the ground truth state of the weather than ERA5. This is because ERA5 uses an NWP model that was operational in 2017 \citepapp{benbouallegue2023it}. In the seven years since, IFS has undergone a large number of improvements. 
\end{enumerate}
We do not use the original WeatherBench2 evaluation protocol because until recently this used comparisons at 1.4\degree of ERA5 as inputs and targets. Not only is this a low spatial resolution, ERA5 inputs are reanalyses and as such do not reflect available information at the time of a forecast because they incorporate observational data from the future. Moreover, as mentioned above, ERA5 is considered less accurate to the ground truth than IFS HRES analysis. 
More recently, WeatherBench2 started to support the evaluation protocol outlined above. 

\subsection{Aurora versus GraphCast at 0.25\degree{}}
\label{sec:graphcast-compare}
\begin{figure}[ht]
    \centering
    \includegraphics[width=\columnwidth]{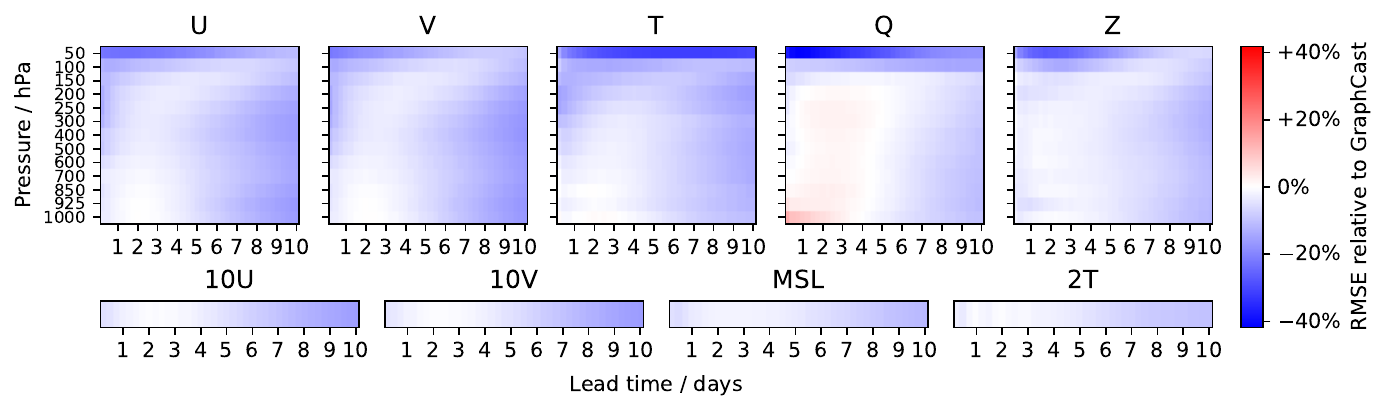}
    \caption{Scorecard comparing Aurora and GraphCast at $0.25^\circ$. Aurora matches or outperforms GraphCast on 94\% of targets. Aurora has the largest gains (40\%) over GraphCast in the upper atmosphere, where GraphCast performance has notable room for improvement. The two models are closest to each other in the lower atmosphere at the 2--3 day lead time, which corresponds to the lead time GraphCast was roll-out fine-tuned on.}
    \label{fig:0.25deg_scorecard}
\end{figure}

Aurora consistently outperforms GraphCast~\citepapp{lam2023graphcast} across the vast majority of target variables, pressure levels, and lead times (\cref{fig:0.25deg_scorecard}). The performance gains are most pronounced at lead times past 3 days, as well as for the upper atmospheric levels where we see a reduction in RMSE up to 40\%. 
Large improvements up to 10-15\% are observed at both short and long lead times, especially for variables such as temperature (T), geopotential height (Z) and wind velocity components (U, V). The smallest gains are observed for specific humidity, where GraphCast tends to yield more accurate predictions at shorter lead times (1--3 days) for the surface and lower atmospheric levels. 

At longer lead times, Aurora's forecasts more closely resemble the ensemble mean than GraphCast (see \cref{sec:power-spectra}), but Aurora also outperforms GraphCast at short lead times on U, V, T, and Z.  Taken together, we postulate that these results are a collective consequence of Aurora's scale, both in terms of architecture design and training data corpus, as well as its pretraining and fine-tuning protocols. We discuss the origin of Aurora's performance in more detail below. 
% \vspace{-2em}

\begin{figure}[ht]
    \centering
    \includegraphics[width=\columnwidth]{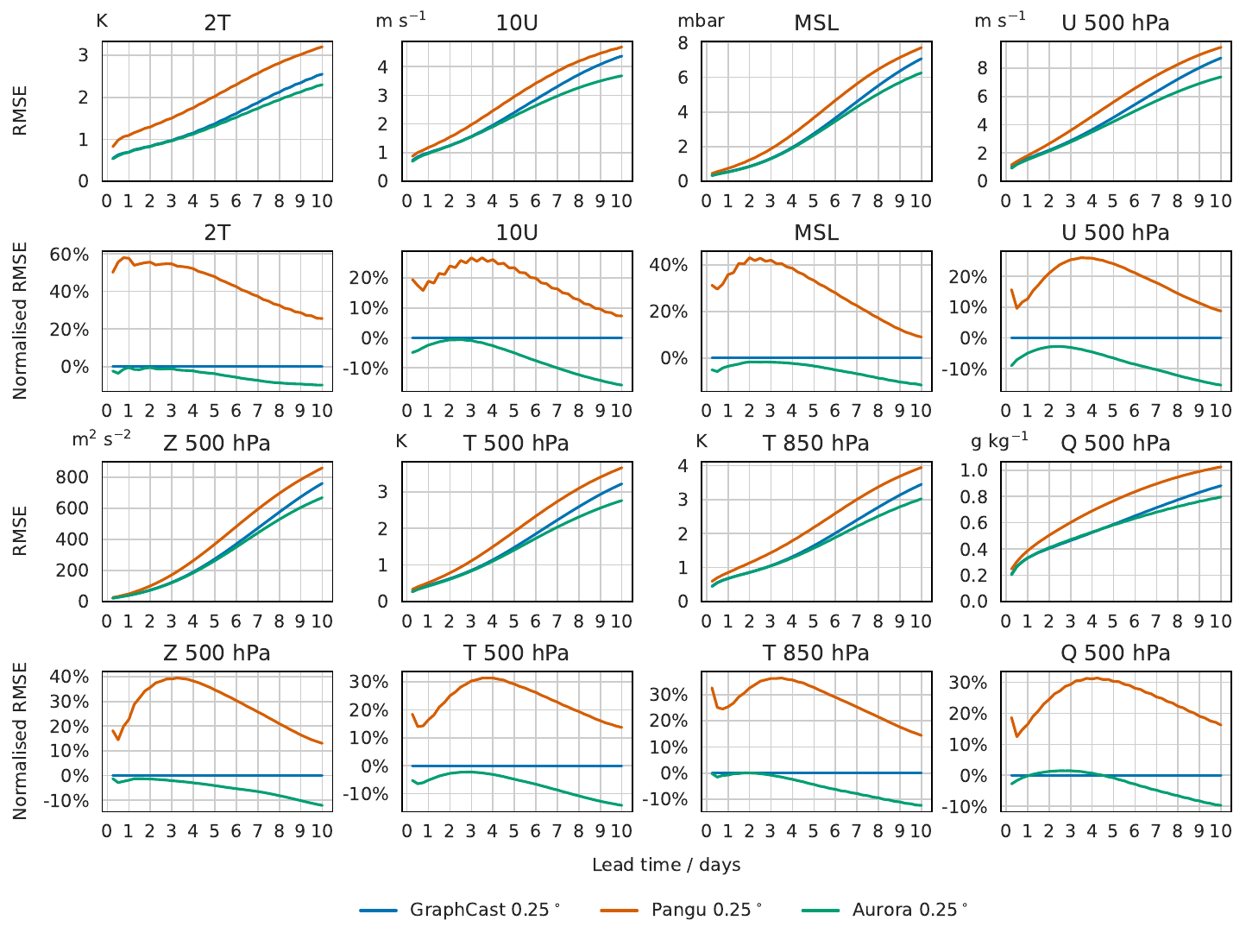}
    \caption{Root mean square error (RMSE) of Aurora \SI{0.25}{\degree} compared to GraphCast for the headline variables. Shows both unnormalised and normalised RMSEs, where GraphCast is used as the baseline. The Pangu weather results at \SI{0.25}{\degree} are included for reference.}
    \label{fig:headline-gc-0.25}
\end{figure}

\begin{figure}[t]
    \centering
    \includegraphics[width=\linewidth]{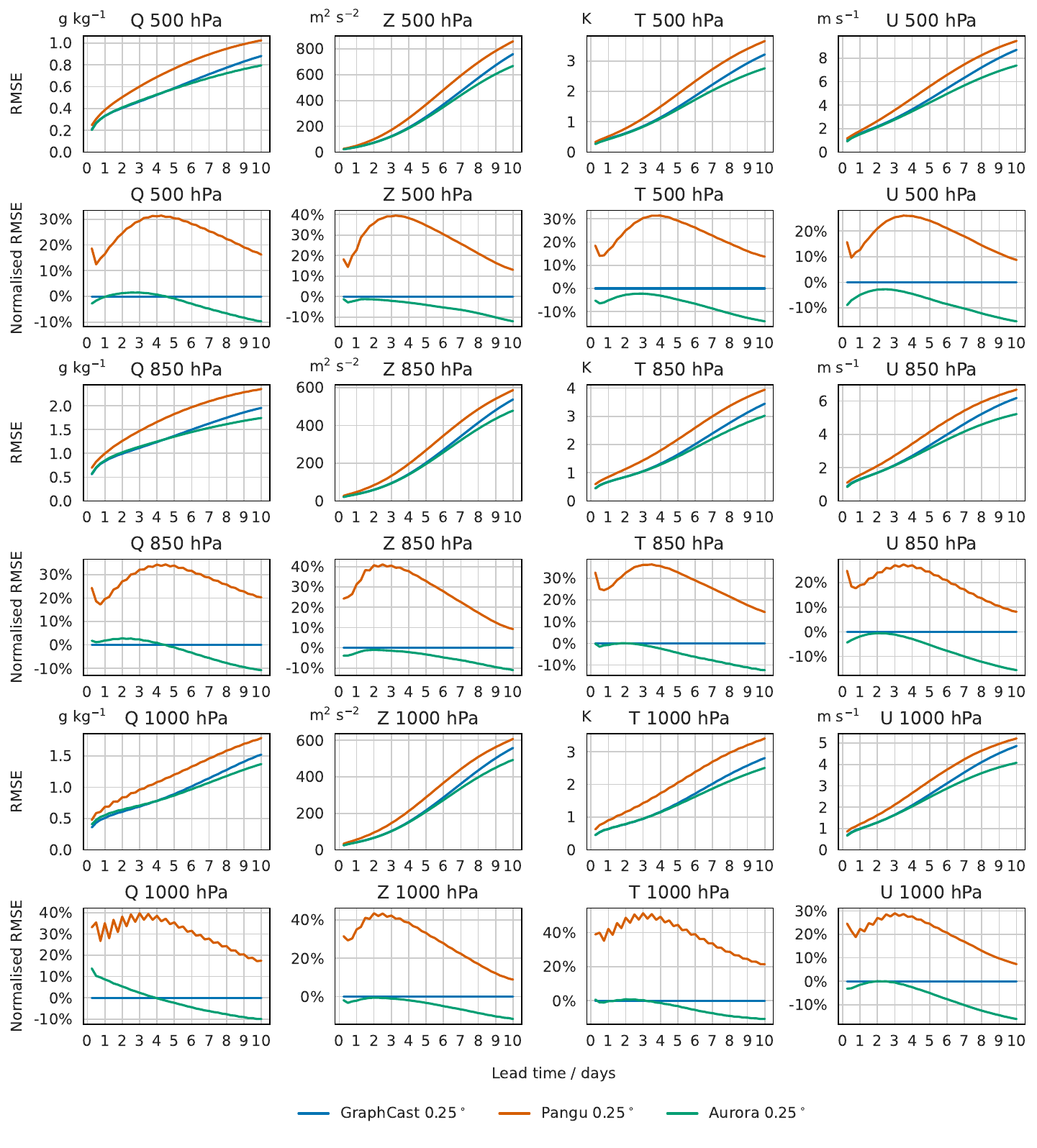}
    \caption{Root mean square error (RMSE) of Aurora \SI{0.25}{\degree} compared to GraphCast in the lower atmosphere. Shows both unnormalised and normalised RMSEs, where GraphCast is used as the baseline. The Pangu weather results at \SI{0.25}{\degree} are included for reference.}
    \label{fig:lower-atmos-gc-0.25}
\end{figure}

\begin{figure}[t]
    \centering
    \includegraphics[width=\linewidth]{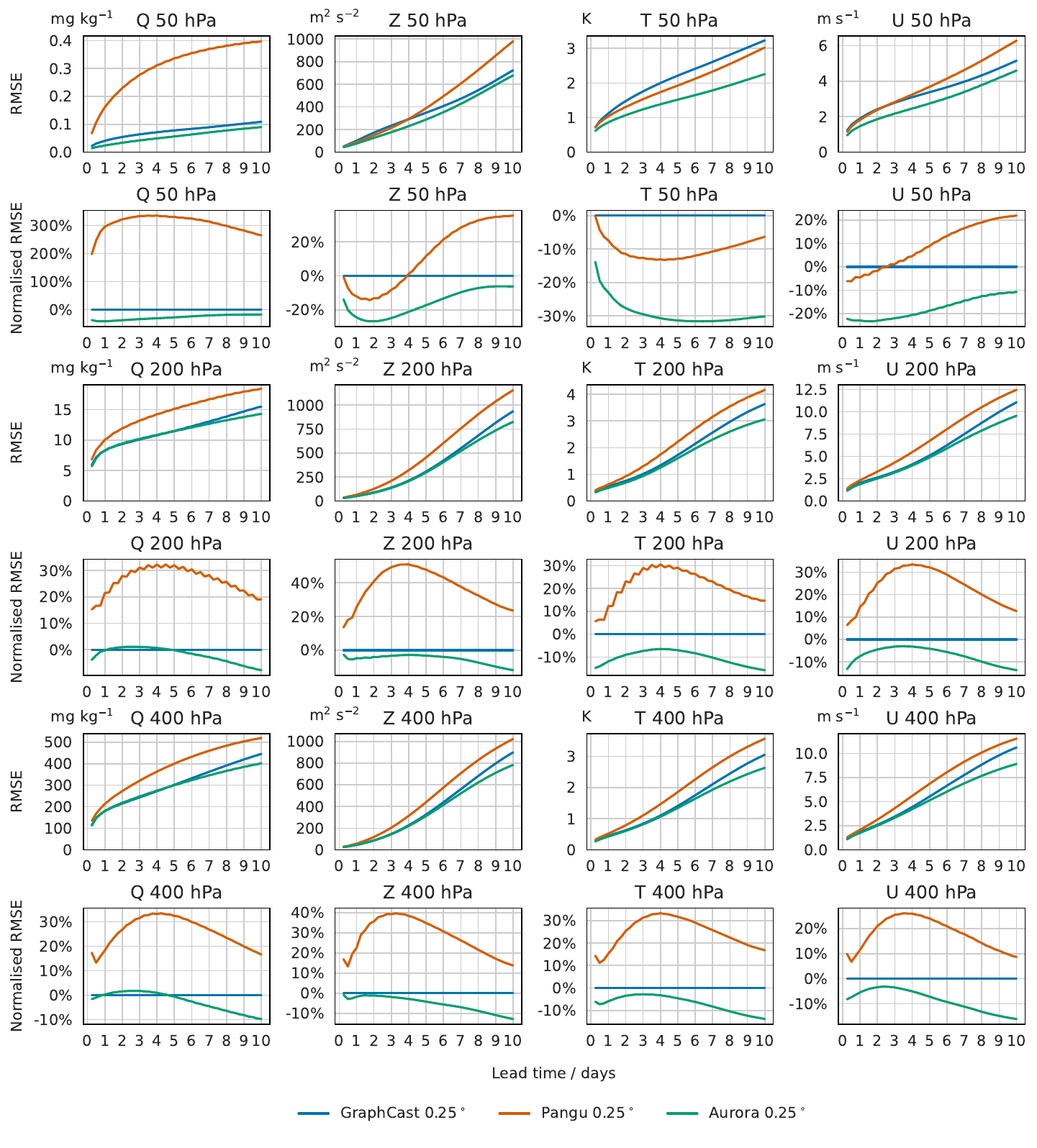}
    \caption{Root mean square error (RMSE) of Aurora \SI{0.25}{\degree} compared to GraphCast in the upper atmosphere. Shows both unnormalised and normalised RMSEs, where GraphCast is used as the baseline. The Pangu weather results at \SI{0.25}{\degree} are included for reference.}
    \label{fig:upper-atmos-gc-0.25}
\end{figure}

\begin{figure}[t]
    \centering
    \includegraphics[width=\linewidth]{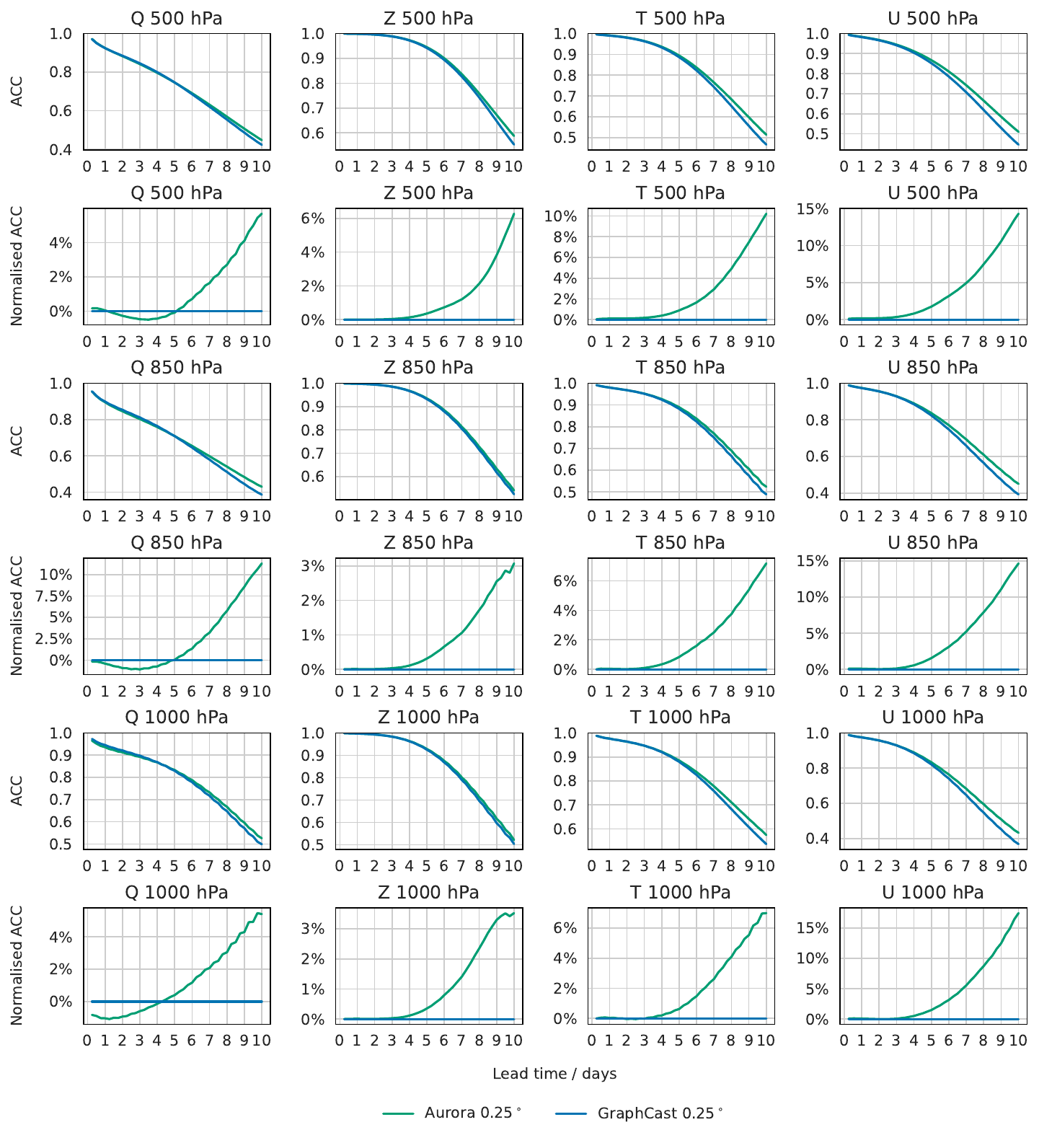}
    \caption{Anomaly correlation coefficient (ACC) of Aurora \SI{0.25}{\degree} compared to GraphCast in the lower atmosphere. Show both unnormalised and normalised ACCs.}
    \label{fig:lower-atmos-gc-0.25-acc}
\end{figure}

\begin{figure}[t]
    \centering
    \includegraphics[width=\linewidth]{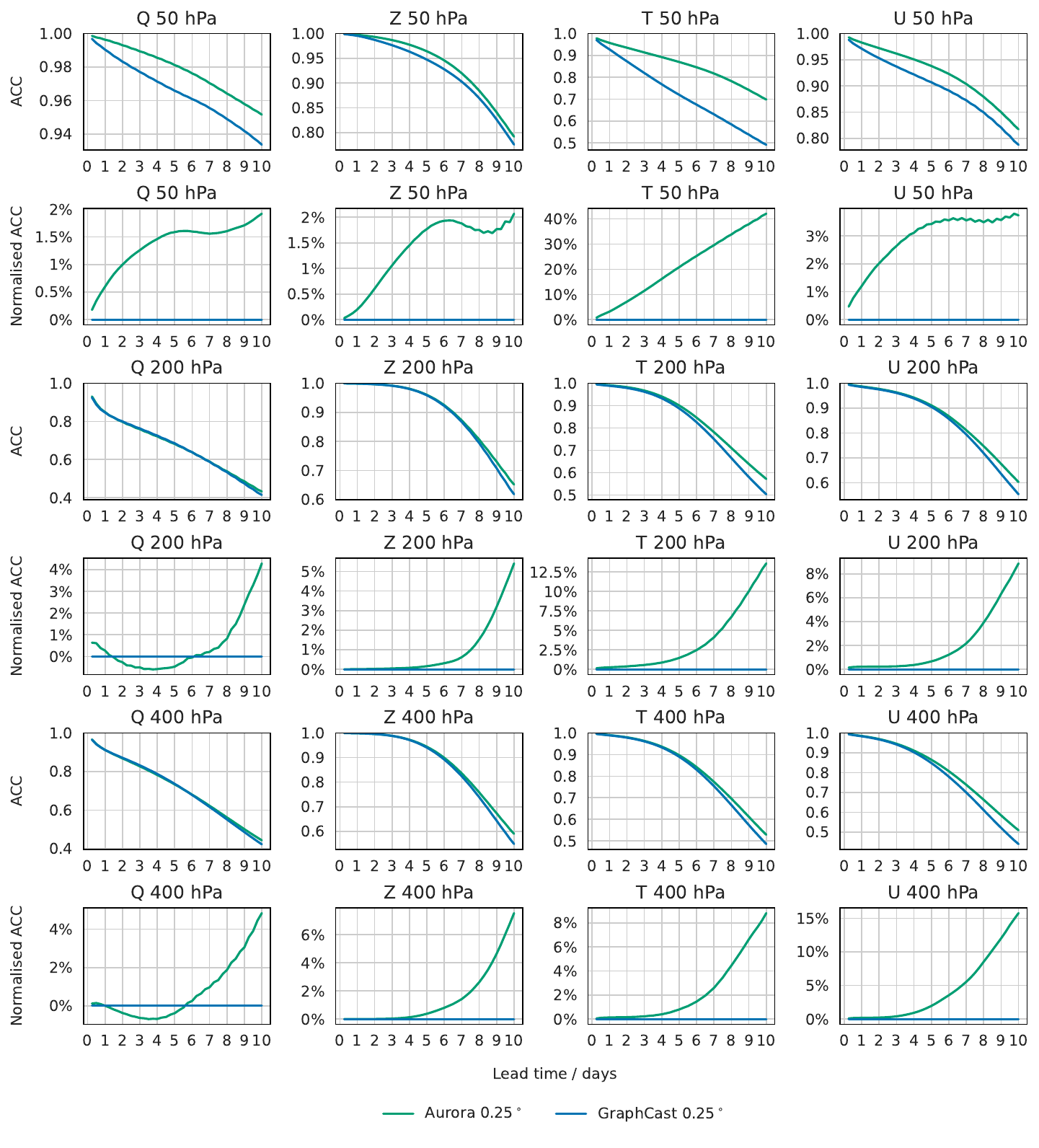}
    \caption{Anomaly correlation coefficient (ACC) of Aurora \SI{0.25}{\degree} compared to GraphCast in the upper atmosphere. Shows both unnormalised and normalised ACCs.}
    \label{fig:upper-atmos-gc-0.25-acc}
\end{figure}

For a more detailed comparison against GraphCast, we include various additional roll-out plots:
RMSEs for the headline metrics (\cref{fig:headline-gc-0.25}),
RMSEs in the lower atmosphere (\cref{fig:lower-atmos-gc-0.25}), ACCs in the lower atmosphere (\cref{fig:lower-atmos-gc-0.25-acc}), RMSEs in the upper atmosphere (\cref{fig:upper-atmos-gc-0.25}), and ACCs in the upper atmosphere (\cref{fig:upper-atmos-gc-0.25-acc}).

We believe Aurora's improved performance over GraphCast can be attributed to several factors:

\begin{enumerate}
    \item {\bf Data diversity:} Our ablation studies in \cref{sec:scaling} demonstrate clear benefits from diverse training data, with \cref{fig:data-scaling-era5-main} showing improved accuracy across variables when including climate simulations and multiple weather datasets. \cref{fig:data-scaling-hres-extremes} provides concrete evidence that this diversity leads to up to 35\% error reduction in extreme value prediction compared to training on ERA5 alone.
    \item {\bf Model size and architecture:} Aurora's larger size (1.3B vs GraphCast's 37M parameters) translates to measurable performance gains, as shown by our scaling experiments in \cref{fig:model-scaling} where validation performance improves systematically with model size. Our flexible encoder architecture (detailed in \cref{sec:encoder}) enables efficient processing of heterogeneous inputs, which is crucial for learning from diverse datasets.
    \item {\bf Foundation model approach:} The effectiveness of our pretraining strategy is demonstrated in Supplementary \cref{sec:app_data_scaling}, where \cref{fig:data-scaling-hres} and \cref{fig:data-scaling-era5} show consistent improvements across variables and pressure levels as we include more diverse datasets in pretraining. This validates our approach of leveraging knowledge from multiple domains to improve overall performance.
    \item {\bf Advanced fine-tuning strategy:} Our two-stage fine-tuning process, including the use of LoRA for roll-out fine-tuning (Supplementary \cref{sec:rollout_finetuning}), provides better control of the model's behavior at different lead times. The benefits of this approach are evident in our comparison against GraphCast (\cref{fig:0.25deg_scorecard}), where Aurora shows improved performance across the vast majority of targets, particularly at longer lead times.
\end{enumerate}

\clearpage

\subsection{Aurora versus Pangu at 0.25\degree{}}
\label{sm:0.25_extra_results:pangu_results}
\begin{figure}[ht]
    \centering
    \includegraphics[width=\textwidth]{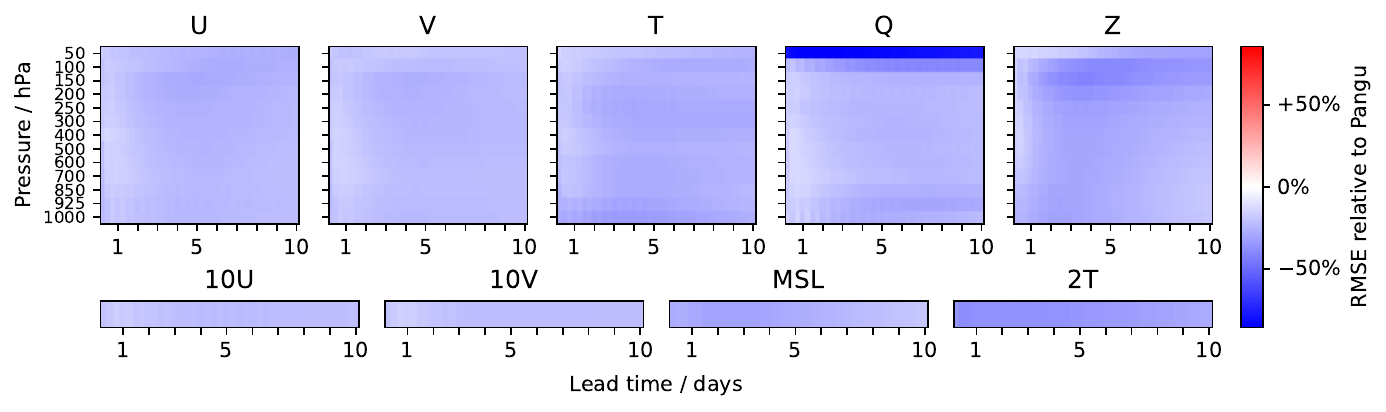}
    \caption{Scorecard of Aurora 0.25\degree{} versus Pangu at 0.25\degree{}. Aurora out-performs Pangu across all variables and pressure levels.}
    \label{fig:scorecard-vs-pangu-0.25}
\end{figure}

To further expand upon the evaluation of Aurora at 0.25\degree{} against existent AI models, we include a scorecard comparison with Pangu weather \citepapp{bi2022pangu} 0.25\degree{} in \cref{fig:scorecard-vs-pangu-0.25}. Aurora consistently out-performs Pangu across all variables and pressure levels.

\subsection{Aurora versus IFS HRES at 0.25\degree{}}
\label{sm:0.25_extra_results:additional_results}
\begin{figure}[ht]
    \centering
    \includegraphics[width=\textwidth]{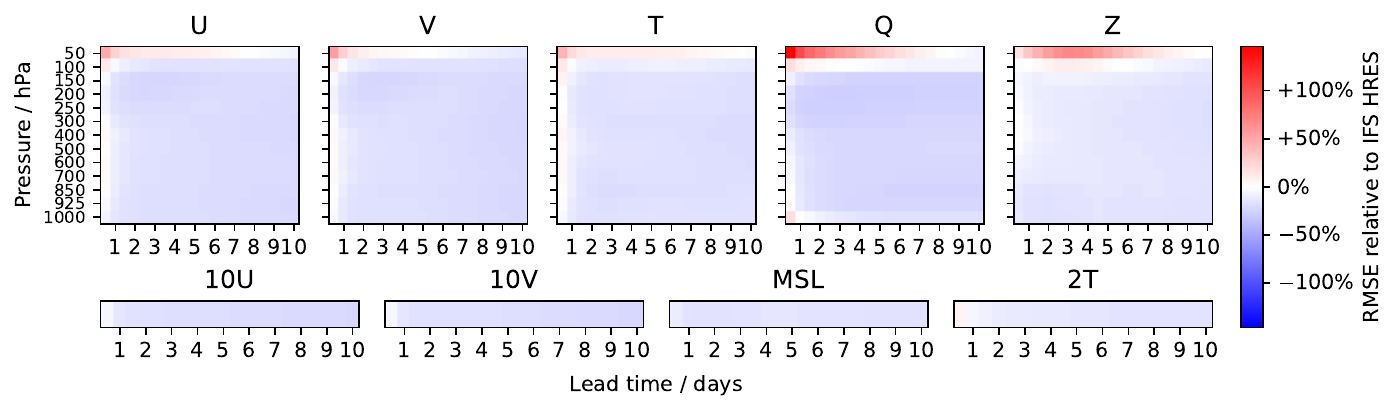}
    \caption{Scorecard of Aurora 0.25\degree{} versus IFS HRES 0.25\degree{}. Despite significantly improving versus GraphCast in the higher atmosphere, Aurora is also worse than HRES at the top of the atmosphere. However, Aurora generally outperforms HRES on other levels. }
    \label{fig:scorecard-vs-hres-0.25}
\end{figure}

\begin{figure}[ht]
    \centering
    \includegraphics[width=\textwidth]{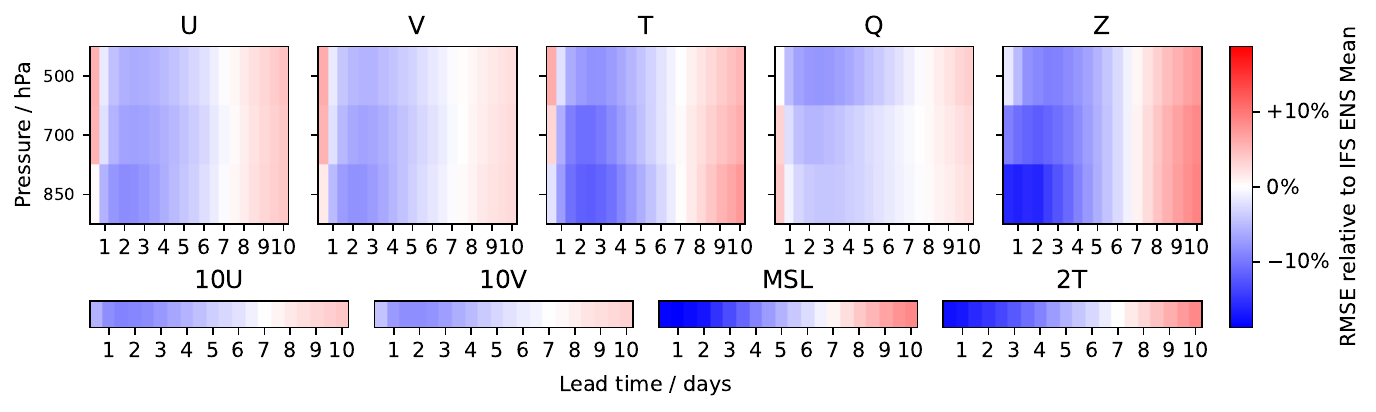}
    \caption{Scorecard of Aurora 0.25\degree{} versus IFS ENS mean 0.25\degree{}. Aurora outperforms the ensemble mean up to around 7 days lead time.}
    \label{fig:scorecard-vs-ifs-ens-mean-0.25}
\end{figure}

To further expand upon the evaluation of Aurora at 0.25\degree{}, we include a scorecard comparison with IFS HRES 0.25\degree{} in \cref{fig:scorecard-vs-hres-0.25}.
Aurora largely outperforms HRES, but despite its improvements in the upper atmosphere compared to GraphCast, it still performs worse than HRES at the top of the atmosphere. 
\Cref{fig:scorecard-vs-ifs-ens-mean-0.25} shows a scorecard comparing Aurora at 0.25\degree{} to the IFS ensemble mean.
Aurora outperforms the ensemble mean up to 7 days lead time.
\vspace{-1em}

\vspace{-2.5em}

\subsection{Verification against weather station measurements at 0.25\degree{}}
\begin{figure}[t]
    \centering
    \includegraphics[width=1.0\textwidth]{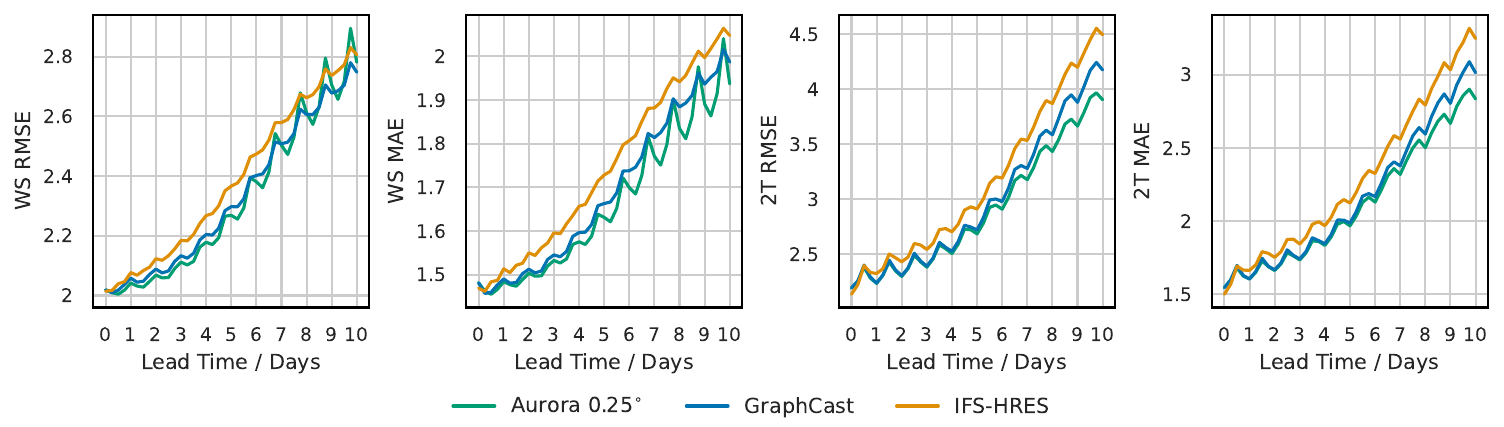}
    \caption{Root mean square error (RMSE) and mean absolute error (MAE) for Aurora, GraphCast, and IFS HRES as measured by global weather stations during 2022 for wind speed (left two panels) and surface temperature (right two panels).}
    \label{fig:stations-rmse}
\end{figure}

As for the high-resolution model, we also evaluate Aurora against station observations for forecasts issued at 00 UTC between 2 January 2022 and 19 December 2022. 
\cref{fig:stations-rmse} shows the RMSE and MAE as a function of lead time for the 0.25\degree~AIWP models versus IFS HRES. The relative error in temperature between Aurora and GraphCast tracks very closely with the scorecard results in \cref{fig:0.25deg_scorecard}. Despite an advantage in resolution, IFS HRES only leads within the first 12 hours. For wind speed, Aurora outperforms GraphCast by a larger margin over the first 6 days compared to evaluations on gridded data, but performs worse than GraphCast after day 7 on RMSE values, particularly on 18 UTC timestamps. These results show that AIWP models such as Aurora are still capable of achieving high forecast accuracy as measured by observations, even at the coarser 0.25\degree~resolution.

\subsection{Extreme event prediction at 0.25\degree{}}
\label{sec:extremes}
\begin{figure}[ht]
    \centering
    \includegraphics[width=0.98\textwidth]{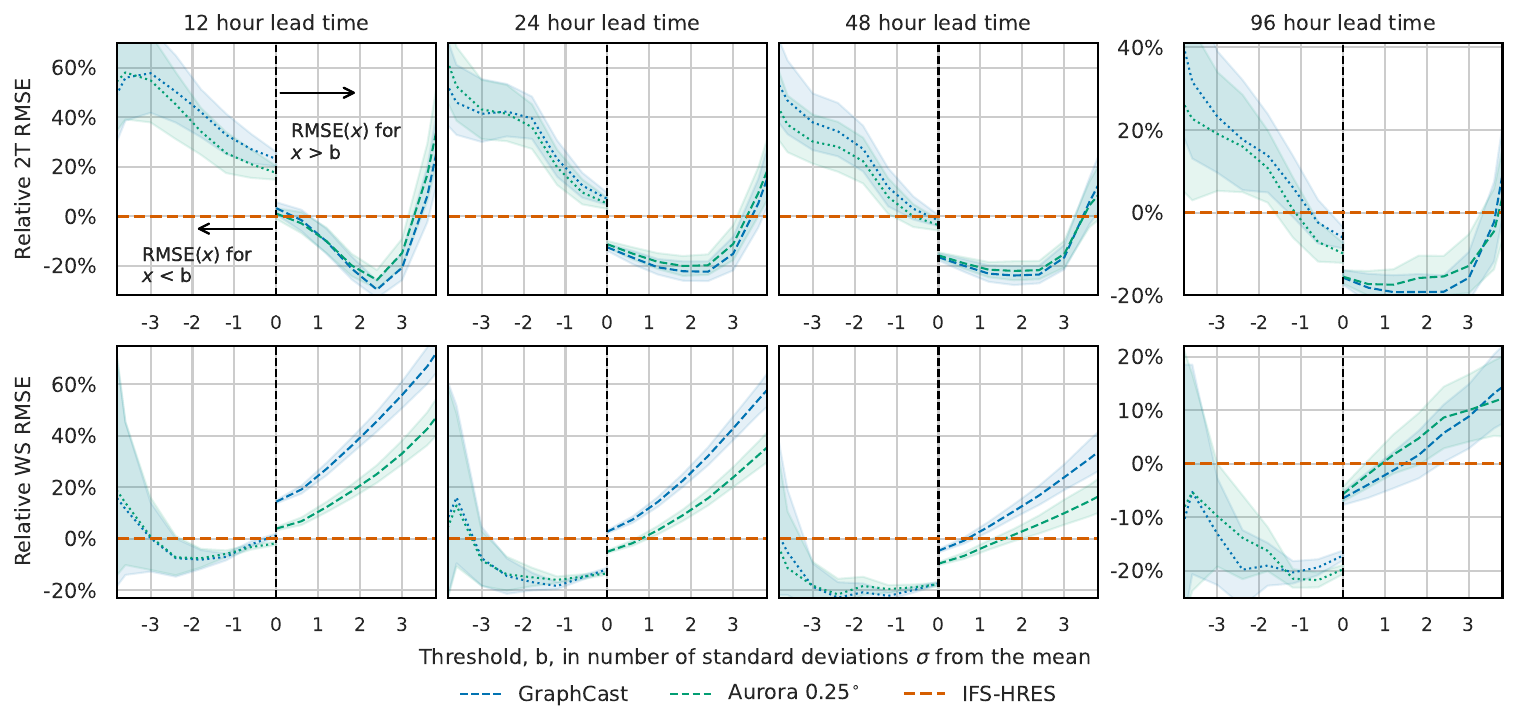}
    \caption{Thresholded RMSE for Aurora, GraphCast and IFS HRES normalized by IFS HRES performance. Aurora demonstrates improved prediction for the extreme values (i.e., tails) of the surface variable distributions. In each plot, values to the right of the centre line are cumulative RMSEs for targets found to sit above the threshold, while those to the left represent target values sitting below the threshold. The full procedure is outlined in~\cref{app:extreme_metrics}. The shading represents 95\% confidence intervals obtained through bootstrapping. Specifically, for each threshold value, we compute RMSE values using \cref{eq:threshold_rmses} in \cref{app:extreme_metrics}, and these values are sampled with replacement to estimate the confidence intervals.}
    \label{fig:extreme_rmses}
\end{figure}

From a meteorological perspective, accurate forecasting of the extremes of surface weather, such as wind speed and temperature, is of critical importance in forward-planning to mitigate the impact on life. To verify the performance of Aurora in predicting the tails of the relevant distributions, we present a comparison with GraphCast and IFS HRES using the 06 UTC and 18 UTC initialisations from the 2022 IFS HRES 0.25\degree~dataset in~\cref{fig:extreme_rmses}. 
Here we show RMSEs computed in the upper and lower tails of the data distribution as described  in~\cref{app:extreme_metrics}, normalized by the RMSEs from IFS HRES. By sweeping a threshold, we can analyse performance as we move further into the extremes.
Values for thresholds above zero are shown on the right of the x-axis in each plot, and values for thresholds below zero are shown on the on the left of the x-axis in each plot. 

We find that Aurora outperforms GraphCast on wind speed prediction at the surface.
At longer lead times, the performance of AIWP models relative to IFS HRES becomes markedly improved across the distribution. 
Note however that 2T  exhibits a difference in behaviour between the warmer and colder sections of the distribution; it is reported that IFS HRES performance for winter extremes is biased, resulting in less accurate forecasting during winter months \citeapp{benbouallegue2023it}. Since AIWP models include such biased forecasting data rather than just forward simulations within physical constraints, there is a tendency to exhibit these biases when compared to NWP models such as IFS HRES. 

\begin{figure}[b]
\begin{subfigure}[t]{.475\linewidth}
  \includegraphics[width=\linewidth]{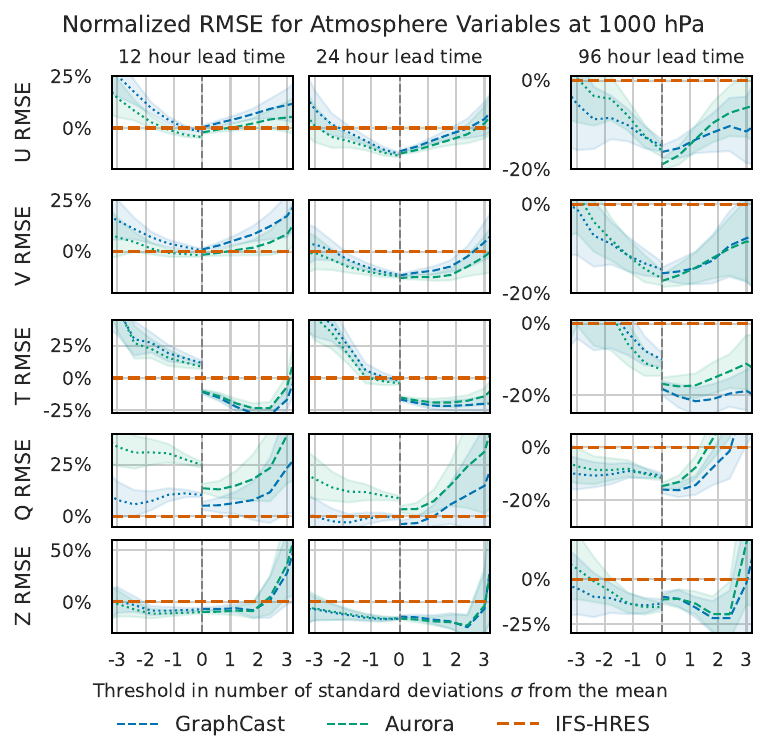}
  \vspace{-230pt}
  \caption{}
  \label{MLEDdet}
\end{subfigure}\hfill % <-- "\hfill"
\begin{subfigure}[t]{.475\linewidth}
  \includegraphics[width=\linewidth]{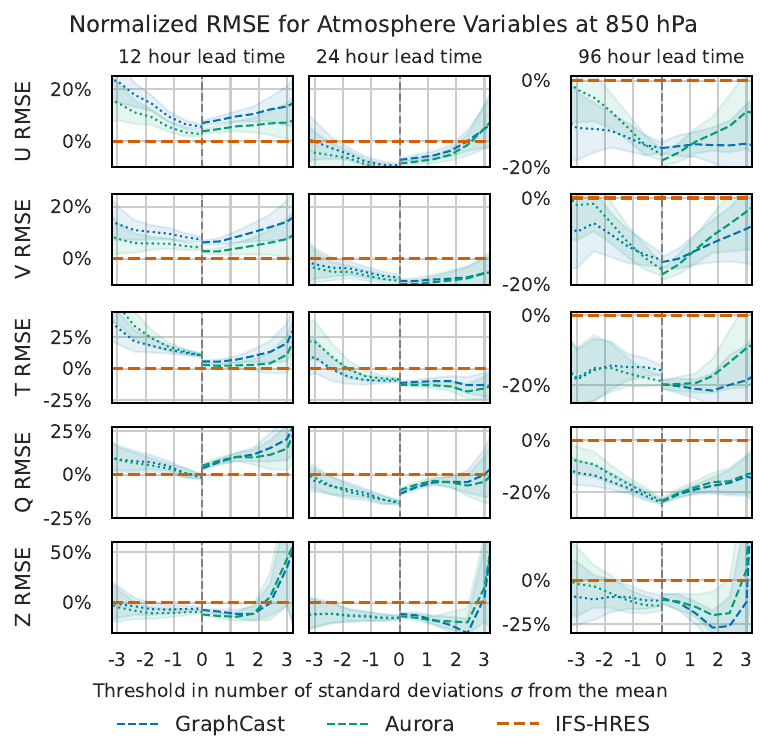}
  \vspace{-230pt}
  \caption{}
  \label{energydetPSK}
\end{subfigure}

\medskip % create some *vertical* separation between the graphs
\begin{subfigure}[t]{.475\linewidth}
  \includegraphics[width=\linewidth]{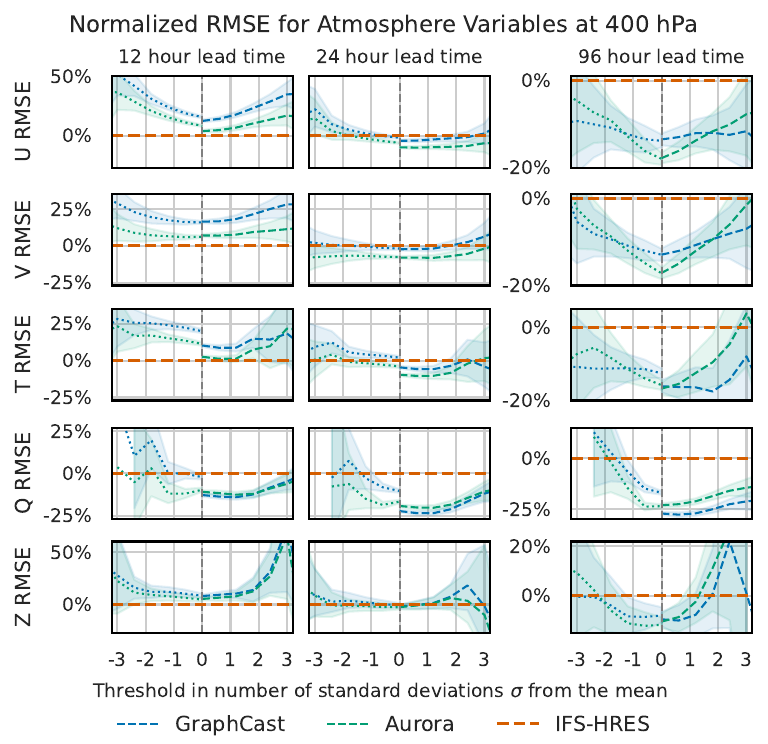}
  \vspace{-230pt}
  \caption{}
  \label{velcomp}
\end{subfigure}\hfill % <-- "\hfill"
\begin{subfigure}[t]{.475\linewidth}
  \includegraphics[width=\linewidth]{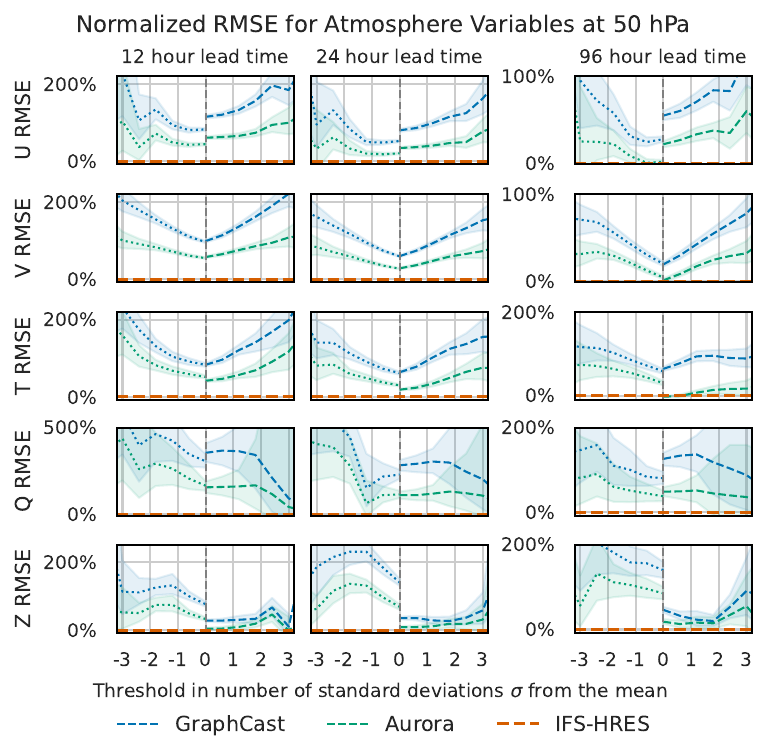}
  \vspace{-230pt}
  \caption{}
  \label{estcomp}
\end{subfigure}

\caption{Thresholded RMSE results for atmospheric variables at four pressure levels distributed throughout the atmosphere, presented as percentages relative to the performance of IFS HRES at 0.25\degree~resolution. The origin of these curves is explained in ~\cref{app:extreme_metrics}. The discontinuity at zero reflects that only one side of the curve is shown; on the right hand side of each figure we show $\text{RMSE}_{g}$ where target values are above the threshold determined by $g$, and on the LHS of the figure, $\text{RMSE}_{g}$ is the cumulative RMSE found against target values below the threshold. The shading represents 95\% confidence intervals obtained through bootstrapping. Specifically, for each threshold value, we compute RMSE values using \cref{eq:threshold_rmses} in \cref{app:extreme_metrics}, and these values are sampled with replacement to estimate the confidence intervals.}
\label{fig:extremes_atmos}
\end{figure}

In~\cref{fig:extremes_atmos}, we show the thresholded RMSES for atmospheric variables distributed over four pressure levels throughout the atmosphere. At the higher pressure levels (i.e., near the surface), results are consistent with those shown for the surface variables in~\cref{sec:extremes}. 
However, the temperature in the atmosphere is more typically distributed as it is less susceptible to bias through near-surface effects. In the higher atmosphere, overall performance is more challenging, although Aurora remains able to outperform GraphCast, consistent with total RMSE results shown in~\cref{fig:0.25deg_scorecard}.

% \subsection{Extreme events forecasting}
% \label{app:sec:extreme-events}

\FloatBarrier
\section{Additional results}
\label{sec:verification_studies}
In this section, we present results that supplement the results in the main body.

\subsection{Comparison against operational CAMS}
\label{app:cams:full-results}
It is important to note that we take CAMS analysis data as the ground truth. 
This is because CAMS analysis is produced by combining CAMS forecasts with observations (i.e., assimilation), the lack of real-time emission data means that CAMS analysis is more biased towards similarity with CAMS forecasts than is the case for meteorological variables, especially for near-surface pollution variables at short lead times. 
For example, the PM values are only weakly constrained via aerosol optic depth satellite measurements as CAMS does not assimilate any real-time emission observations. This may disadvantage Aurora in this comparison. Below are the details of the various experiments we conduct. 

\begin{figure}[t]
    \centering
    \includegraphics[width=\columnwidth]{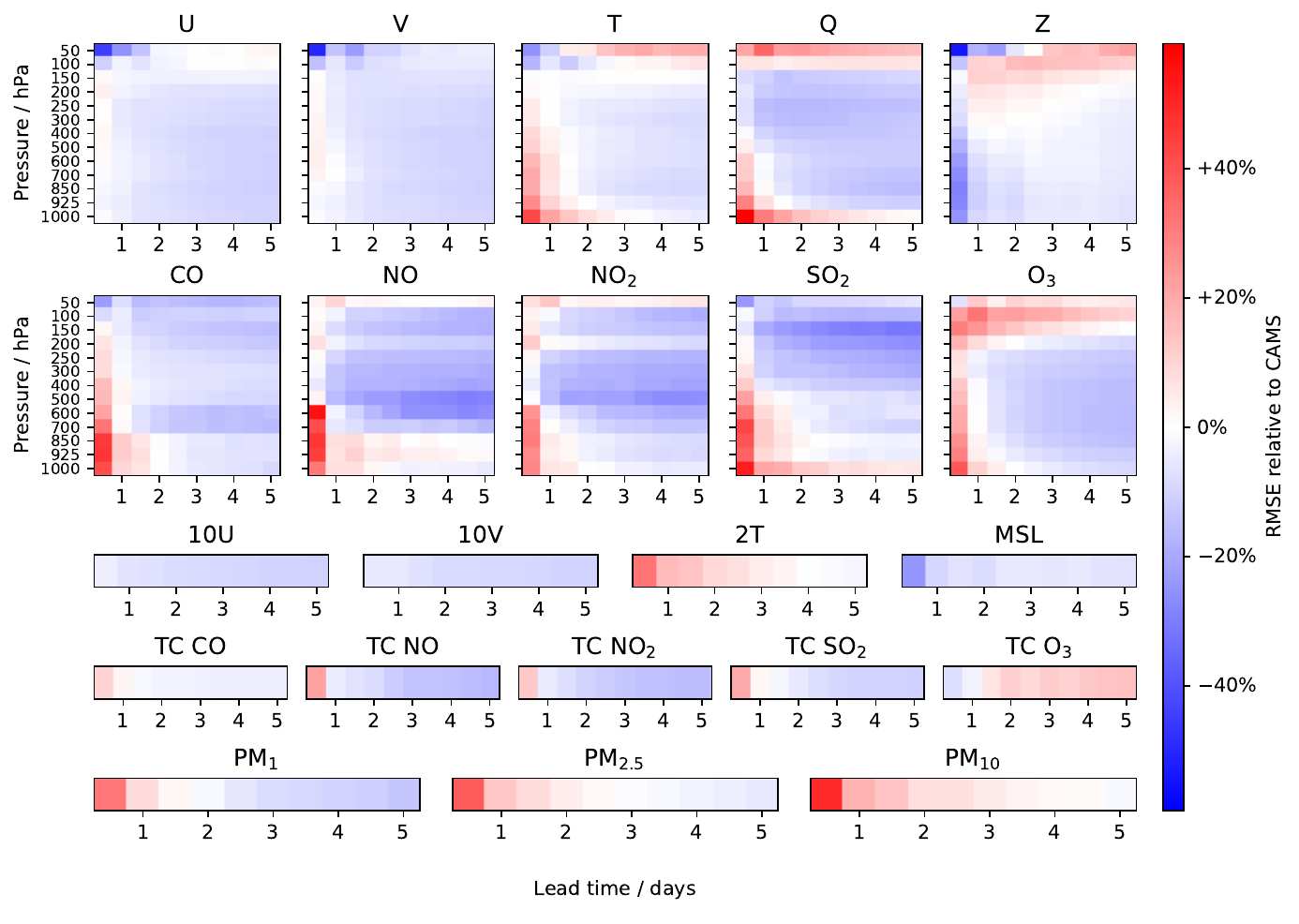}
    \caption{
        Latitude-weighted root mean square error (RMSE) of Aurora relative to CAMS, where negative values (blue) mean that Aurora is better.
        The RMSEs are computed over the period Jun 2022 to Nov 2022 inclusive. 
        Aurora is competitive to CAMS (within 20\% RMSE) on 96\% of all targets, and Aurora matches or outperforms CAMS on 76\% of all targets. 
    }
    \label{fig:cams-scorecard-full}
\end{figure}

\begin{figure}[t]
    \centering
    \includegraphics[width=\columnwidth]{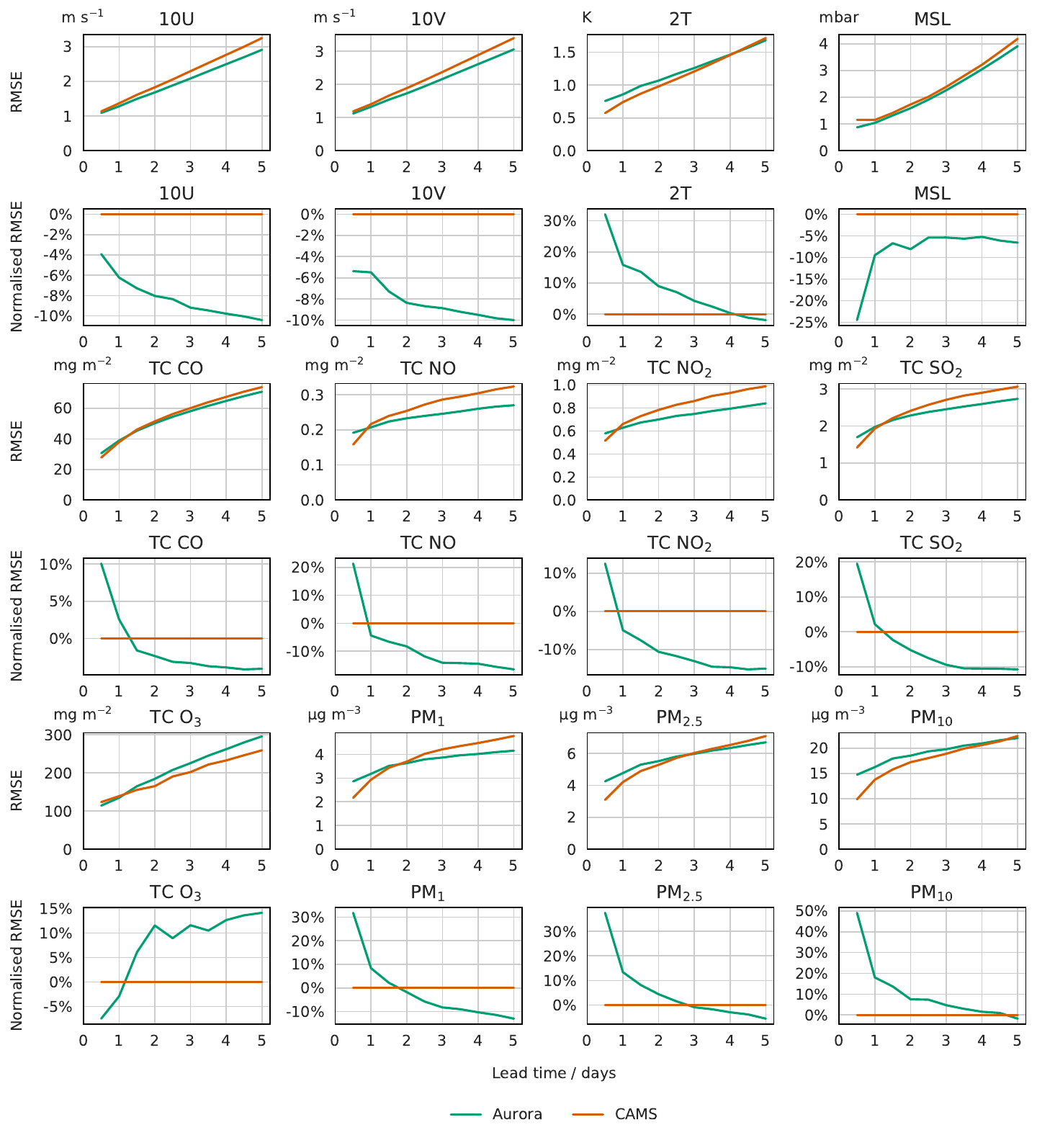}
    \caption{
        Latitude-weighted root mean square error (RMSE) of the surface-level variables compared to CAMS.
        The RMSEs are computed over the period Jun 2022 to Nov 2022 inclusive.
    }
    \label{fig:cams-surf-vars}
\end{figure}

\begin{figure}[t]
    \centering
    \includegraphics[width=\columnwidth]{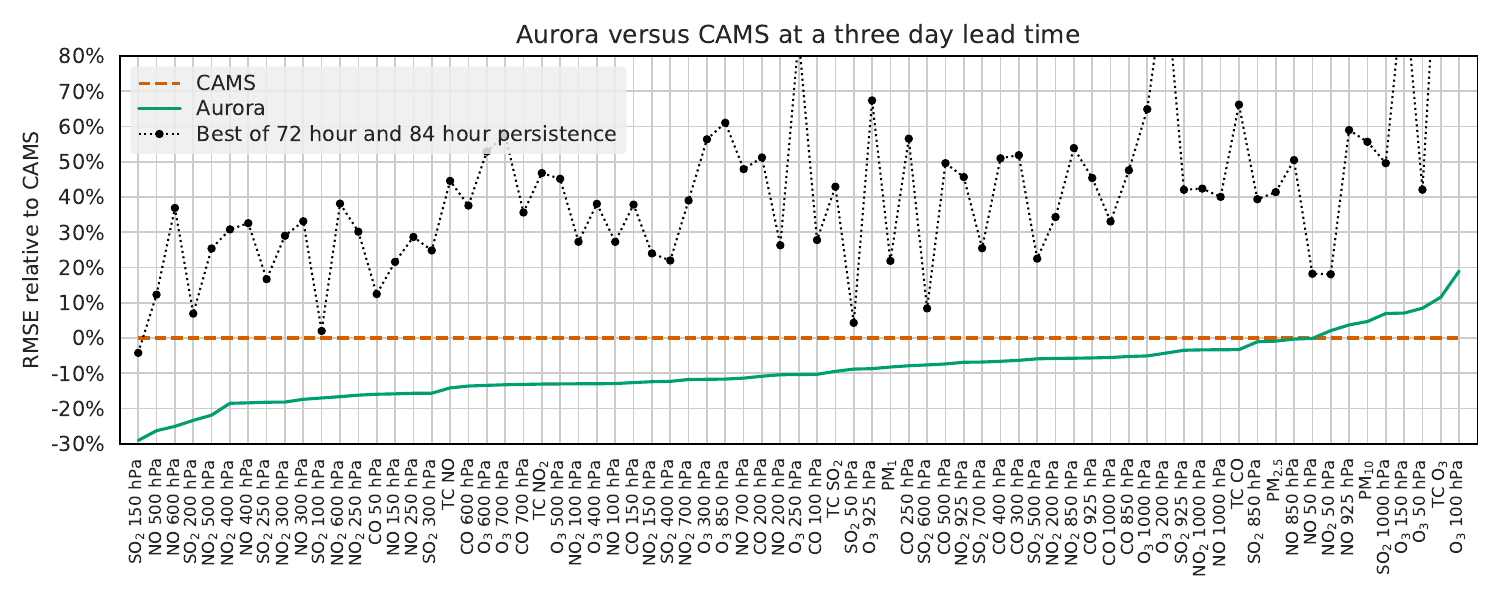}
    \caption{
        Latitude-weighted root mean square error (RMSE) of Aurora to CAMS at three days lead time, where negative values mean that Aurora is better.
        Also shows the RMSE of the persistence prediction relative to CAMS.
        The RMSEs are computed over the period Jun 2022 to Nov 2022 inclusive.
        Aurora matches or outperforms CAMS on 89\% of all variables and is strictly significantly better than the persistence prediction. 
    }
    \label{fig:cams-three-days-full}
\end{figure}

\Cref{fig:cams-scorecard-full} shows a scorecard for the full collection of CAMS variables.
Aurora is competitive to CAMS (within 20\% RMSE) on 96\% of all targets, and Aurora matches or outperforms CAMS on 76\% of all targets. 
\cref{fig:cams-surf-vars} compares Aurora RMSEs to CAMS for only the surface-level variables, 
\Cref{fig:cams-three-days-full} compares the RMSE of Aurora, CAMS, and the persistence prediction at three days lead time. 
Aurora matches or outperforms CAMS on 89\% of all variables and outperforms the persistence prediction for all variables and all lead times. 

\begin{figure}[t]
    \centering
    \includegraphics[width=\columnwidth]{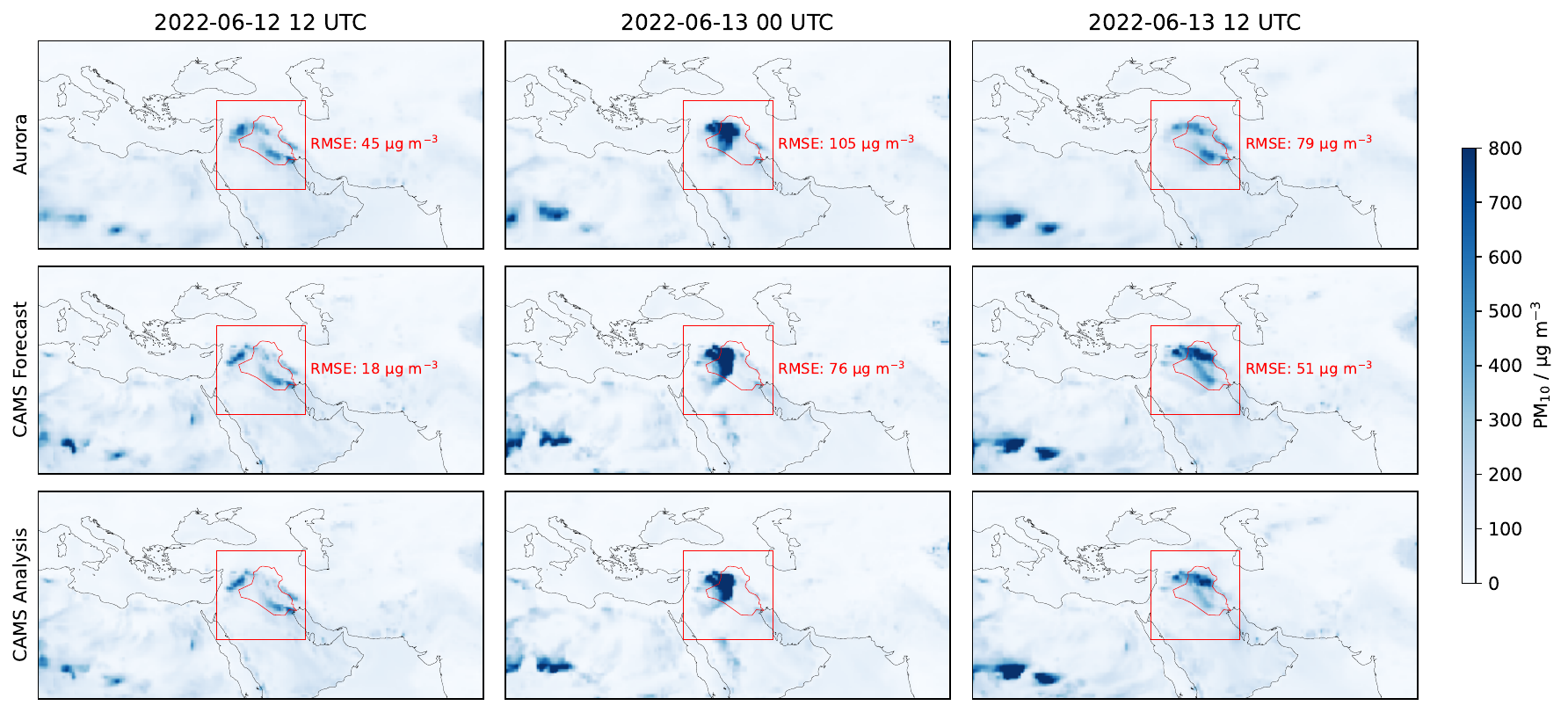}
    \caption{
        Predictions for \PMc{} by Aurora and CAMS for the sandstorm that hit Iraq (highlighted in red) on 13 June 2022.
        Aurora and CAMS were initialised with CAMS analysis at 12 Jun 2022 00 UTC.
        Aurora and CAMS both predict the sandstorm a day in advance.
        The figure also shows the RMSE computed over the red rectangle.
    }
    \label{fig:cams-iraq}
\end{figure}

\paragraph{Iraq sandstorm.}
\cref{fig:cams-iraq} visualises predictions for \PMc{} by Aurora and CAMS for the sandstorm that hit Iraq on 13 June 2022.

\begin{figure}[t]
    \begin{subfigure}[t]{\linewidth}
        \includegraphics[width=\linewidth]{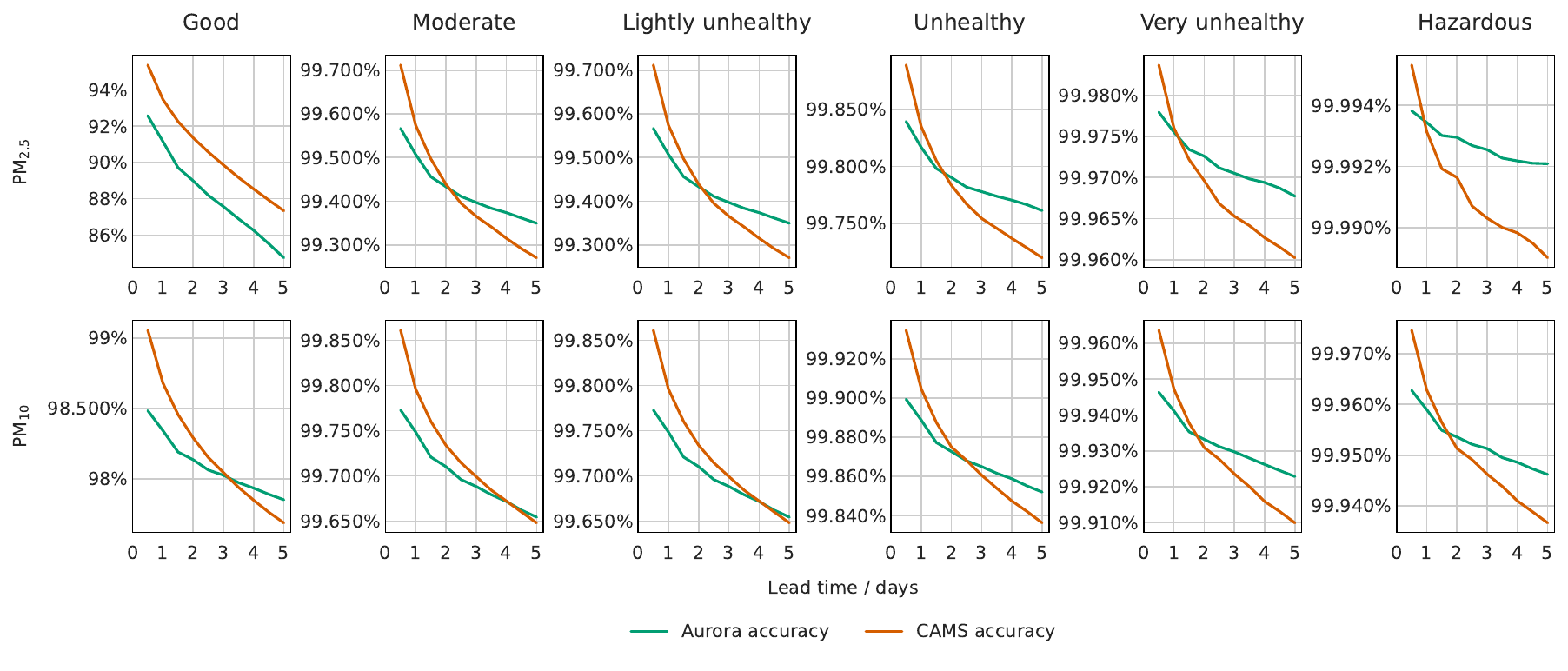}
        \vspace{-220pt}
        \caption{}
    \end{subfigure}
    \begin{subfigure}[t]{\linewidth}
        \includegraphics[width=\linewidth]{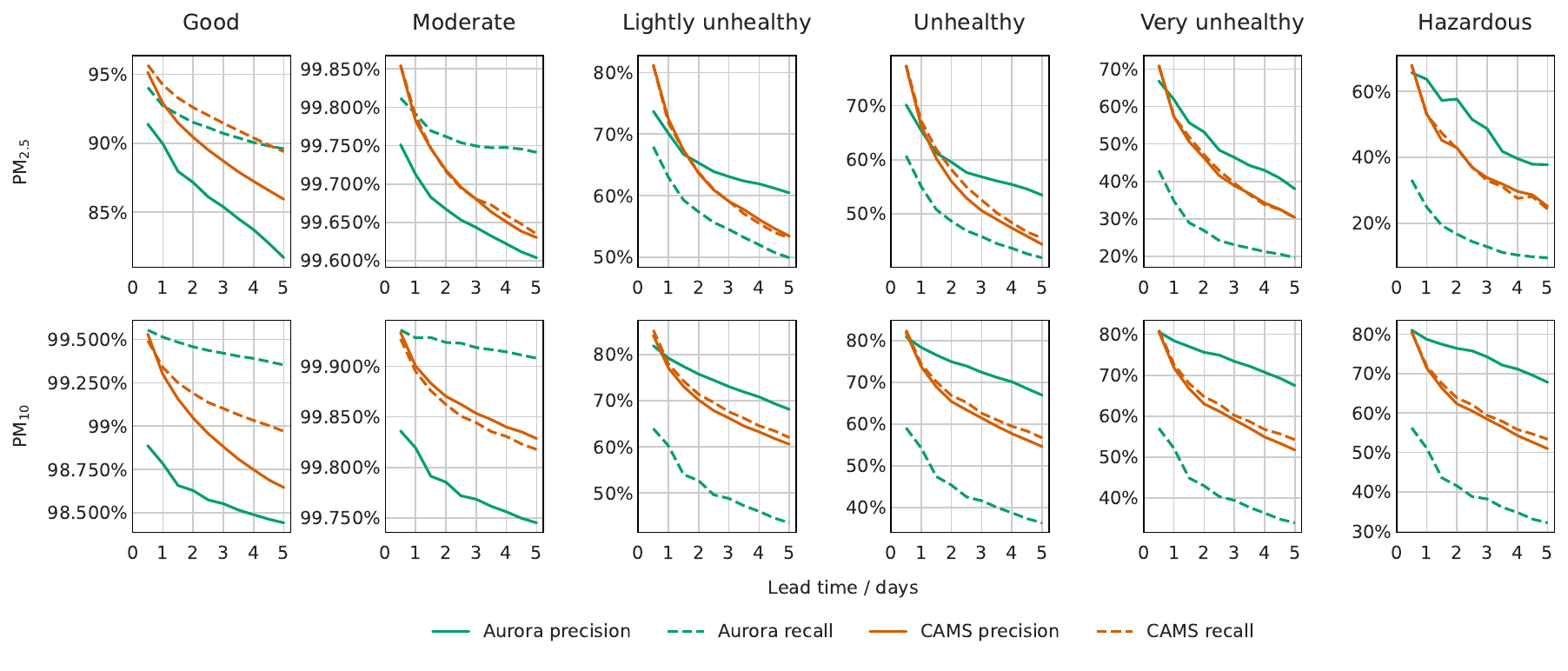}
        \vspace{-220pt}
        \caption{}
    \end{subfigure}
    \caption{%
        \textbf{a:}
            Classification accuracy of levels of \PMb{} according to the following categories from AirNow:
                good (\SI{9.1}{\micro g \, m^{-3}} or lower),
                moderate (\SI{35}{\micro g \, m^{-3}} or lower),
                lightly unhealthy (\SI{35}{\micro g \, m^{-3}} or higher),
                unhealthy (\SI{55}{\micro g \, m^{-3}} or higher),
                very unhealthy (\SI{125}{\micro g \, m^{-3}} or higher),
                and hazardous (\SI{225}{\micro g \, m^{-3}} or higher).
            Also shows the classification accuracy of levels of \PMc{} according to the same categories from AirNow:
                good (\SI{55}{\micro g \, m^{-3}} or lower),
                moderate (\SI{155}{\micro g \, m^{-3}} or lower),
                lightly unhealthy (\SI{155}{\micro g \, m^{-3}} or higher),
                unhealthy (\SI{255}{\micro g \, m^{-3}} or higher),
                very unhealthy (\SI{355}{\micro g \, m^{-3}} or higher),
                and hazardous (\SI{425}{\micro g \, m^{-3}} or higher).
            The accuracies are computed over the period Jun 2022 to Nov 2022 inclusive.
        \textbf{b:}
            Like \textbf{a}, but shows precision and recall.
    }
    \label{fig:cams-threshold-results}
\end{figure}

\paragraph{Classification of levels of \PMb{} and \PMc{}.}
\Cref{fig:cams-threshold-results} compares the accuracy, precision, and recall of CAMS and Aurora for classifying levels of \PMb{} and \PMc{} according to the categories of AirNow \citepapp{airnow}.
Aurora and CAMS both generally obtain very high accuracy, although Aurora predictions come at a fraction of the cost. 
At longer lead times, Aurora tends to obtain higher precision and lower recall, meaning that predictions by Aurora are generally more conservative.
For example, for unhealthy (or worse) levels of \PMb{}, on average Aurora has 4\% higher precision but 8\% lower recall.

\begin{figure}[t]
    \centering
    \includegraphics[width=\linewidth,trim={0 1cm 0 0},clip]{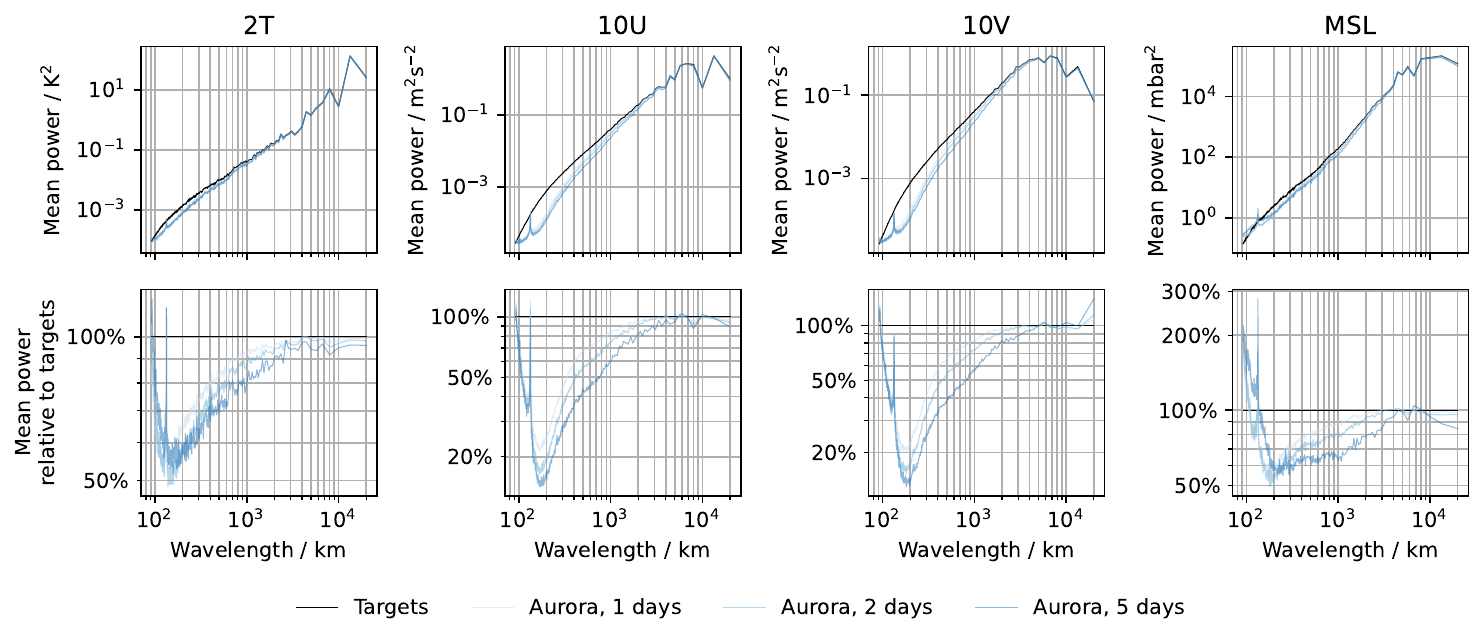}
    \includegraphics[width=\linewidth,trim={0 1cm 0 0},clip]{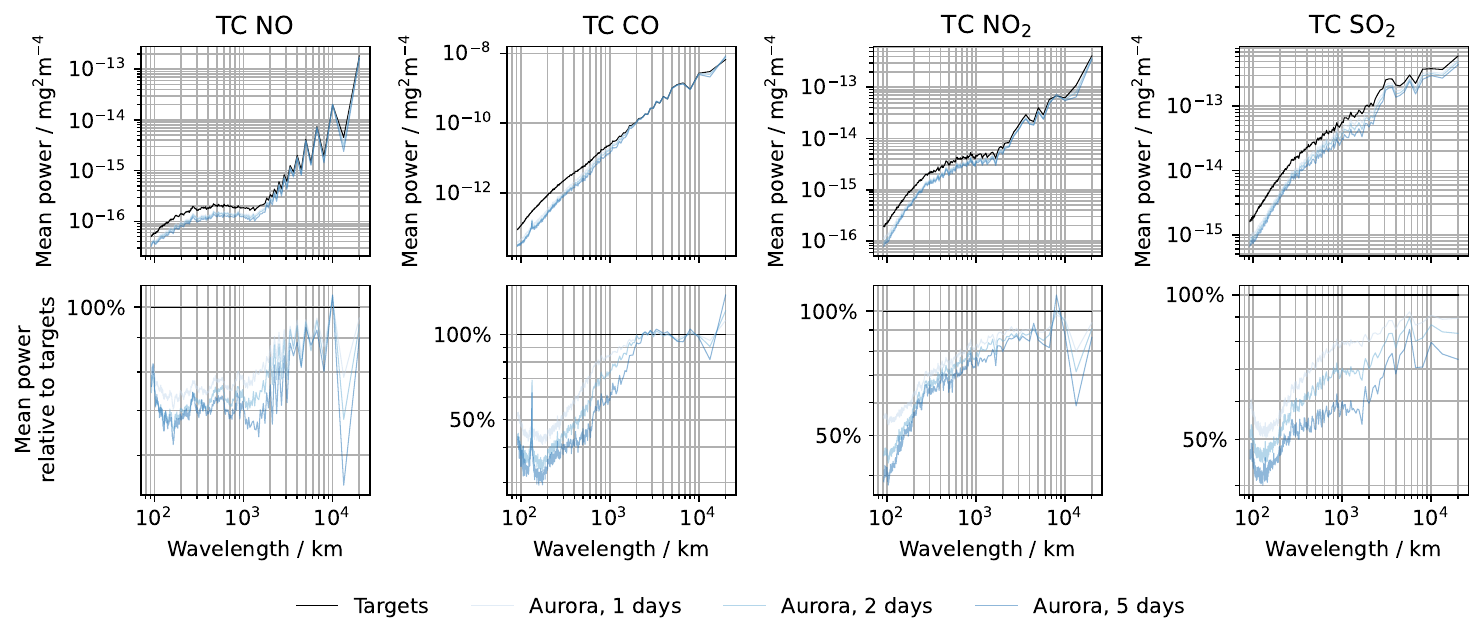}
    \includegraphics[width=\linewidth]{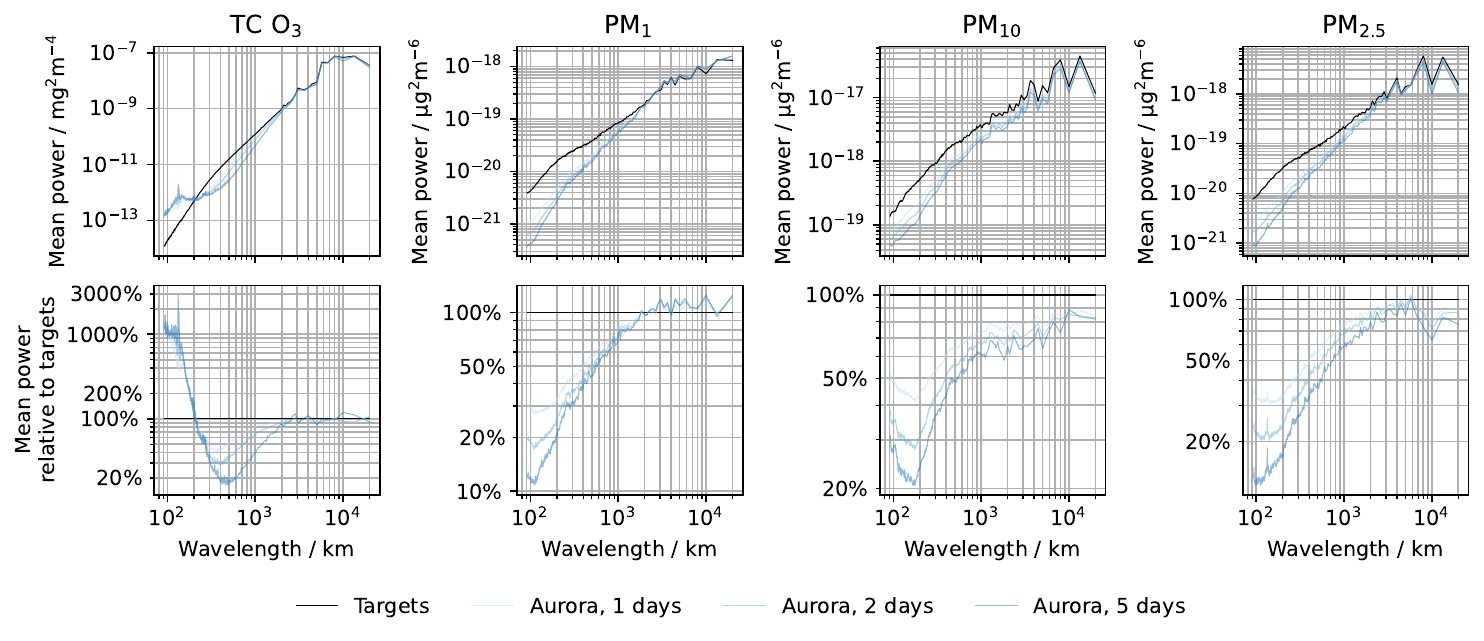}
    \caption{
        Spectral content of the predictions for the surface-level variables by Aurora, compared to CAMS analysis.
        The computation uses the period Jun 2022 to Nov 2022 inclusive.
    }
    \label{fig:cams-spectral-plots}
\end{figure}

\paragraph{Spectral analysis.}
\cref{fig:cams-spectral-plots} includes a spectral analysis of Aurora's predictions for all surface-level variables, performed according to the method in \citetapp{lam2023graphcast}.
For longer lead times, power at smaller wave lengths tends to decrease, meaning that the predictions tend to become more smooth.
For TC NO, TC CO, TC \NOb{}, TC \SOb{}, and TC \Oc{}, this decrease is comparable to that for the meteorological variables 2T, U10, V10, and MSL.
For \PMa{}, \PMb{}, and \PMc{}, the decrease is larger.

\begin{figure}[t]
    \centering
    \includegraphics[width=\columnwidth]{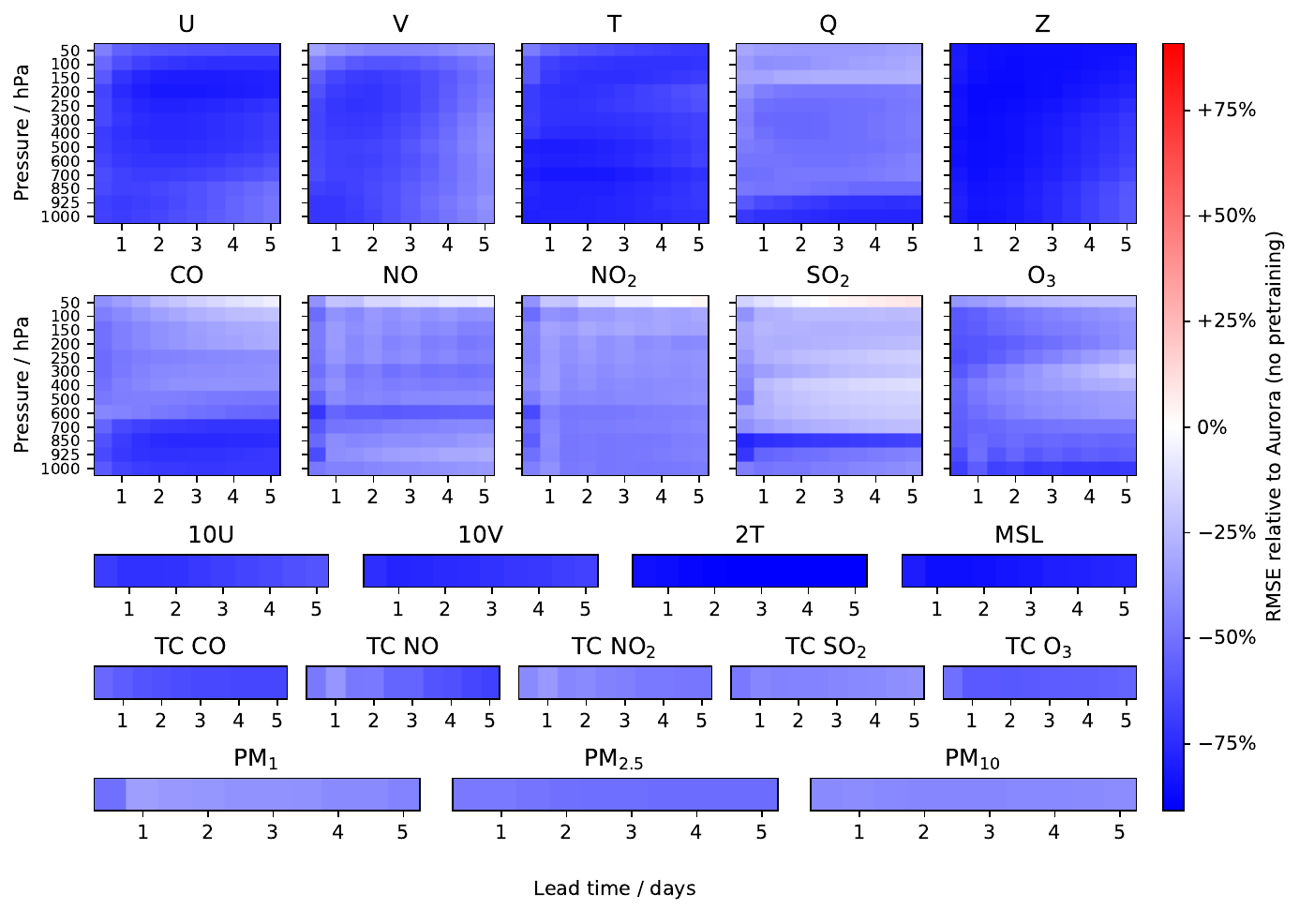}
    \caption{
        Scorecard comparing performance of Aurora model fine-tuned on CAMS data to a model trained from scratch on the same data. Fine-tuned Aurora consistently outperforms any attempt to train a model to predict these variables from scratch.
        On average, pretraining improves performance by 54\%.
    }
    \label{fig:cams-scratch-scorecard}
\end{figure}

\paragraph{Comparison against a model trained from scratch on CAMS data.}
We train the model from scratch on the CAMS analysis training dataset detailed in \cref{app:sec:splits} to clarify the impact of pretraining on our ability to predict atmospheric variables. The scorecard results are presented in \cref{fig:cams-scratch-scorecard}. Fine-tuned Aurora consistently outperforms any attempt to train a model to predict these variables from scratch, likely due to the inherently challenging distributional behaviour of these variables, and the small amount of data available. 

\FloatBarrier
\subsection{Comparison against operational IFS HRES-WAM}
\label{app:wave:full-results}

\begin{figure}[b]
    \centering
    \includegraphics[width=\textwidth]{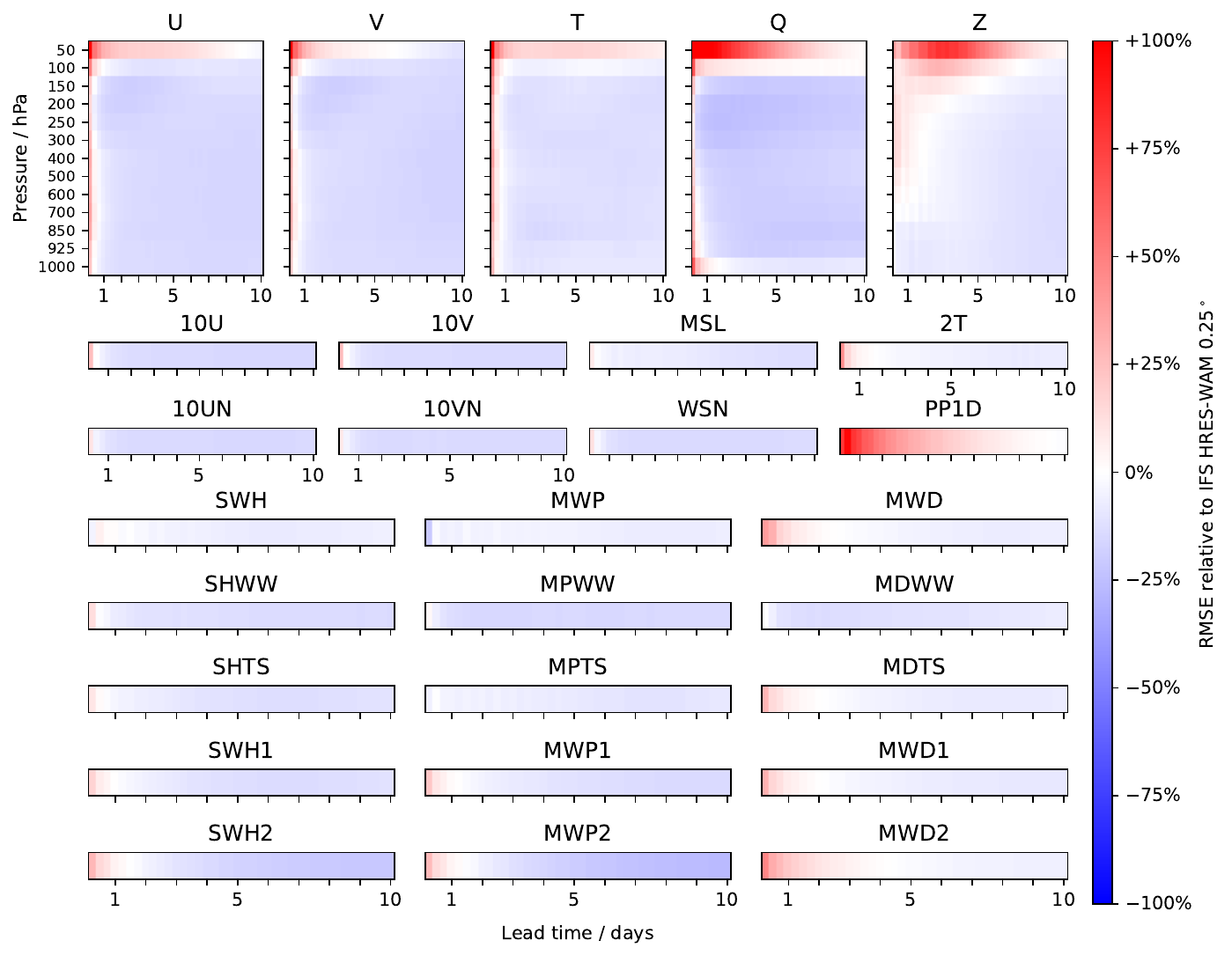}
    \caption{Latitude-weighted root mean square error (RMSE) of Aurora \SI{0.25}{\degree} relative to HRES-WAM \SI{0.25}{\degree}, where negative values (blue) mean that Aurora is better. Only for Q at \SI{50}{hPa}, the RMSE initially exceeds 100\% and peaks at 306\%; no other variable exceeds 100\%. Aurora is competitive (within 20\% RMSE) to HRES-WAM on 95\% of all targets, and Aurora matches or outperforms HRES-WAM on 85\% of all targets.
    Note that the meteorological variables are from HRES T0, and the surface-level wave variables from HRES-WAM.
    See \cref{section:waves} in the main body.
    }
    \label{fig:scorecard-wave-full}
\end{figure}

\begin{figure}[t]
    \centering
    \includegraphics[width=\linewidth]{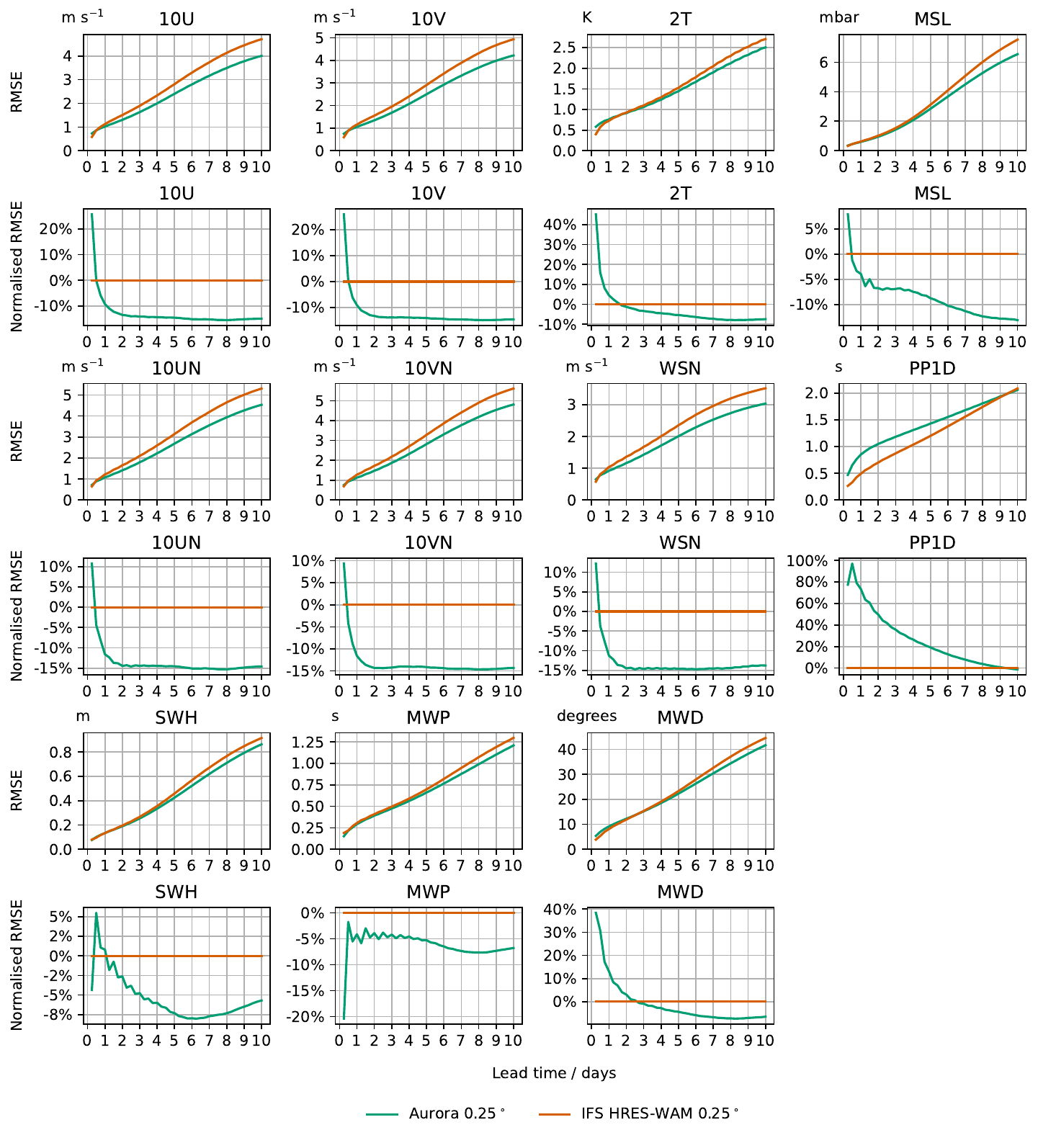}
    \caption{Latitude-weighted root mean square error (RMSE) of Aurora \SI{0.25}{\degree} compared to HRES-WAM \SI{0.25}{\degree} on the meteorological surface-level variables and main wave variables. Shows both unnormalised and normalised RMSEs.}
    \label{fig:wave-rollout-main}
\end{figure}

\begin{figure}[t]
    \centering
    \includegraphics[width=\linewidth]{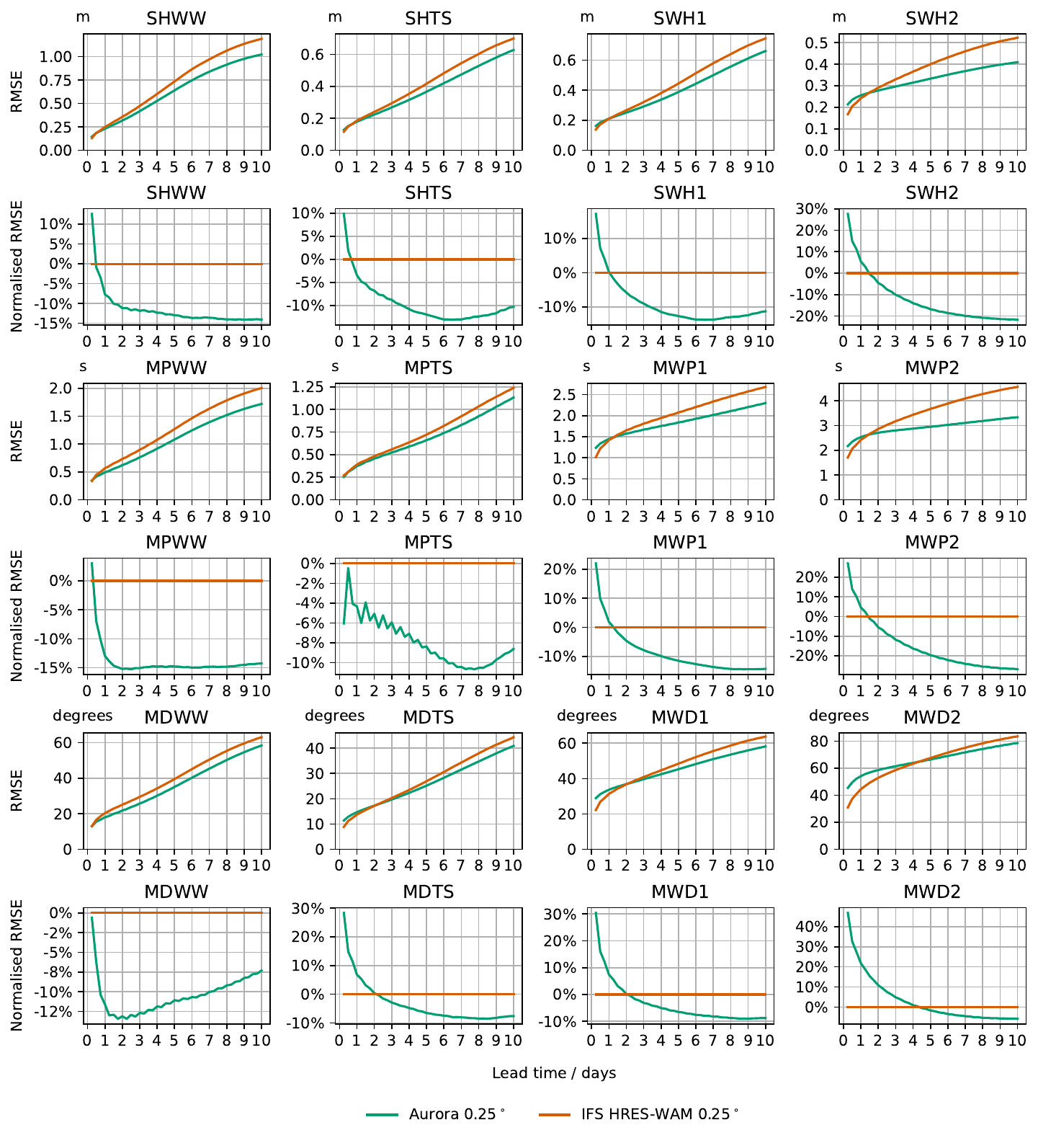}
    \caption{Latitude-weighted root mean square error (RMSE) of Aurora \SI{0.25}{\degree} compared to HRES-WAM \SI{0.25}{\degree} on the wave variables of the various wave components. Shows both unnormalised and normalised RMSEs.}
    \label{fig:wave-rollout-component}
\end{figure}

\Cref{fig:scorecard-wave-full} shows a scorecard for all wave and meteorological variables.
Aurora is competitive (within 20\% RMSE) to HRES-WAM on 95\% of all targets, and Aurora matches or outperforms HRES-WAM on 85\% of all targets.
\Cref{fig:wave-rollout-main,fig:wave-rollout-component} compare Aurora's RMSEs to HRES-WARM for all meteorological and wave surface-level variables.

\FloatBarrier
\subsection{Details for tropical cyclone forecasting}
\label{app:tcs}

In this appendix, we describe the tropical cyclone tracker and provide more detail on how the comparison against various agencies in various regions was performed.

\subsubsection{Baselines and ground truth}
All baseline models from \cref{fig:tcs-scorecard} are summarised in \Cref{tab:tc-baselines}.
For the North Atlantic and East Pacific we compare Aurora to both the official forecast and multiple forecast models used by human forecasters to create these products. In all cases we compare the early version of Aurora (described below) to official and early models, ensuring that the comparison is between products available at the forecast time. Data for the official forecasts and all guidance models is taken from the Automated Tropical Cyclone Forecasting System A-Decks \citepapp{sampson2000automated}. For the Northwest Pacific, multiple agencies issue official forecasts. We compare Aurora to forecasts from the Joint Typhoon Warning Centre, Central Weather Administration Taiwan, China Meteorological Administration, Japan Meteorological Agency and Korea Meteorological Administration. Finally, for the Australian Region, we compare to the offical predictions from the Australian Bureau of Meteorology. 

For ground truth we use the International Best Track Archive for Climate Stewardship (IBTrACS). IBTrACS is a global database that consolidates tropical cyclone data from multiple meteorological agencies worldwide. It provides comprehensive historical records on storm tracks, intensities, and characteristics, supporting climate research and storm impact analysis.

\begin{table}[t]
   \centering
    % \captionsetup{width=\textwidth}
     \caption{
         Baselines for the tropical cyclone tracking experiment.
         NHC refers to National Hurricane Centre.
     }
     \begin{tabular}{l>{\raggedright}p{1cm}p{9cm}}
         \toprule
         Model & Agency & Description \\ 
          \midrule
          \multicolumn{3}{c}{\textsc{North Atlantic and East Pacific}} \\[0.25em]
          OFCL & NHC & Official forecast \\
          OCD5 & NHC & Statistical baseline: climatology and persistence model   \\
          HWFI & NHC &  Multi-layer regional dynamical: Hurricane Weather and Research Forecasting Model    \\
          HMNI & NHC & Multi-layer regional dynamical: Hurricanes in a Multi-Scale Ocean-Coupled Non-Hydrostatic Model   \\
          CTCI & NHC & Multi-layer regional dynamical: COAMPS-TC using GFS initial and boundary conditions  \\
          NVGI & NHC & Multi-layer global dynamical: Navy Global Environmental Model    \\
          EMXI & NHC & Multi-layer global dynamical: ECMWF global model  \\
          CMCI & NHC & Multi-layer global dynamical: Environment Canada global model  \\
          AEMI & NHC & Consensus: GFS ensemble mean \\
          FSSE & NHC & Corrected consensus: FSU super-ensemble   \\
          TVCA & NHC & Consensus: average of at least two of GFSI, EGRI, HFAI, HFBI, EMXI, CTCI, HWFI HMNI EMNI \\
          HCCA & NHC & Corrected consensus: weighted average of AEMI, CTCI, EGRI, EMNI, EMXI, GFSI, HFAI, HFBI, HWFI, and UEMI \\
          \midrule
    \end{tabular}
    \begin{tabular}{l>{\raggedright}p{6cm}p{4cm}}
            \multicolumn{3}{c}{\textsc{Northwest Pacific}} \\[0.25em]
            PGTW & Joint Typhoon Warning Centre & Official forecast \\
            CWA & Central Weather Administration Taiwan & Official forecast \\ 
            BABJ & China Meteorological Administration & Official forecast \\
            RJTD & Japan Meteorological Agency & Official forecast \\
            RKSL & Korea Meteorological Administration & Official forecast \\
            \midrule
            \multicolumn{3}{c}{\textsc{Australian Region (Aus)}} \\[0.25em]
            BoM & Australian Bureau of Meteorology & Official forecast \\
         \bottomrule
     \end{tabular}
     \label{tab:tc-baselines}
 \end{table}

\subsubsection{Considerations for fair comparison to baselines}
In making comparisons to the baselines, we are careful not to advantage Aurora over the baselines in any way.
Two key considerations need to be made: dealing with early versus late models and instances where the tracker fails.

\textbf{Early versus late models.} In tropical cyclone forecasting, models are classified as either early or late depending on when they become available in the forecast cycle.
Track forecast models are initialised at UTC 00, UTC 06, UTC 12 and UTC 18.
Official track forecasts are due three hours after these times.
\emph{Early models} are models whose predictions are available before the end of this three hour window.
Predictions of early models can therefore be used to guide the official prediction.
In contrast, predictions of \emph{late models} become available after the three hour window.
Predictions of late models cannot be used to guide the official prediction.
To use late models to guide the official prediction, predictions from one forecast cycle earlier must be used.
For example, the prediction for the initialisation at UTC 06 can be used guide the official track prediction at UTC 12.

Although Aurora is able to be run in minutes, the assimilation window for the initial conditions used in this study, HRES T0, is up to three hours past the analysis time. As this places Aurora right on the divide between early and late models, out of an abundance of caution we opt to delay all predictions by six hours.
For example, the UTC 12 track prediction from Aurora is generated by running Aurora from the HRES T0 initial conditions from UTC 06.
This ensures that Aurora is an early model and suitable for comparison with the official forecasts. 

\textbf{Tracker failure.}
The tracker used in this study is much simpler than those used operationally. Whilst we manually verified that this simple tracker achieves excellent performance in the vast majority of cases, it occasionally fails, resulting in incorrect predictions.
To not advantage Aurora in any way, we never reject any track prediction, meaning that we also retain the track prediction whenever the tracker fails.
In fact, this slightly disadvantages Aurora in the comparison, as the baselines and IBTrACS use state-of-the-art trackers. 

\subsubsection{Tracker}
\label{app:tcs:tracker}
Our heuristic tracker is based on forecasts by Aurora \SI{0.25}{\degree} (\cref{sm:0.25_extra_results}) initialised with HRES T0.
The tracker attempts to follow a local minimum in predictions for MSL. A challenge in this approach is that MSL may be plagued by multiple local minima due to local topological features like mountains. We therefore adopt a multi-step procedure that resorts to Z700 when detection of the MSL local minimum fails. 

When we say that
the tracker attempts to find the local minimum closest to a reference position in a $x^\circ \times x^\circ$ box, the tracker performs the following procedure.
It first selects a $x^\circ \times x^\circ$ box centred around the reference position.
Within this box,
a point is classified as a local minimum if it attains the lowest value within a further $2^\circ \times 2^\circ$ box centred around that point. 
The tracker then finds the local minimum (according to this classification) in the $x^\circ \times x^\circ$ box closest to the reference position.
If this closest local minimum lies on the edge of the $x^\circ \times x^\circ$ box, then we consider this a failure and say that no local minimum can be found.

The tracker is initialised with the current latitude--longitude position of the TC.
Let the \emph{current position} be the initialisation.
If a $5^\circ\times5^\circ$ box around the current position is clear of land, the tracker attempts to update the current position to the closest local minimum in this box for the \SI{6}{h} prediction of MSL.
If the current position was not updated, the tracker attempts a $4^\circ\times4^\circ$ box, a $3^\circ\times3^\circ$ box, a $2^\circ\times2^\circ$ box, and eventually a $1.5^\circ\times1.5^\circ$ box.
If the current position is then still not updated, the tracker attempts to update the current position to the closest local minimum in a $5^\circ \times 5^\circ$ box for the \SI{6}{h} prediction of Z700, and then repeats the procedure with varying box sizes for MSL.
The resulting current position is the track prediction at lead time \SI{6}{h}.

For all following lead times, a guess $\tilde x_{t+1}$ is made for the new latitude--longitude position by linearly extrapolating the sequence of tracked latitudes and longitudes based on the previous eight estimates (corresponding to two days).
Now call this linear extrapolation the current position.
Using predictions \SI{6}{h} further into the future,
the current position is then updated exactly as for the initialisation.
The resulting current position is the track prediction \SI{6}{h} further into the future.
We repeat this procedure until a track up to five days lead time has been produced.

\subsubsection{Details for Figure \ref{fig:tcs-scorecard}}
\label{app:tcs:details-scorecard}

To compute a column in \cref{fig:tcs-scorecard}, we consider all TCs in 2022 and 2023 in the relevant region.
For every storm, we consider all times when a latitude--longitude position is available in IBTrACS.
Call such a time a \emph{forecast time}.
From all forecast times, we only consider the times when the associated storm is classified as disturbance or tropical.
For every such forecast time, we generate a five-day track forecast with Aurora and consider the track forecast by the model on the $x$-xis.
We then compute the track forecast errors at every lead time where the storm is classified as disturbance or tropical.
This restriction to forecast times and lead times where the the storm is classified as disturbance or tropical is conform the NHC evaluation procedure \citepapp{NHC_Verification_2022}.
By averaging over all storms and forecast times, we obtain a mean absolute error (MAE) at lead times one to five days for both Aurora and the model on the $x$-axis. 
If the model on the $x$-axis happens to have no prediction at a certain lead time for a particular forecast time, then we ignore that particular lead time for that particular forecast time in the mean computation for both Aurora and the model on the $x$-axis.
We note that this slightly advantages the baselines in this comparison, since Aurora is forced to always produce a forecast.
The MAEs at all lead times between Aurora and the model on the $x$-axis are finally compared to produce the column of \cref{fig:tcs-scorecard}.

Call a forecast time for a particular storm a \emph{storm-time} pair.
Following the above procedure, for the North Atlantic and East Pacific, we consider 1333 storm-time pairs from 76 different storms; for the West Pacific, 737 storm-time pairs from 47 different storms; and for the Australian region, 213 storm-time pairs from 35 different storms.

Finally, for every square of \cref{fig:tcs-scorecard}, we compute a 95\% uncertainty interval by resampling entire tropical cyclones with replacement \citepapp{lam2023graphcast}.
Whenever a storm is resampled, all associated storm-time pairs are added to the sample. 
This resampling procedure is conform the statistical structure of the data and approximates the error w.r.t.\ taking the average over infinitely many storms.
If the 95\% uncertainty interval for a square includes zero, ``$\approx$'' is depicted in the middle of the square.

\FloatBarrier
\subsection{Comparison against operational IFS HRES at 0.1\degree{}}
\label{sec:supp_0.1_results}
\begin{figure}[ht]
    \centering
    \includegraphics[width=\linewidth]{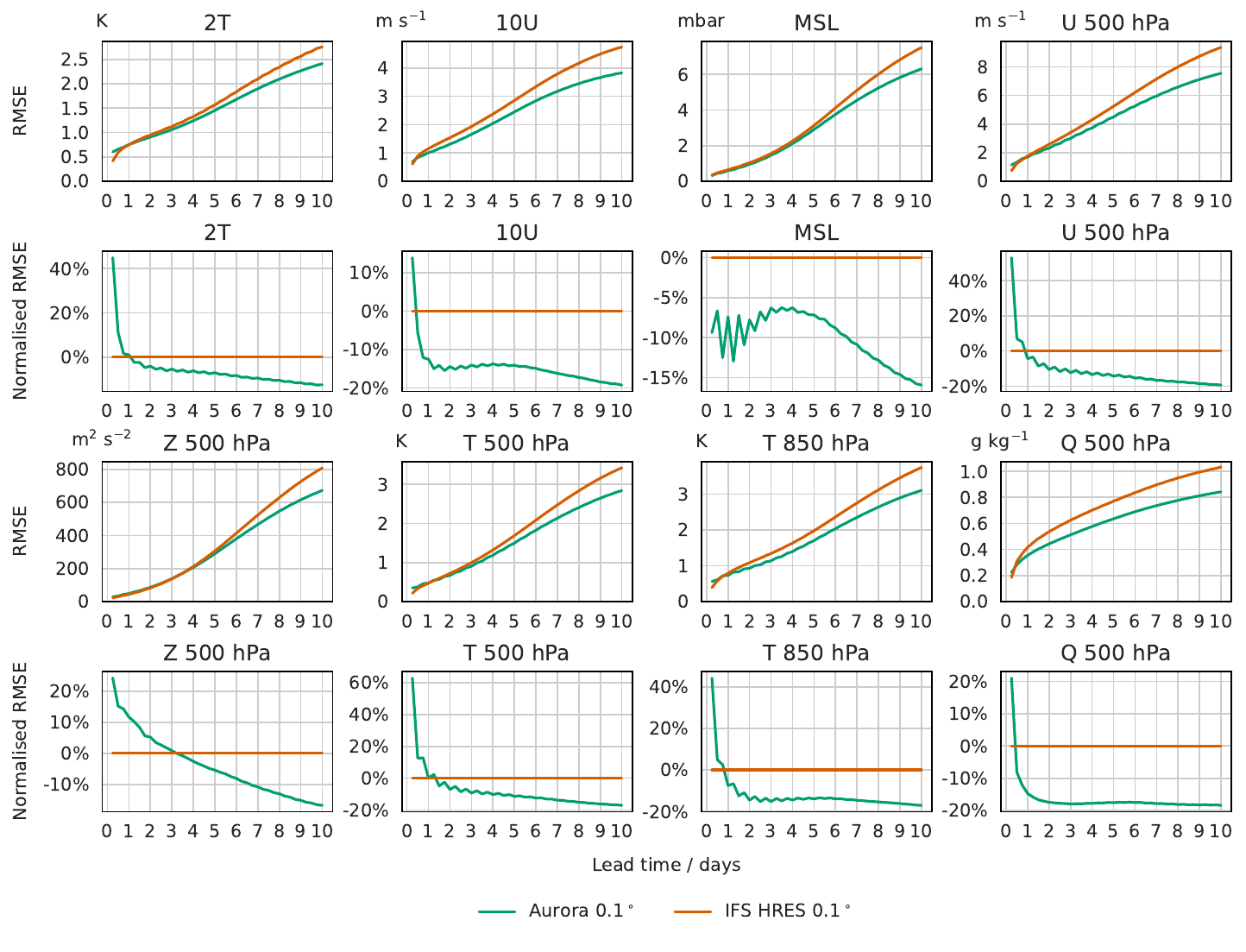}
    \caption{Root mean square error (RMSE) of Aurora \SI{0.1}{\degree} compared to HRES \SI{0.1}{\degree} for headline variables. Shows both unnormalised and normalised RMSEs.}
    \label{fig:headline-hres-0.1}
\end{figure}

\begin{figure}[b]
    \centering
    \includegraphics[width=\linewidth]{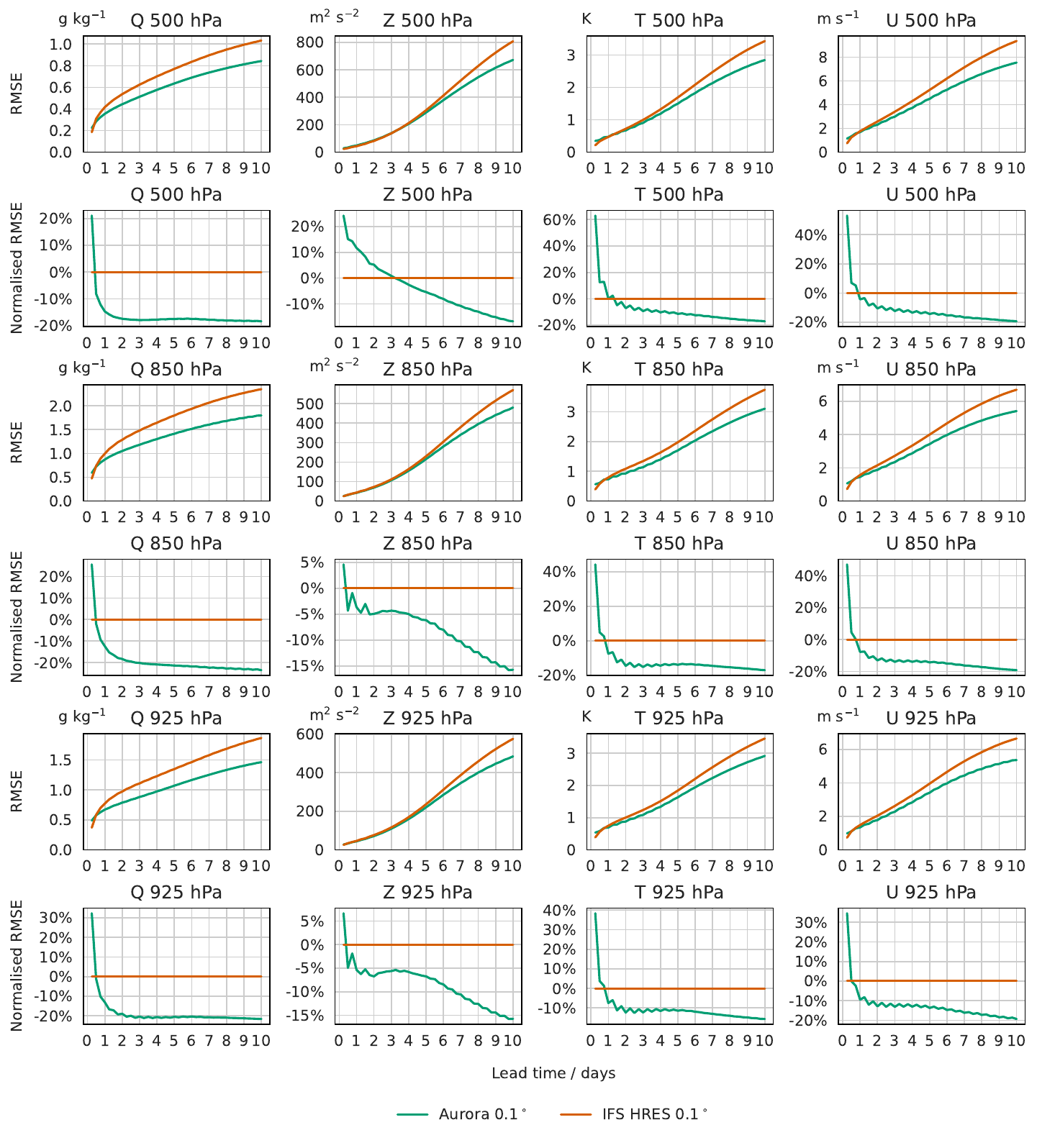}
    \caption{Root mean square error (RMSE) of Aurora \SI{0.1}{\degree} compared to HRES \SI{0.1}{\degree} in the lower atmosphere. Shows both unnormalised and normalised RMSEs.}
    \label{fig:lower-atmos-hres-0.1}
\end{figure}

\begin{figure}[t]
    \centering
    \includegraphics[width=\textwidth]{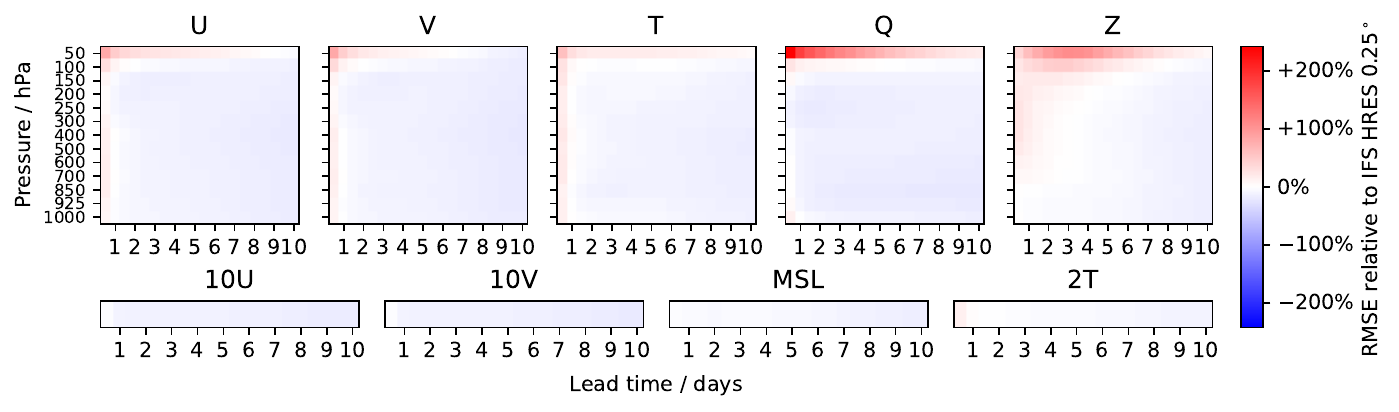}
    \caption{Scorecard comparing Aurora 0.1\degree~vs HRES 0.25\degree~across all levels. We note that the RMSEs of 0.25\degree~HRES are generally lower than those of 0.1\degree~HRES. Nonetheless, Aurora shows good performance and generally outperforms HRES also in the higher atmosphere. }
    \label{fig:scorecard-aurora-01-vs-hres25}
\end{figure}

We include roll-out plots of Aurora at 0.1\degree~compared to HRES forecasts at 0.1\degree~for headline variables (\cref{fig:headline-hres-0.1}) and the lower atmosphere (\cref{fig:lower-atmos-hres-0.1}).
Due to the lack of 0.1\degree~forecasts in the upper atmosphere, we include as a preliminary result a scorecard versus HRES 0.25\degree~in \cref{fig:scorecard-aurora-01-vs-hres25}. Although this is not an apples-to-apples comparison, HRES 0.25\degree~RMSEs tend to be lower than HRES 0.1\degree~RMSEs since higher-resolution models have more fine-grained details that they need to get right. Therefore, the scorecard provides a good approximation to the performance of Aurora 0.1\degree~in the upper atmosphere. 

\begin{figure}[t]
    \centering
    \includegraphics[width=\textwidth]{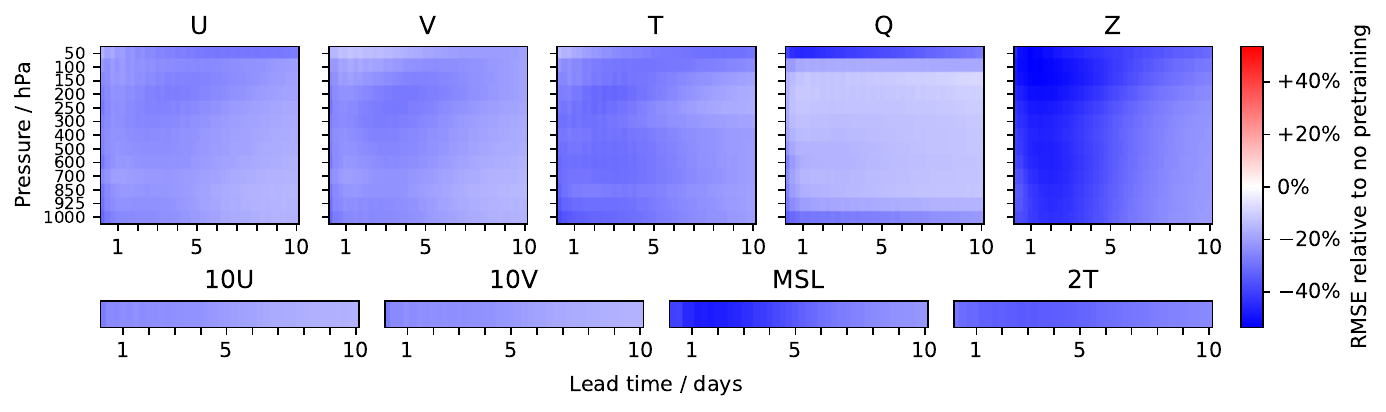}
    \caption{
        Scorecard comparing Aurora 0.1\degree~vs Aurora 0.1\degree~trained from scratch with no pretraining. Aurora 0.1\degree~consistently outperforms the model with no pretraining.
        On average, pretraining improves performance by 25\%.
    }
    \label{fig:scorecard-aurora-01-scratch}
\end{figure}

\paragraph{Aurora 0.1\degree~model trained with no pretraining.}
To demonstrate the impact of pretraining we also include a scorecard (\cref{fig:scorecard-aurora-01-scratch}) comparing the performance of Aurora 0.1\degree~fine-tuned versus. the model trained from scratch on the 0.1\degree~data. The fine-tuned model consistently outperforms a model trained from scratch despite considerable effort being expended to optimise model training without first pretraining. 

\vspace{-2em}

\FloatBarrier
\begin{figure}[ht]
    \centering
    \includegraphics[width=.6\linewidth]{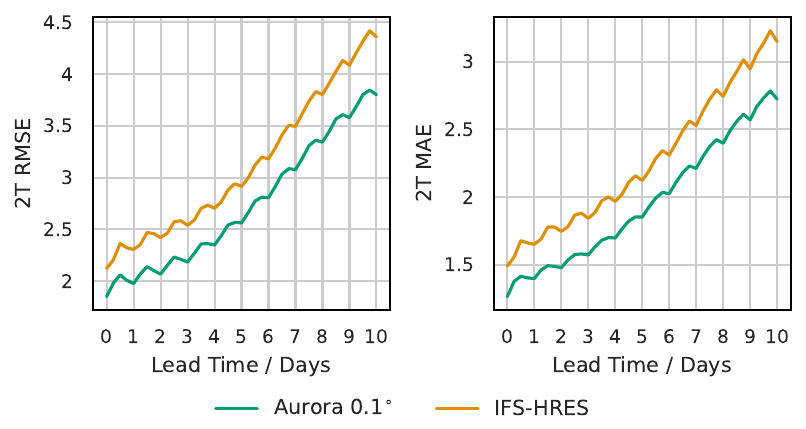}
    \caption{\SI{2}{m} temperature performance at weather station observation points for Aurora 0.1\degree~versus IFS HRES 0.1\degree. RMSE and MAE. Due to the different initialisations, Aurora has an advantage. Nonetheless, Aurora's forecast performance degrades at roughly the same rate as HRES as the lead time increases. }
    \label{fig:app:weather_station_obs}
\end{figure}
\subsection{Comparison against weather station measurements}

\label{app:weather_station_obs}
Most verification efforts for AIWP models have focused on gridded reanalysis data such as ERA5 \citepapp{rasp2023weatherbench}. Not only are weather station observations more reflective of weather experienced by people, but they can also reveal deficiencies in models relying on diagnostic parameters generated by the NWP models producing the analysis. Therefore, we also evaluate Aurora's forecasts on raw observations from the WeatherReal-ISD dataset \citeapp{weatherreal}. This evaluation includes daily forecasts issued at 00 UTC between 2 January 2023 and 19 December 2023.
Aurora outperforms IFS HRES for \SI{10}{m} wind speed forecasts across all lead times up to 10~days (\cref{fig:weather_obs_0.1}). Further results for \SI{2}{m} temperature are available in~\cref{app:extreme_metrics}. This result shows how a global high-resolution AIWP model can set a new state-of-the-art as measured by real weather observations.

Aurora is initialised from the official analysis product, while forecasts are initialised from HRES T0. The former includes an additional assimilation step for the surface, which improves the 2T estimate. Therefore, Aurora has an unfair advantage in terms of initial conditions when compared on 2T (\cref{fig:app:weather_station_obs}). Nonetheless, the errors of HRES and Aurora remain parallel as the lead time increases which shows that the forecast quality of Aurora does not degrade over time compared to HRES. 

\FloatBarrier
\subsection{Benefits of modelling at 0.1\degree~resolution}
\label{sec:0.1_vs_0.25}
In \cref{section:0.1}, we fine-tune Aurora for high-resolution predictions at \SI{0.1}{\degree}.
Instead of fine-tuning at \SI{0.1}{\degree} resolution, one could ask why not simply fine-tune at \SI{0.25}{\degree} and simply interpolate the predictions to \SI{0.1}{\degree} resolution?
The benefit of modelling at \SI{0.1}{\degree} resolution is that the higher resolution can better represent physical processes at small scales, and hence predictions at \SI{0.1}{\degree} can capture these effects, whereas predictions at \SI{0.25}{\degree} could struggle.
The meso-$\beta$ (\SI{20}{km} to \SI{200}{km}) scale exactly represents physical processes that \SI{0.1}{\degree} should be able to capture but \SI{0.25}{\degree} might struggle with, such as mesoscale convective systems and complexes or mountain and valley breezes.
In this appendix, we provide numerical evidence that fine-tuning at \SI{0.1}{\degree} produces superior predictability for physical processes at the meso-$\beta$ scale.

\paragraph{Methodology.}
We compare the predictability of Aurora \SI{0.1}{\degree} to Aurora \SI{0.25}{\degree} regridded to \SI{0.1}{\degree} using the PyPI package \texttt{xesmf} on three different meteorological scales:
the meso-$\beta$ scale (\SI{20}{km} to \SI{200}{km}), the meso-$\alpha$ scale (\SI{200}{km} to \SI{2000}{km}), and the synoptic scale (\SI{2000}{km} and larger).
Predictability is measured with the coefficient of determination, $R^2$, which quantifies precisely which proportion of the data explained by the predictions.
We take HRES analysis at \SI{0.1}{\degree} as the ground truth.
To disentangle $R^2$ at different meteorological scales, we use the spherical harmonic transform from \texttt{torch-harmonics} \citepapp{bonev2023spherical} to decompose the prediction error for both models into errors at different wavelengths.
We then select the appropriate wavelengths for every meteorological scale, and sum the squared amplitudes back to a total MSE for that scale.
The MSE is divided by the total variance of the ground truth in the same wavelength band, and the resulting ratio is equal to $1 - R^2$.
The difference in $R^2$, $R^2_{\text{Aurora \SI{0.1}{\degree}}} - R^2_{\text{Aurora \SI{0.25}{\degree}}}$, finally quantifies the difference in predictability between Aurora \SI{0.1}{\degree} and Aurora \SI{0.25}{\degree} regridded to \SI{0.1}{\degree}.

\begin{figure}[tp]
    \centering
    \includegraphics[width=\linewidth]{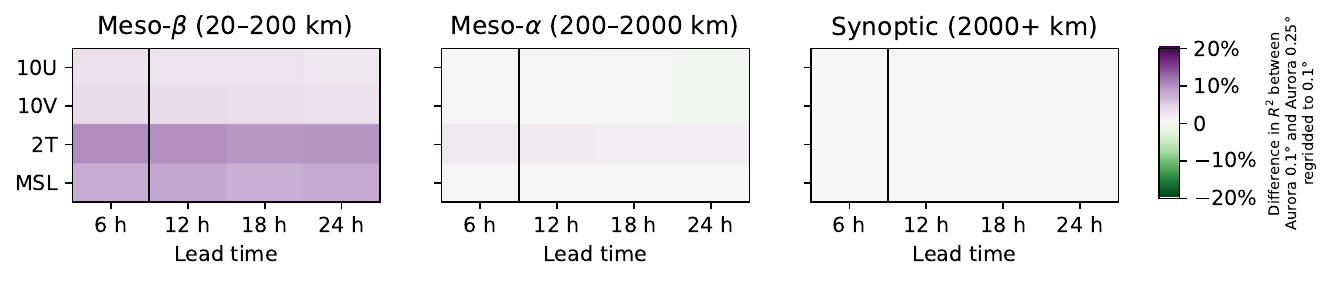}
    \caption{
        At various meteorological scales, comparison of $R^2$, a measure of predictability, of Aurora 0.1\degree~to Aurora 0.25\degree~regridded to 0.1\degree~resolution. IFS HRES analysis at \SI{0.1}{\degree} is used as the ground truth.
        $R^2$ is estimated from predictions for UTC 00 and UTC 12 from the period Jan 2023 to Mar 2023 inclusive.
        Purple means that Aurora \SI{0.1}{\degree} has better predictability,
        and green means that Aurora \SI{0.25}{\degree} regridded to \SI{0.1}{\degree} has better predictability.
    }
    \label{fig:0.1_vs_0.25_spectra}
\end{figure}

\paragraph{Results and conclusion.}
Aurora \SI{0.1}{\degree} demonstrates substantially improved predictability for meso-$\beta$-scale physical processes  (a 5.4\% increase in $R^2$ on average), whereas predictability for meso-$\alpha$-scale and synoptic physical processes remain roughly unchanged (differences of 0.3\% and 0.03\% on average) (see \cref{fig:0.1_vs_0.25_spectra}).
That Aurora \SI{0.1}{\degree} shows substantially improved predictability on the meso-$\beta$ scale is not a surprising finding, 
because this is exactly the scale that a resolution of \SI{0.25}{\degree} cannot effectively capture, but a resolution of \SI{0.1}{\degree} can.

\FloatBarrier
\subsection{Storm Ciar\'an} 
\label{section:sm-ciaran}

\begin{figure}[th]
    \centering
     \includegraphics[width=0.49\textwidth]{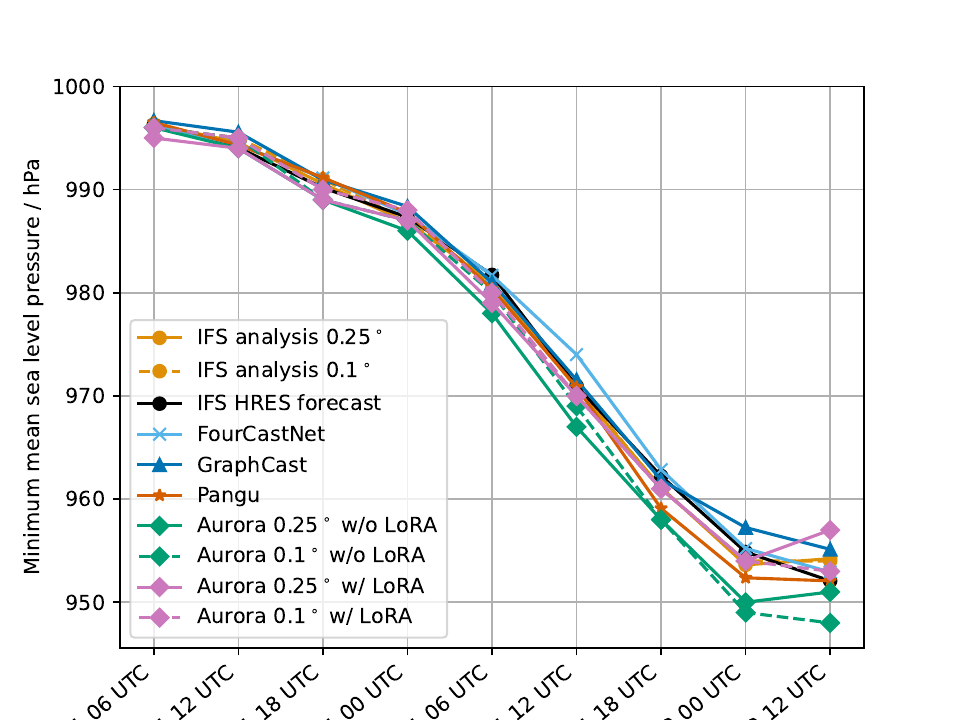}
    \includegraphics[width=0.49\textwidth]{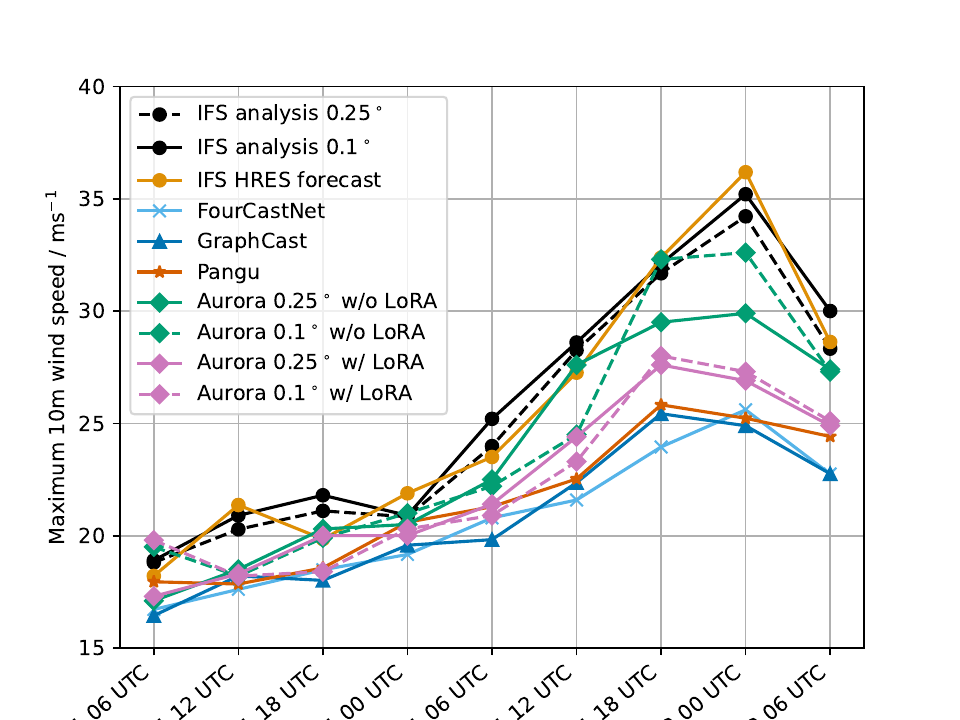}
    \caption{
        Minimum mean sea level pressure (left) and maximum \SI{10}{m} wind speed (right) for Storm Ciar\'an, at 0.25\degree~and 0.1\degree~resolution, both with and without LoRA fine-tuning.}
    \label{fig:ciaran-sm-allmodels}
\end{figure}
% \vspace{-2em}

\begin{table}[th]
    \centering
    \captionsetup{width=\linewidth}
    \caption{
        Bounding boxes that track Storm Ciar\'an.
        We used these bounding boxes to compute the maximum \SI{10}{m} wind speed and minimum MSL.
    }
    \label{tab:ciaran-bounding-boxes}
    \begin{tabular}{lll}
        \toprule
        Time range & Latitudes & Longitudes \\ \midrule
        2023-10-30 18 UTC to 2023-10-31 06 UTC
            & $[35\degree, 48\degree]$
            & $[-75\degree, -48\degree]$ \\ 
        2023-10-31 12 UTC to 2023-11-01 00 UTC
            & $[40\degree, 48\degree]$
            & $[-50\degree, -25\degree]$ \\ 
        2023-11-01 06 UTC to 2023-11-01 12 UTC
            & $[35\degree, 50\degree]$
            & $[-65\degree, -10\degree]$ \\ 
        2023-11-01 18 UTC to 2023-11-02 00 UTC
            & $[35\degree, 60\degree]$
            & $[-25\degree, 10\degree]$ \\ 
        2023-11-02 06 UTC and later
            & $[35\degree, 60\degree]$
            & $[-10\degree, 10\degree]$ \\ 
        \bottomrule
    \end{tabular}
\end{table}

Storm Ciar\'an was a high-impact windstorm which took place across North-West Europe in late 2023. 
In \cref{fig:ciaran-maxwindspeed-main}, we reproduce the findings by \citeapp{charltonperez2024ciaran} and include predictions from Aurora at 0.1\degree~resolution. 
All models are initialized on 31 Oct 00 UTC using the latest cycle of IFS analysis, CY48R1. 
We observe that among the AI models tested, Aurora stands out as the only one capable of accurately predicting the abrupt rise in maximum \SI{10}{m} wind speed. 
It forecasts approximately 7 ms$^{-1}$ higher than the other AI techniques, closely matching IFS analysis, which we take to be the ground truth. 
Notably, FourCastNet \citepapp{pathak2022fourcastnet}, GraphCast \citepapp{lam2023graphcast}, and Pangu-Weather \citepapp{bi2022pangu} are constrained to a 0.25\degree~resolution, potentially hindering their ability to capture rapid onset phenomena. However, even at 0.25\degree~resolution, Aurora demonstrates some capability in capturing this sharp increase, albeit to a lesser degree than its 0.1\degree~counterpart; see \cref{fig:ciaran-sm-allmodels}. 

\cref{fig:ciaran-accuracy} shows Storm Ciar\'an's intensification over a \SI{24} hour period, where we compare Aurora forecasts (top) to IFS analysis (bottom) visually in terms of \SI{10}{m} wind speed. We also plot the minimum MSL, corresponding to the center of the storm, and find that the Aurora's forecasts are similar to IFS HRES analysis. 
The primary difference lies in the level of detail: IFS analysis captures higher spatial frequency content, whereas Aurora's forecasts are smoother, although less so than those from the 0.25\degree~model. Despite this disparity, the forecasts are visually similar.

\paragraph{Two versions of Aurora.}
We compare Aurora variants with and without LoRA fine-tuning for predicting Storm Ciar\'an. 
\cref{fig:ciaran-sm-allmodels} shows minimum mean sea level pressure (left) and maximum \SI{10}{m} wind speed (right), including the two variants of Aurora, namely, with and without LoRA fine-tuning, at both 0.25\degree~and 0.1\degree~resolution. 
We evaluate all methods based on how closely they resemble IFS analysis at the corresponding resolution, and find that, as expected, the Aurora variants at 0.1\degree~resolution clearly outperform those at 0.25\degree~resolution. 

For minimum mean sea level pressure, the variants with LoRA fine-tuning perform better (i.e., align more closely with IFS analysis). However, for maximum 10m wind speed, the variants without LoRA fine-tuning perform better, capturing the peak wind speeds more accurately. 
We see that Aurora at 0.1\degree~(with LoRA fine-tuning) aligns almost perfectly with IFS analysis at 0.1\degree. 
In \cref{section:0.1}, we showed that Aurora is the only AIWP model capable of capturing the peak in maximum \SI{10}{m} wind speed on 2 November 2023 at 00 UTC, where the best version of Aurora is at 0.1\degree, without LoRA fine-tuning. 
\Cref{tab:ciaran-bounding-boxes} tabulates the bounding boxes used to compute the maximum \SI{10}{m} wind speed and minimum MSL.

We postulate that LoRA is useful when computing statistics of smooth quantities with less extreme activity (e.g., MSL), where the loss of detail and additional smoothing is less critical; while turning LoRA off is better for computing statistics such as wind speed, where the loss of detail can substantially affect the statistics by suppressing the extrema. This behavior is further supported by our spectral analysis in \cref{sec:power-spectra}, where we show how LoRA affects the power spectrum of the model's predictions at different lead times (\cref{fig:power-lora}).

In summary, the use of LoRA during roll-out fine-tuning provides us with a mechanism to control the trade-off between RMSE performance and the retention of high-frequency details in long-term predictions. This flexibility is valuable in operational settings where different types of predictions may be required for different applications.\\
% One could reasonably compare Aurora without LoRA to IFS HRES and Aurora with LoRA to the IFS ENS mean.

% \FloatBarrier
\subsection{Power spectra}
\label{sec:power-spectra}

To assess the quality of predictions, a useful characteristic is the power spectrum. This section contains plots of power spectra for predictions and targets calculated using the method described in \citeapp{lam2023graphcast}. 

\begin{figure}[t]
    \centering
    \includegraphics[width=\columnwidth]{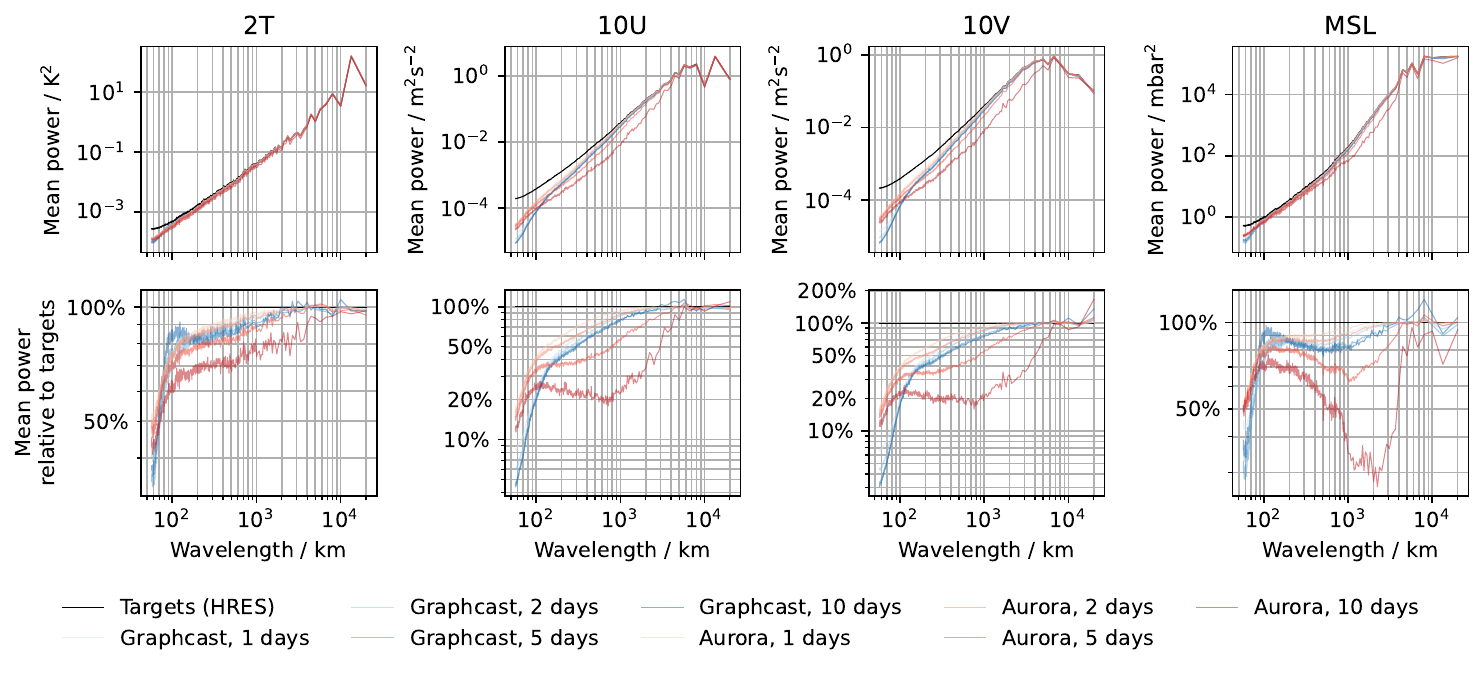}
    \caption{Power spectra of Aurora's predictions and Graphcast's predictions for 2022. The first row plots the mean power for predictions. The columns represent 2T, 10U, 10V, and MSL, respectively. The second row plots the mean power relative to the HRES targets. Each individual chart illustrates the change in the power spectrum for lead times of 1, 2, 5, and 10 days.}
    \label{fig:power-gc-and-aurora}
\end{figure}

\begin{figure}[t]
    \centering
    \includegraphics[width=\columnwidth]{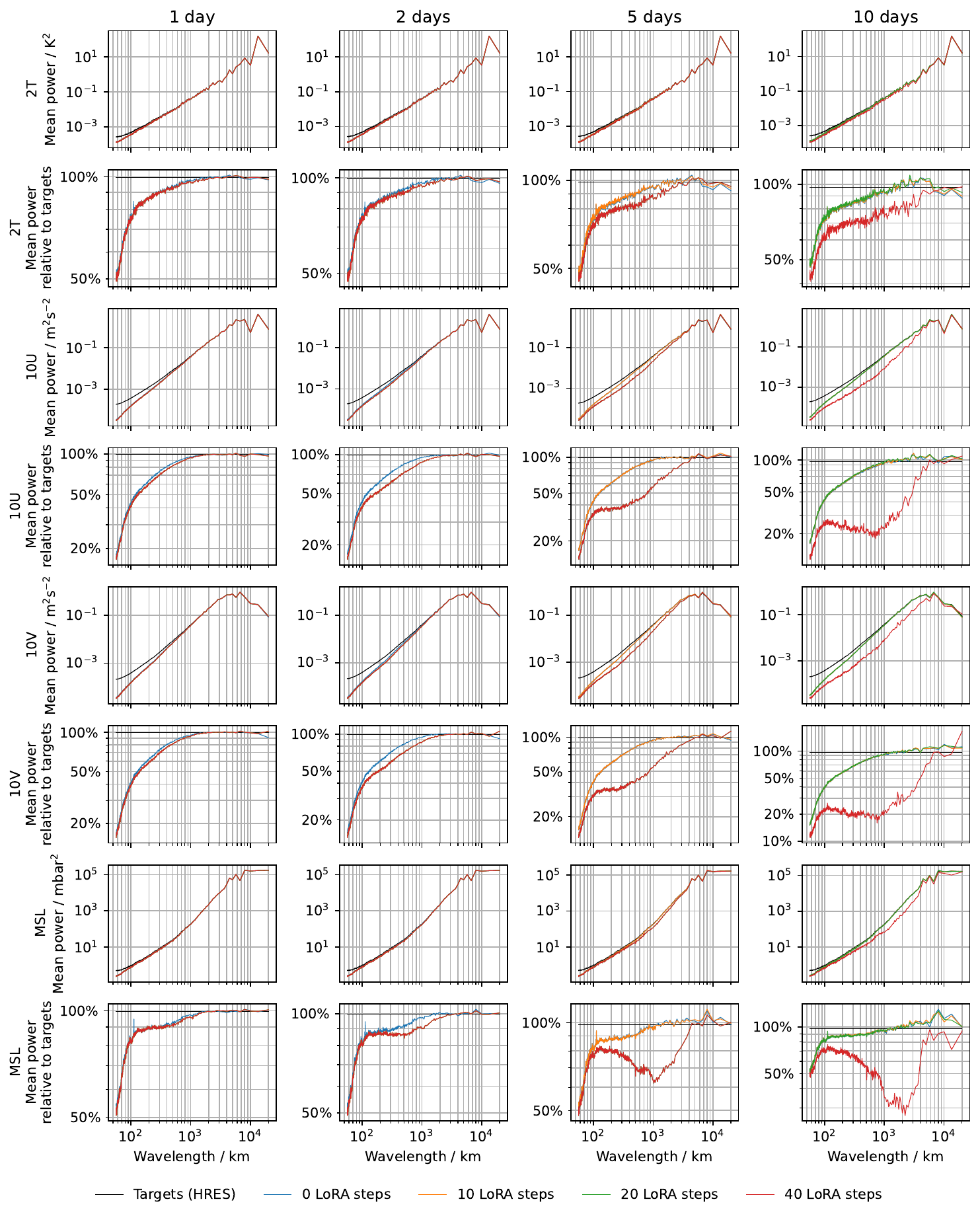}
    \caption{Power spectra of Aurora's predictions with different numbers of LoRA steps for 2022. The first row plots the mean power for 2T predictions. Each column represents a different lead time of 1, 2, 5, or 10 days, respectively. The second row plots the mean power for 2T relative to the HRES targets. Similarly, subsequent pairs of plots represent, respectively, the power and relative power of predictions of 10U, 10V, and MSL.}
    \label{fig:power-lora}
\end{figure}

At short wavelengths (i.e.\ high frequencies) Aurora's forecasts are uniformly higher in power than those of GraphCast, which means less blurring (\cref{fig:power-gc-and-aurora}). In the mid-range of the spectrum, Aurora's forecasts have higher power than those of GraphCast for lead times of 1 and 2 days, but then drop in power below GraphCast's forecasts with 5 and 10 days lead times. In general, as more LoRA steps are added, the power in the forecast is reduced, especially in the mid-range of the spectrum (\cref{fig:power-lora}). For the longest lead time of 10 days and with 40 LoRA steps, there is a clear reduction in power at the short wavelengths (i.e.\ high frequencies) which results in blurring.

When comparing the spectrum of Aurora between 0.25\degree~and 0.1\degree, the power of the 0.1\degree~forecasts is generally lower than the power of the 0.25\degree~forecasts. However, the power of the 0.1\degree~forecasts extends to shorter wavelengths (i.e.\ higher frequencies) than 0.25\degree~forecasts, reflecting the higher resolution of the latter. These observations are in agreement with the discussion provided in \cref{section:sm-ciaran}, where we pointed out that the difference between Aurora forecasts with and without LoRA fine-tuning can be viewed analogously to the difference between the forecasts produced by the IFS ensemble mean and IFS HRES.

\begin{figure}[t]
    \centering
    \includegraphics[width=\columnwidth]{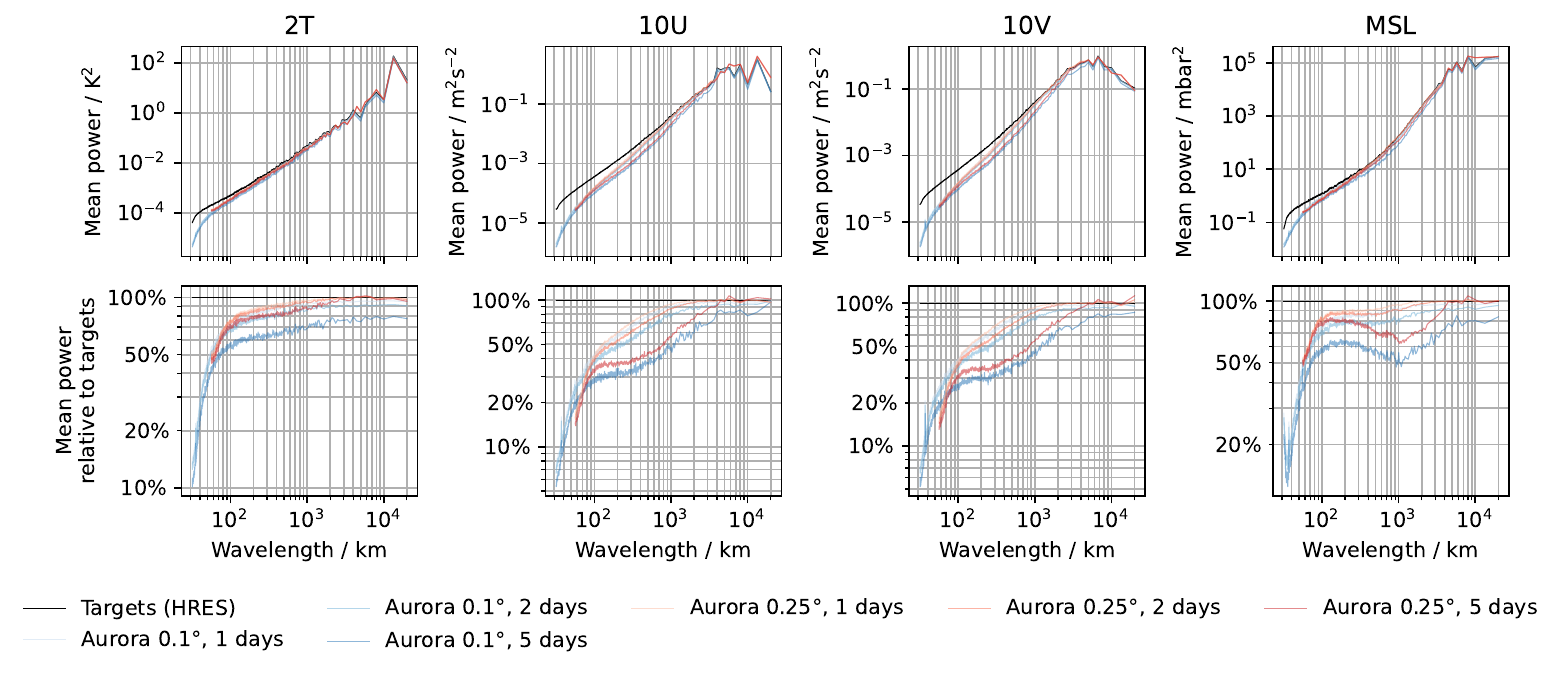}
    \caption{Power spectra of Aurora's predictions at both \SI{0.1}{\degree} and \SI{0.25}{\degree} resolutions for January 2022. The first row plots the mean power for predictions. The columns represent 2T, 10U, 10V, and MSL, respectively. The second row plots the mean power relative to the HRES targets.}
    \label{fig:power-9km-25km}
\end{figure}

\FloatBarrier

% An appendix contains supplementary information that is not an essential part of the text itself but which may be helpful in providing a more comprehensive understanding of the research problem or it is information that is too cumbersome to be included in the body of the paper.




\end{appendices}


\bibliographystyleapp{sn-mathphys-num}
\bibliographyapp{bibliography}% common bib file


\end{document}